\documentclass[10pt,conference,letterpaper]{IEEEtran}

\usepackage{subfigure}
\usepackage{graphicx}
\usepackage{rotating}
\usepackage{boxedminipage}
\usepackage{amssymb}
\usepackage{pifont}
\usepackage{booktabs}
\usepackage{color}
\usepackage{colortbl}
\usepackage{array}
\usepackage{tabularx}
\usepackage{url}

\definecolor{darkgray}{rgb}{0.88,0.88,0.88}
\definecolor{lightgray}{rgb}{0.96,0.96,0.96}

\IEEEoverridecommandlockouts

\begin{document}

\title{SP{\LARGE ${\mathbf{^2}}$}Bench: A SPARQL Performance Benchmark}
\author{
{Michael Schmidt$^*$\thanks{$^*$The work of this author was funded by DFG grant GRK 806/2}{\small $~^{\sharp}$}, Thomas Hornung{\small $~^{\sharp}$}, Georg Lausen{\small $~^{\sharp}$}, Christoph Pinkel{\small $~^{\natural}$} }%
\vspace{1.6mm}\\
\fontsize{10}{10}\selectfont\itshape
$~^{\sharp}$Freiburg University\\
Georges-Koehler-Allee 51, 79110 Freiburg, Germany\\
\fontsize{9}{9}\selectfont\ttfamily\upshape
\{mschmidt|hornungt|lausen\}@informatik.uni-freiburg.de
\vspace{1.6mm}\\
\fontsize{10}{10}\selectfont\rmfamily\itshape
$^{\natural}$MTC Infomedia OHG\\
Kaiserstrasse 26, 66121 Saarbr\"ucken, Germany\\
\fontsize{9}{9}\selectfont\ttfamily\upshape
c.pinkel@mtc-infomedia.de
}

\maketitle

\begin{abstract}
Recently, the SPARQL query language for RDF has reached the W3C
recommendation status. In response to this emerging standard, the
database community is currently exploring efficient storage techniques
for RDF data and evaluation strategies for SPARQL queries. A meaningful
analysis and comparison of these approaches necessitates a comprehensive
and universal benchmark platform. To this end, we have developed SP$^2$Bench,
a publicly available, language-specific SPARQL performance benchmark.
SP$^2$Bench is settled in the DBLP scenario and comprises both a data
generator for creating arbitrarily large DBLP-like documents and a set
of carefully designed benchmark queries. The generated documents mirror
key characteristics and social-world distributions encountered in the
original DBLP data set, while the queries implement meaningful requests
on top of this data, covering a variety of SPARQL operator constellations
and RDF access patterns. As a proof of concept, we apply SP$^2$Bench to
existing engines and discuss their strengths and weaknesses that
follow immediately from the benchmark results.
\end{abstract} 

\section{Introduction}
\label{sec:introduction}

The Resource Description Framework~\cite{rdfconceptsw3c} (RDF)
has become the standard format for encoding machine-readable information
in the Semantic Web~\cite{semanticweb}. RDF databases can be represented
by labeled directed graphs, where each edge connects a so-called
{\it subject} node to an {\it object} node under label {\it predicate}.
The intended semantics is that the {\it object} denotes the value of
the {\it subject}'s property {\it predicate}. Supplementary to RDF,
the W3C has recommended the declarative SPARQL~\cite{sparqlw3c} query
language, which can be used to extract information from RDF graphs.
SPARQL bases upon a powerful graph matching facility, allowing to
bind variables to components in the input RDF graph. In addition,
operators akin to relational joins, unions, left outer joins, selections,
and projections can be combined to build more expressive queries.

By now, several proposals for the efficient evaluation of SPARQL
have been made. These approaches comprise a wide range of optimization
techniques, including normal forms~\cite{pag2006}, graph pattern reordering
based on selectivity estimations~\cite{ssbkr2008} (similar to relational
join reordering), syntactic rewriting~\cite{hh2007}, specialized
indices~\cite{ggl2007,hd2005} and storage
schemes~\cite{ackp2001,bkh2002,hg2003,ammh2007,wkb2008} for RDF, and Semantic
Query Optimization~\cite{lms2008}. Another viable option is
the translation of SPARQL into SQL~\cite{c2005,cljf2005} or
Datalog~\cite{p2007}, which facilitates the evaluation with
traditional engines, thus falling back on established optimization
techniques implemented in conventional engines.

As a proof of concept, most of these approaches have been evaluated
experimentally either in user-defined scenarios, on top of the
LUBM benchmark~\cite{gph2004}, or using the Barton Library
benchmark~\cite{barton}. We claim that none of these scenarios
is adequate for testing SPARQL implementations in a general and
comprehensive way: On the one hand, user-defined scenarios are
typically designed to demonstrate very specific properties and,
for this reason, lack generality. On the other hand, the Barton
Library Benchmark is application-oriented, while LUBM was primarily
designed to test the reasoning and inference mechanisms of Knowledge
Base Systems. As a trade-off, in both benchmarks central SPARQL operators
like \textsc{Optional} and \textsc{Union}, or solution modifiers
are not covered.


With the {\bf SP}ARQL {\bf P}erformance {\bf Bench}mark (SP$^2$Bench)
we propose a language-specific benchmark framework specifically designed
to test the most common SPARQL constructs, operator constellations,
and a broad range of RDF data access patterns. The SP$^2$Bench data
generator and benchmark queries are available for download in a
ready-to-use
format.\footnote{http://dbis.informatik.uni-freiburg.de/index.php?project=SP2B}

In contrast to application-specific benchmarks, SP$^2$Bench aims at
a comprehensive performance evaluation, rather than assessing the
behavior of engines in an application-driven scenario. Consequently,
it is not motivated by a single use case, but instead covers a broad
range of challenges that SPARQL engines might face in different contexts.
In this line, it allows to assess the generality of optimization approaches
and to compare them in a universal, application-independent setting.
We argue that, for these reasons, our benchmark provides excellent support
for testing the performance of engines in a comprising way, which might
help to improve the quality of future research in this area. We emphasize
that such language-specific benchmarks (e.g., XMark~\cite{swkcmb2002})
have found broad acceptance, in particular in the research community.

It is quite evident that the domain of a language-specific benchmark
should not only constitute a representative scenario that captures
the philosophy behind the data format, but also leave room for
challenging queries. With the choice of the DBLP~\cite{dblp}
library we satisfy both desiderata. First, RDF has
been particularly designed to encode metadata, which makes DBLP an
excellent candidate. Furthermore, DBLP reflects interesting
social-world distributions (cf.~\cite{el2005}), and hence captures
the social network character of the Semantic Web, whose idea is to
integrate a great many of small databases into a global semantic
network. In this line, it facilitates the design of interesting
queries on top of these distributions. 

Our data generator supports the creation of arbitrarily large DBLP-like
models in RDF format, which mirror vital key characteristics and
distributions of DBLP. Consequently, our framework combines the
benefits of a data generator for creating arbitrarily large documents
with interesting data that contains many real-world characteristics,
i.e.~mimics natural correlations between entities, such as power
law distributions
(found in the citation system or the distribution of papers among
authors) and limited growth curves (e.g., the increasing number of
venues and publications over time). For this reason our generator
relies on an in-depth study of DBLP, which comprises the analysis of
entities (e.g.~articles and authors), their properties, frequency,
and also their interaction.

Complementary to the data generator, we have designed~17~meaningful
queries that operate on top of the generated documents. They cover
not only the most important SPARQL constructs and operator constellations, 
but also vary in their characteristics, such as complexity and result
size. The detailed knowledge of data characteristics plays a crucial
role in query design and makes it possible to predict the challenges
that the queries impose on SPARQL engines. This, in turn,
facilitates the interpretation of benchmark results.

The key contributions of this paper are the following.\vspace{-0.05cm}

\begin{itemize}

\item We present SP$^2$Bench, a comprehensive benchmark for
the SPARQL query language, comprising a data generator and
a collection of~17~benchmark queries.

\item Our generator supports the creation of arbitrarily large
DBLP documents in RDF format, reflecting key characteristics
and social-world relations found in the original DBLP database. The
generated documents cover various RDF constructs, such as blank
nodes and containers.

\item The benchmark queries have been carefully designed to test a
variety of operator constellations, data access patterns, and optimization
strategies. In the exhaustive discussion of these queries we also
highlight the specific challenges they impose on SPARQL engines.  

\item As a proof of concept, we apply SP$^2$Bench to selected SPARQL
engines and discuss their strengths and weaknesses that follow from
the benchmark results. This analysis confirms that our benchmark
is well-suited to identify deficiencies in SPARQL implementations.

\item We finally propose performance metrics that capture different
aspects of the evaluation process.
\end{itemize}

{\bf Outline.} 
We next discuss related work and design decisions in
Section~\ref{sec:relatedwork}. The analysis of DBLP in
Section~\ref{sec:dblp} forms the basis for our data generator in
Section~\ref{sec:datagen}. Section~\ref{sec:queries} gives an
introduction to SPARQL and describes the benchmark queries. The
experiments in Section~\ref{sec:experiments} comprise a short evaluation
of our generator and benchmark results for existing SPARQL
engines. We conclude with some final remarks in Section~\ref{sec:conclusion}.

\section{Benchmark Design Decisions}
\label{sec:relatedwork}

{\bf Benchmarking.}
The Benchmark Handbook~\cite{g1993} provides a summary of important
database benchmarks. Probably the most ``complete'' benchmark suite
for relational systems is TPC\footnote{See http://www.tpc.org.}, which defines
performance and correctness benchmarks for a large variety of scenarios.
There also exists a broad range of benchmarks for
other data models, such as object-oriented databases
(e.g., OO7~\cite{cdn1993}) and XML (e.g., XMark~\cite{swkcmb2002}).

Coming along with its growing importance, different benchmarks for
RDF have been developed. The Lehigh University Benchmark~\cite{gph2004}
(LUBM) was designed with focus on inference and reasoning capabilities of
RDF engines. However, the SPARQL specification~\cite{sparqlw3c}
disregards the semantics of RDF and RDFS~\cite{rdfschemaw3c,rdfsemanticsw3c},
i.e.~does not involve automated reasoning on top of RDFS constructs such
as subclass and subproperty relations. With this regard, LUBM does not
constitute an adequate scenario for SPARQL performance evaluation.
This is underlined by the fact that central SPARQL operators,
such as \textsc{Union} and \textsc{Optional}, are not addressed in LUBM.

The Barton Library benchmark~\cite{barton} queries implement a
user browsing session through the RDF Barton online catalog. By design,
the benchmark is application-oriented. All queries are encoded
in SQL, assuming that the RDF data is stored in a relational DB.
Due to missing language support for aggregation, most queries cannot
be translated into SPARQL. On the other hand, central SPARQL features
like left outer joins (the relational equivalent of SPARQL operator
\textsc{Optional}) and solution modifiers are missing. In summary,
the benchmark offers only limited support for testing native SPARQL engines.

The application-oriented Berlin SPARQL Benchmark~\cite{bsbm} (BSBM)
tests the performance of SPARQL engines in a prototypical e-commerce
scenario. BSBM is use-case driven and does not particularly address
language-specific issues. With its focus, it is supplementary to the
SP$^2$Bench framework. 

The RDF(S) data model benchmark in~\cite{macp2002} focuses on structural
properties of RDF Schemas. In~\cite{ttkc2008} graph features of RDF Schemas
are studied, showing that they typically exhibit power law distributions
which constitute a valuable basis for synthetic schema generation.
With their focus on schemas, both~\cite{macp2002} and~\cite{ttkc2008}
are complementary to our work.

A synthetic data generation approach for OWL based on test data is
described in~\cite{DBLP:conf/semweb/WangGQH05}. There, the focus is on rapidly
generating large data sets from representative data of a fixed domain. Our data
generation approach is more fine-grained, as we analyze the development
of entities (e.g.~articles) over time and reflect many characteristics found
in social communities.

{\bf Design Principles.} In the Benchmark Handbook~\cite{g1993}, four key
requirements for domain specific benchmarks are postulated, i.e.~it
should be (1)~{\it relevant}, thus testing typical operations within the
specific domain, (2)~{\it portable}, i.e.~should be executable on different
platforms, (3)~{\it scalable}, e.g.~it should be possible to run the
benchmark on both small and very large data sets, and last but not least
(4)~it must be {\it understandable}.

For a language-specific benchmark, the relevance requirement~(1) suggests
that queries implement realistic requests on top of the data.
Thereby, the benchmark should not focus on correctness verification,
but on common operator constellations that impose particular challenges.
For instance, two SP$^2$Bench queries test negation, which (under
closed-world assumption) can be expressed in SPARQL through a combination
of operators \textsc{Optional}, \textsc{Filter}, and~\textsc{bound}.

Requirements (2) portability and (3) scalability bring along
technical challenges concerning the implementation of the data generator.
In response, our data generator is deterministic,
platform independent, and accurate w.r.t.~the desired size of generated documents.
Moreover, it is very efficient and gets by with a constant amount of main memory,
and hence supports the generation of arbitrarily large RDF documents.

From the viewpoint of engine developers, a benchmark should give
hints on deficiencies in design and implementation. This is where
(4)~understandability comes into play, i.e.~it is
important to keep queries simple and understandable. At the same time,
they should leave room for diverse optimizations. In this regard, the
queries are designed in such a way that they are amenable to a wide
range of optimization strategies.

{\bf DBLP.}
We settled SP$^2$Bench in the DBLP~\cite{dblp} scenario.
The DBLP database contains bibliographic information about the field of
Computer Science and, particularly, databases.

In the context of semi-structured data one often distinguishes
between data- and document-centric scenarios. Document-centric design
typically involves large amounts of free-form text, while data-centric
documents are more structured and usually processed by machines rather
than humans. RDF has been specifically designed for encoding
information in a machine-readable way, so it basically follows the
data-centric approach. DBLP, which contains structured data and little
free text, constitutes such a data-centric scenario.

As discussed in the Introduction, our generator mirrors vital real-world
distributions found in the original DBLP data. This constitutes
an improvement over existing generators that create purely synthetic data,
in particular in the context of a language-specific
benchmark. Ultimately, our generator might also be useful in other
contexts, whenever large RDF test data is required. We point out that
the DBLP-to-RDF translation of the original DBLP data in~\cite{d2rdblp}
provides only a fixed amount of data and, for this reason, is not
sufficient for our purpose.

We finally mention that sampling down large, existing data sets such as 
U.S.~Census\footnote{http://www.rdfabout.com/demo/census/}
(about $1$ billion triples) might be another reasonable option to
obtain data with real-world characteristics. The disadvantage, however,
is that sampling might destroy more complex distributions in
the data, thus leading to unnatural and ``corrupted'' RDF graphs.
In contrast, our decision to build a data generator from scratch allows us
to customize the structure of the RDF data, which is in line with the
idea of a comprehensive, language-specific benchmark. This way, we
easily obtain documents that contain a rich set of RDF constructs, such
as blank nodes or containers.

\section{The DBLP Data Set}
\label{sec:dblp}
The study of the DBLP data set in this section lays the foundations
for our data generator.
The analysis of frequency distributions
in scientific production has first been discussed in~\cite{l1926}, and
characteristics of DBLP have been investigated in~\cite{el2005}. The
latter work studies a subset of DBLP, restricting DBLP to publications
in database venues. It is shown that (this subset of) DBLP
reflects vital social relations, forming a ``small world'' on its
own. Although this analysis forms valuable groundwork, our approach
is of more pragmatic nature, as we approximate distributions by concrete
functions.

We use function families that naturally reflect the scenarios,
e.g.~logistics curves for modeling limited growth or power equations for
power law distributions. All approximations have been done with
the {\it ZunZun}\footnote{http://www.zunzun.com} data modeling tool
and the {\it gnuplot}\footnote{http://www.gnuplot.info} curve fitting module.
Data extraction from the DBLP XML data was realized with the 
MonetDB/XQuery\footnote{http://monetdb.cwi.nl/XQuery/} processor.

An important objective of this section is also to provide
insights into key characteristics of DBLP data. Although it is
impossible to mirror all relations found in the original data,
we work out a variety of interesting relationships, considering
entities, their structure, or the citation system. The insights that
we gain establish a deep understanding of the benchmark queries
and their specific challenges. As an example, $Q3a$, $Q3b$, and
$Q3c$ (see Appendix) look similar, but pose different challenges
based on the probability distribution of article properties
discussed within this section; $Q7$, on the other
hand, heavily depends on the DBLP citation system.

Although the generated data is very similar to the original DBLP
data for years up to the present, we can give no guarantees
that our generated data goes hand in hand with the original DBLP
data for future years. However, and this is much more important,
even in the future the generated
data will follow reasonable (and well-known) social-world distributions.
We emphasize that the benchmark queries are designed to primarily
operate on top of these relations and distributions, which makes them
realistic, predictable and understandable. For instance, some
queries operate on top of the citation system, which is mirrored by
our generator. In contrast, the distribution of article release months
is ignored, hence no query relies on this property.

\begin{figure}
\hspace{-0.1cm}
\begin{boxedminipage}{8.75cm}
{\scriptsize
\begin{verbatim}
<!ELEMENT dblp
  (article|inproceedings|proceedings|book|
   incollection|phdthesis|mastersthesis|www)*>
<!ENTITY % field 
  "author|editor|title|booktitle|pages|year|address|
   journal|volume|number|month|url|ee|cdrom|cite|
   publisher|note|crossref|isbn|series|school|chapter">
<!ELEMENT article (%field;)*>...<!ELEMENT www (%field;)*>
\end{verbatim}
}
\end{boxedminipage}
\caption{Extract of the DBLP DTD}
\vspace{-0.3cm}
\label{fig:dblpdtd} 
\end{figure}


\subsection{Structure of Document Classes}
\label{subsec:structure}
Our starting point for the discussion is the DBLP DTD and the
February 25, 2008 version of DBLP. An extract of
the DTD is provided in Figure~\ref{fig:dblpdtd}. The {\small \sf dblp}
element defines eight child entities, namely~\textsc{Article},
\textsc{Inproceedings}, $\dots$, and \textsc{WWW}
resources. We call these entities {\it document classes}, and instances
thereof {\it documents}. Furthermore, we distinguish between
\textsc{Proceedings} documents, called {\it conferences}, and instances
of the remaining classes, called {\it publications}. 

The DTD defines~$22$ possible child tags, such as {\small \sf author}
or {\small \sf url}, for each document class. They {\it describe}
documents, and we call them {\it attributes} in the following. According
to the DTD, each document might be described by arbitrary combination of
attributes. Even repeated occurrences of the same attribute are allowed,
e.g.~a document might have several authors. However, in practice only a subset of all
document class/attribute combinations occurs. For instance,
(as one might expect) attribute {\small \sf pages} is never associated
with \textsc{WWW} documents, but typically associated with \textsc{Article}
entities. In Table~\ref{tbl:attributes} we show, for selected document
class/attribute pairs, the probability that the attribute describes
a document of this class\footnote{The full correlation matrix can be found
in Table \ref{tbl:attributes-tr} in the Appendix.}. To give an example, about
$92.61\%$ of all \textsc{Article} documents are described by the attribute {\small \sf pages}.

This probability distribution forms the basis for generating document
class instances. Note that we simplify and assume that the presence of
an attribute does not depend on the presence of other attributes,
i.e.~we ignore conditional probabilities. We will elaborate on
this decision in Section~\ref{sec:conclusion}.

{\bf Repeated Attributes.}
A study of DBLP reveals that, in practice, only few attributes occur
repeatedly within single documents. For the majority of them, the number of
repeated occurrences is diminishing, so we restrict ourselves on the most
frequent {\it repeated attributes} {\small \sf cite},
{\small \sf editor}, and {\small \sf author}.

\begin{figure*}[t]
\hspace{-0.2cm}
\includegraphics[scale=0.48]{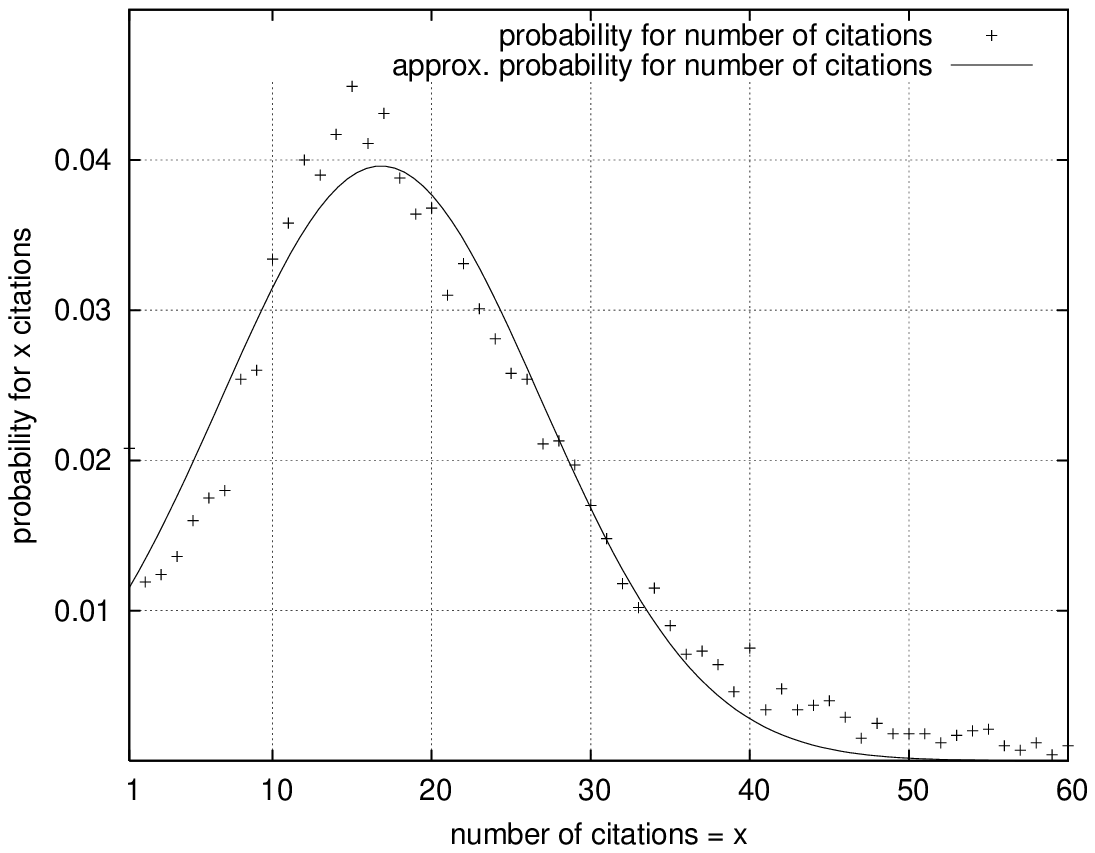}
\includegraphics[scale=0.48]{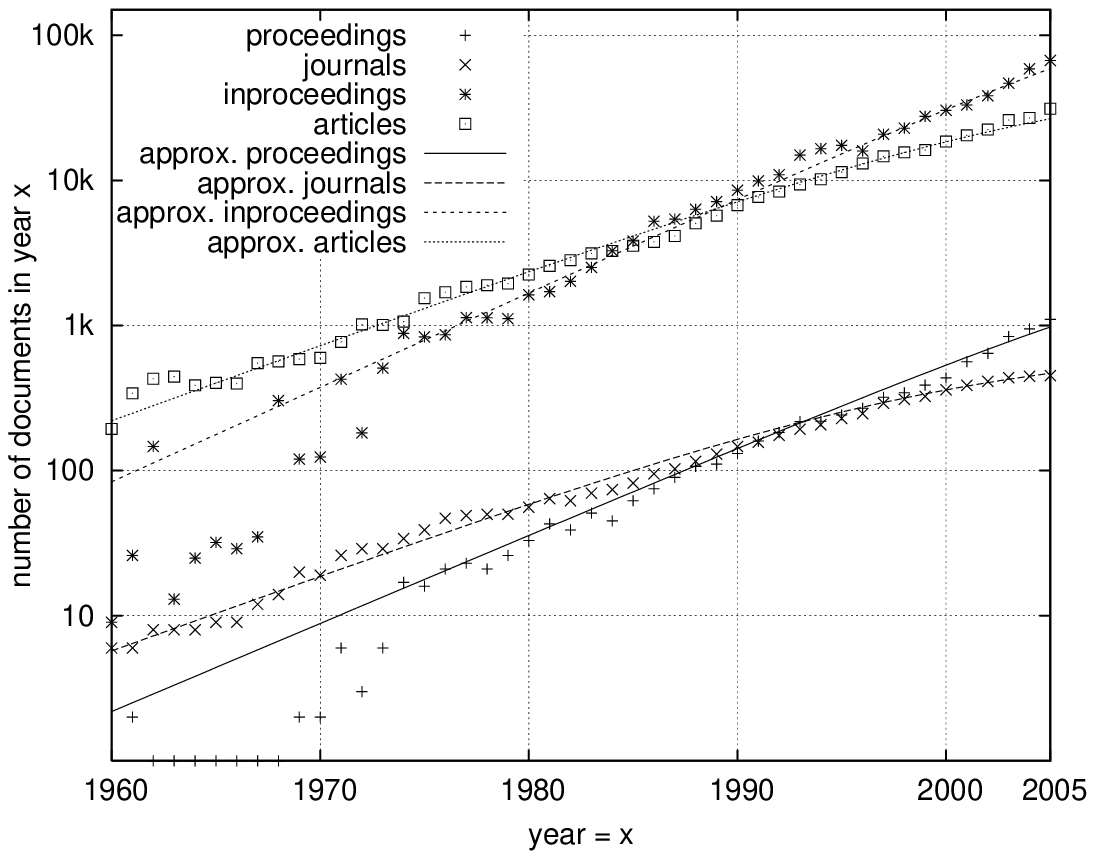}
\includegraphics[scale=0.48]{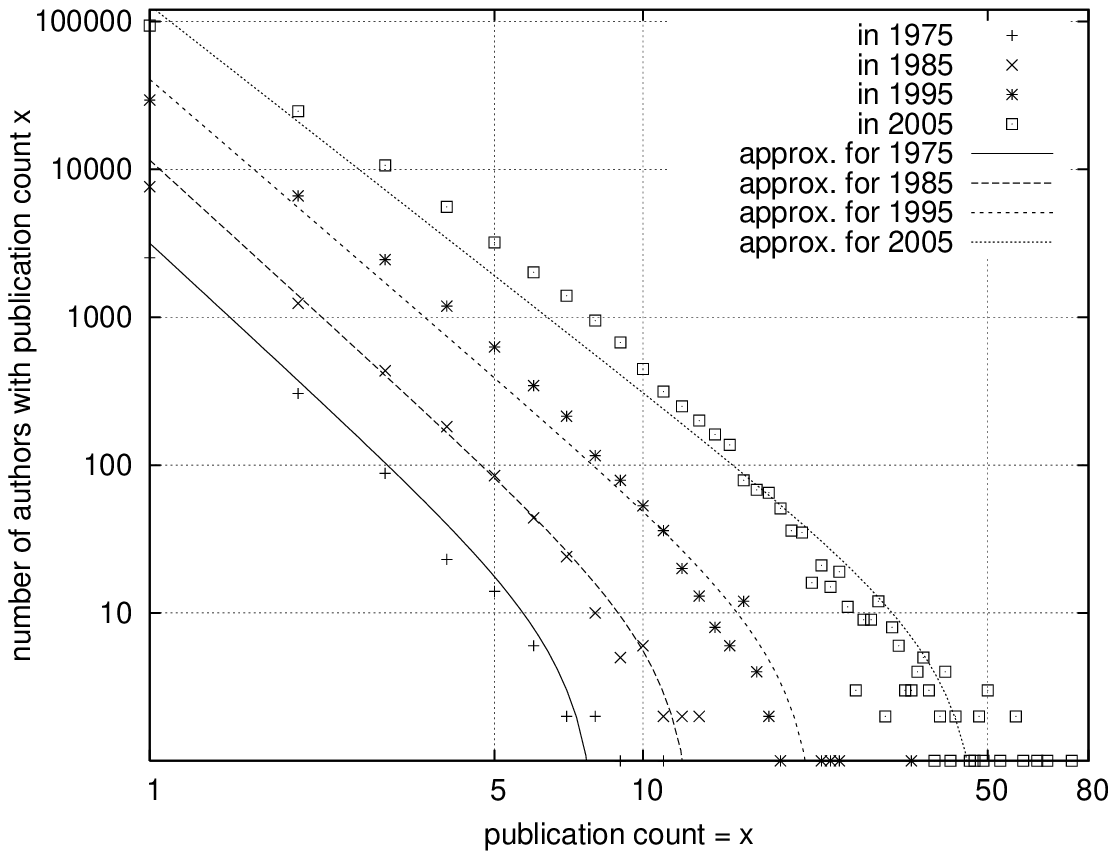}
\vspace{-0.6cm}
\caption{(a) Distribution of citations, (b) Document class instances, and (c) Publication counts}
\vspace{-0.2cm}
\label{fig:dblpanalysis}
\end{figure*}

Figure~\ref{fig:dblpanalysis}(a) exemplifies our analysis
for attribute {\small \sf cite}. It shows, for each documents with at
least one {\small \sf cite} occurrence, the probability ($y$-axis) that
the document has exactly $n$ {\small \sf cite} attributes ($x$-axis).
According to Table~\ref{tbl:attributes}, only a small
fraction of documents are described by {\small \sf cite} (e.g.~$4.8\%$ of
all \textsc{Article} documents). This value should be close to $100\%$ in real
world, meaning that DBLP contains only a fraction of all citations.
This is also why, in Figure~\ref{fig:dblpanalysis}(a), we consider
only documents with at least one outgoing citation; when assigning
citations later on, however, we first use the probability distribution
of attributes in Table~\ref{tbl:attributes} to estimate the number of
documents with at least one outgoing citation and afterwards apply the
citation distribution in Figure~\ref{fig:dblpanalysis}(a). This way,
we exactly mirror the distribution found in the original DBLP data. 

\begin{table}[t]
\caption{Probability distribution for selected attributes}
\vspace{-0.1cm}
\hspace{0.2cm}
{\small
\begin{tabular}{lccccc}
\toprule[.11em]
\rowcolor{darkgray}
& {\bf Article} & {\bf Inproc.} & {\bf Proc.} & {\bf Book} & {\bf WWW}\\
\midrule
{\bf author} & 0.9895 & 0.9970 & 0.0001 & 0.8937 & 0.9973\\
\rowcolor{lightgray}
{\bf cite} & 0.0048 & 0.0104 & 0.0001 & 0.0079 & 0.0000\\
{\bf editor} & 0.0000 & 0.0000 & 0.7992 & 0.1040 & 0.0004\\
\rowcolor{lightgray}
{\bf isbn} & 0.0000 & 0.0000 & 0.8592 & 0.9294 & 0.0000\\
{\bf journal} & 0.9994 & 0.0000 & 0.0004 & 0.0000 & 0.0000\\
\rowcolor{lightgray}
{\bf month} & 0.0065 & 0.0000 & 0.0001 & 0.0008 & 0.0000\\
{\bf pages} & 0.9261 & 0.9489 & 0.0000 & 0.0000 & 0.0000\\
\rowcolor{lightgray}
{\bf title} & 1.0000 & 1.0000 & 1.0000 & 1.0000 & 1.0000\\
\bottomrule[.11em]
\end{tabular}
\label{tbl:attributes}
}
\vspace{-0.3cm}
\end{table}


Based on experiments with different function families, we decided to 
use bell-shaped Gaussian curves for data approximation. Such functions
are typically used to model normal distributions. Strictly
speaking, our data is not normally distributed (i.e.~there is the left
limit~$x=1$), however, these curves nicely fit the data for
$x\geq1$ (cf.~Figure~\ref{fig:dblpanalysis}(a)). Gaussian
curves are described by functions

\vspace{-0.05cm}
\begin{center}
$p_{gauss}^{(\mu,\sigma)}(x) = \frac{1}{\sigma\sqrt{2\pi}}e^{-0.5(\frac{x-\mu}{\sigma})^2}$,
\end{center}
\vspace{-0.05cm}

where $\mu \in \mathbb{R}$ fixes the $x$-position of the peak
and $\sigma \in \mathbb{R}_{>0}$ specifies the statistical spread.
For instance, the approximation function for the {\small \sf cite}
distribution in Figure~\ref{fig:dblpanalysis}(a) is defined by 
$d_{cite}(x) \stackrel{def}{:=} p_{gauss}^{(16.82,10.07)}(x)$.
The analysis and the resulting distribution of repeated
{\small \sf editor} attributes is structurally similar and
is described by the function
$d_{editor}(x) \stackrel{def}{:=} p_{gauss}^{(2.15,1.18)}(x)$.

The approximation function for repeated {\small \sf author} attributes
bases on a Gaussian curve, too. However, we observed that the
average number of authors per publication has increased over
the years. The same observation was made in~\cite{el2005} and explained
by the increasing pressure to publish and the proliferation of new
communication platforms. Due to the prominent role of authors,
we decided to mimic this property. As a consequence, parameters
$\mu$ and $\sigma$ are not fixed (as it was the case for the distributions
$d_{cite}$ and $d_{editor}$), but modeled as functions over time.
More precisely, $\mu$ and $\sigma$ are realized  by limited growth
functions\footnote{We make the reasonable assumption that the number
of coauthors will eventually stabilize.} (so-called logistic curves)
that yield higher values for later years. The distribution is
described by 

\begin{tabbing}
xxx \= \kill
\>$d_{auth}(x,yr) \stackrel{def}{:=} p_{gauss}^{(\mu_{auth}(yr),\sigma_{auth}(yr))}(x)$, where\\
\>$\mu_{auth}(yr) \stackrel{def}{:=} \frac{2.05}{1+17.59e^{-0.11(yr-1975)}}+1.05$, and\\
\>$\sigma_{auth}(yr) \stackrel{def}{:=} \frac{1.00}{1+6.46e^{-0.10(yr-1975)}}+0.50$.
\end{tabbing}

We will discuss the logistic curve function type in more detail in the
following subsection.

\subsection{Key Characteristics of DBLP}
\label{subsec:keycharacteristics}

We next investigate the quantity of document class instances over time. We
noticed that DBLP contains only few and incomplete information in its early
years, and also found anomalies in the final years, mostly in form of lowered
growth rates. It might be that, in the coming years, some more conferences
for these years will be added belatedly (i.e.~data might not yet be totally
complete), so we restrict our discussion to DBLP data ranging from 1960 to 2005.

Figure~\ref{fig:dblpanalysis}(b) plots the number of
\textsc{Proceedings}, \textsc{Journal}, \textsc{Inproceedings},
and \textsc{Article} documents as a function of time. The $y$-axis is in
log scale. Note that \textsc{Journal} is not an explicit document class,
but implicitly defined by the {\small \sf journal} attribute of \textsc{Article}
documents. We observe that inproceedings and articles are closely coupled
to the proceedings and journals. For instance, there are always about
$50$-$60$ times more inproceedings than proceedings, which indicates the
average number of inproceedings per proceeding.

Figure~\ref{fig:dblpanalysis}(b) shows exponential growth for all 
document classes, where the growth rate of \textsc{Journal}
and \textsc{Article} documents decreases in the final years. This
suggests a limited growth scenario. Limited growth is typically modeled
by logistic curves, which describe functions with a lower and an
upper asymptote that either continuously increase or decrease
for increasing $x$. We use curves of the form

\vspace{-0.05cm}
\begin{center}
$f_{logistic}(x) = \frac{a}{1+be^{-cx}}$,
\end{center}
\vspace{-0.05cm}

where $a, b, c \in \mathbb{R}_{>0}$.
For this parameter setting, $a$ constitutes the upper asymptote
and the $x$-axis forms the lower asymptote. The curve
is ``caught'' in-between its asymptotes and increases continuously,
i.e.~it is $S$-shaped. The approximation function for the number of
\textsc{Journal} documents, which is also plotted in
Figure~\ref{fig:dblpanalysis}(b), is defined by the formula

\begin{center}
$f_{journal}(yr) \stackrel{def}{:=} \frac{740.43}{1+426.28e^{-0.12(yr-1950)}}$.
\end{center}

Approximation functions for \textsc{Article}, \textsc{Proceedings},
\textsc{Inproceedings}, \textsc{Book}, and \textsc{Incollection} documents
differ only in the parameters. \textsc{PhD Theses},
\textsc{Masters Theses}, and \textsc{WWW} documents were distributed unsteadily,
so we modeled them by random functions. It is worth mentioning that 
the number of articles and inproceedings per year clearly dominates the number
of instances of the remaining classes. The concrete formulas look as follows.

\medskip

\begin{tabbing}
xxx \= xxxxxxxxxxl \= \kill
\>$f_{article}(yr)$\>$\stackrel{def}{:=} \frac{58519.12}{1+876.80e^{-0.12(yr-1950)}}$\\
\>$f_{proc}(yr)$\>$\stackrel{def}{:=} \frac{5502.31}{1+1250.26e^{-0.14(yr-1965)}}$\\
\>$f_{inproc}(yr)$\>$\stackrel{def}{:=} \frac{337132.34}{1+1901.05e^{-0.15(yr-1965)}}$\\
\>$f_{incoll}(yr)$\>$\stackrel{def}{:=} \frac{3577.31}{196.49e^{-0.09(yr-1980)}}$\\
\>$f_{book}(yr)$\>$\stackrel{def}{:=} \frac{52.97}{40739.38e^{-0.32(yr-1950)}}$\\
\>$f_{phd}(yr)$\>$\stackrel{def}{:=} \textit{random}[0..20]$\\
\>$f_{masters}(yr)$\>$\stackrel{def}{:=} \textit{random}[0..10]$\\
\>$f_{www}(yr)$\>$\stackrel{def}{:=} \textit{random}[0..10]$
\end{tabbing}

\subsection{Authors and Editors}
\label{subsec:persons}
Based on the previous analysis, we can estimate the number
of documents $f_{docs}$ in $yr$ by summing up the individual counts:

\begin{tabbing}
x \= xxxxxxxl \= xxx \= \kill
\>$f_{docs}(yr)$\>$\stackrel{def}{:=}$\>$f_{journal}(yr) + f_{article}(yr) + f_{proc}(yr) +$\\
\>\>\>$f_{inproc}(yr) + f_{incoll} + f_{book}(yr) +$\\
\>\>\>$f_{phd}(yr) + f_{masters}(yr) + f_{www}(yr)$,
\end{tabbing}

The {\it total number of authors}, which we define as the number of
{\small \sf author} attributes in the data set, is computed as follows.
First, we estimate the number of documents described by attribute
{\small \sf author} for each document class individually (using the
distribution in Table~\ref{tbl:attributes}). All these counts are
summed up, which gives an estimation for the total number of documents
with one or more {\small \sf author} attributes. Finally, this
value is multiplied with the expected average number of authors per
paper in the respective year (implicitly given by $d_{auth}$ in
Section~\ref{subsec:structure}).

To be close to reality, we also consider the number 
of distinct persons that appear as authors (per year),
called {\it distinct authors}, and the number of {\it new authors}
in a given year, i.e.~those persons that publish for the first time.

We found that the number of distinct authors $f_{dauth}$ per year
can be expressed in dependence of $f_{auth}$ as follows. 

\begin{center}
$f_{dauth}(yr) \stackrel{def}{:=} (\frac{-0.67}{1+169.41e^{-0.07(yr-1936)}}+0.84) * f_{auth}(yr)$
\end{center}

The equation above indicates that the number of distinct authors relative to the
total authors decreases steadily, from $0.84\%$ to $0.84\% - 0.67\% = 0.17\%$.
Among others, this reflects the increasing productivity of authors over time.

The formula for the number $f_{new}$ of new authors builds on the
previous one and also builds upon a logistic curve:

\begin{center}
$f_{new}(yr) \stackrel{def}{:=} (\frac{-0.29}{1749.00e^{-0.14(yr-1937)}}+0.628)*f_{dauth}(yr)$
\end{center}

{\bf Publications.}
In Figure~\ref{fig:dblpanalysis}(c) we plot, for selected year and publication
count $x$, the number of authors with exactly $x$ publications in this year.
The graph is in log-log scale. We observe a typical power law distribution,
i.e.~there are only a couple of authors having a large number of publications,
while lots of authors have only few publications.

Power law distributions are modeled by functions of the form
$f_{powerlaw}(x) = ax^k + b$, with constants $a \in \mathbb{R}_{>0}$,
exponent $k \in \mathbb{R}_{<0}$, and $b \in \mathbb{R}$.
Parameter $a$ affects the $x$-axis intercept, exponent $k$
defines the gradient, and $b$ constitutes a shift in
$y$-direction. For the given parameter restriction, the functions
decrease steadily for increasing $x\geq 0$.

Figure~\ref{fig:dblpanalysis}(c) shows that, throughout the years, the
curves move upwards. This means that the publication count of the leading
author(s) has steadily increased over the
last~$30$ years, and also reflects an increasing number of authors. We
estimate the number of authors with $x$ publications in year $yr$ as

\begin{tabbing}
xxxx \= xxxxxxxxxxxxxl \= xxl \= \kill
\>$f_{awp}(x,yr) \stackrel{def}{:=} 1.50f_{publ}(yr)x^{-f_{awp}'(yr)}-5$, where\\
\>$f_{awp}'(yr) \stackrel{def}{:=} \frac{-0.60}{1+216223e^{-0.20(yr-1936)}}+3.08$, and
\end{tabbing}

$f_{publ}(yr)$ returns the total number of publications in $yr$.

{\bf Coauthors.} In analyzing coauthor characteristics, we investigated
relations between the publication count of authors and the number of
its total and distinct coauthors. Given a number $x$ of publications,
we (roughly) estimate the average number of total coauthors by
$\mu_{coauth} := 2.12*x$ and the number of its distinct coauthors
by $\mu_{dcoauth} := x^{0.81}$. We take these values into consideration
when assigning coauthors.

{\bf Editors.}
The analysis of authors is complemented by a study of their relations to
editors. We associate editors with authors by investigating
the editors' number of publications in (earlier) venues. As one might
expect, editors often have published before, i.e.~are persons that are
known in the community. The concrete formula is rather technical and
omitted.

\subsection{Citations}
\label{subsec:citations}

In Section~\ref{subsec:structure} we have studied repeated occurrences
of attribute {\small \sf cite}, i.e.~outgoing citations. Concerning the
{\it incoming} citations (i.e.~the count of incoming references for papers),
we observed a characteristic power law distribution: Most papers have few
incoming citations, while only few are cited often. We omit the concrete
power law approximation function.

We also observed that the number of incoming citations is smaller than
the number of outgoing citations. This is because DBLP contains many
untargeted citations (i.e.~empty {\small \sf cite} tags). Recalling
that only a fraction of all papers have outgoing citations
(cf.~Section~\ref{subsec:structure}), we conclude that the DBLP
citation system is very incomplete.

\begin{figure*}[t]
\hspace{-0.2cm}
{\scriptsize
\begin{minipage}{6cm}
\begin{tabular}{lll}
\toprule[.11em]
\rowcolor{darkgray}
{\bf attribute}     & {\bf mapped to prop.}  & {\bf refers to}\\
\midrule
{\tt address}       & {\tt swrc:address}        & {\tt xsd:string}\\
\rowcolor{lightgray}
{\tt author}        & {\tt dc:creator}          &   {\tt foaf:Person}\\
{\tt booktitle}     & {\tt bench:booktitle}     & {\tt xsd:string}\\
\rowcolor{lightgray}
{\tt cdrom}         & {\tt bench:cdrom}         & {\tt xsd:string}\\
{\tt chapter}       & {\tt swrc:chapter}        & {\tt xsd:integer}\\
\rowcolor{lightgray}
{\tt cite}          & {\tt dcterms:references}  & {\tt foaf:Document}\\
{\tt crossref}      & {\tt dcterms:partOf}      & {\tt foaf:Document}\\
\rowcolor{lightgray}
{\tt editor}        & {\tt swrc:editor}         & {\tt foaf:Person}\\
{\tt ee}            & {\tt rdfs:seeAlso}        & {\tt xsd:string}\\
\rowcolor{lightgray}
{\tt isbn}          & {\tt swrc:isbn}           & {\tt xsd:string}\\
{\tt journal}       & {\tt swrc:journal}        & {\tt bench:Journal}\\
\rowcolor{lightgray}
{\tt month}         & {\tt swrc:month}          & {\tt xsd:integer}\\
{\tt note}          & {\tt bench:note}          & {\tt xsd:string}\\
\rowcolor{lightgray}
{\tt number}        & {\tt swrc:number}         & {\tt xsd:integer}\\
{\tt page}          & {\tt swrc:pages}          & {\tt xsd:string}\\
\rowcolor{lightgray}
{\tt publisher}     & {\tt dc:publisher}        & {\tt xsd:string}\\
{\tt school}        & {\tt dc:publisher}        & {\tt xsd:string}\\
\rowcolor{lightgray}
{\tt series}        & {\tt swrc:series}         & {\tt xsd:integer}\\
{\tt title}         & {\tt dc:title}            & {\tt xsd:string}\\
\rowcolor{lightgray}
{\tt url}           & {\tt foaf:homepage}       & {\tt xsd:string}\\
{\tt volume}        & {\tt swrc:volume}         & {\tt xsd:integer}\\
\rowcolor{lightgray}
{\tt year}          & {\tt dcterms:issued}      & {\tt xsd:integer}\\
\bottomrule
\end{tabular}
\end{minipage}
}
\hspace{1.2cm}
\begin{minipage}{10cm}
\fbox{
\includegraphics[scale=0.425]{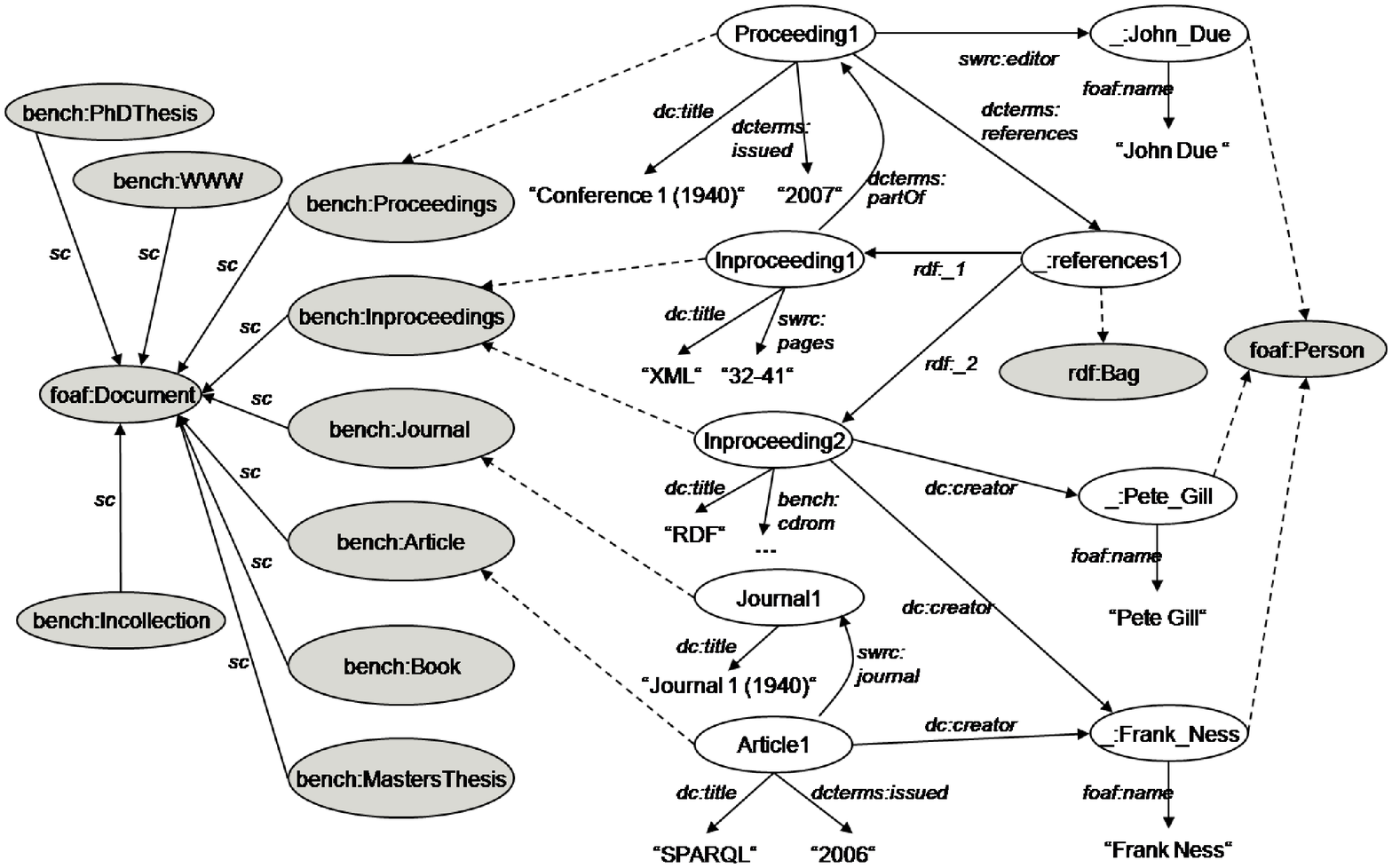}
}
\end{minipage}
\vspace{-0.1cm}
\caption{(a) Translations of attributes, and (b) DBLP sample instance in RDF format}
\label{fig:rdfinstanceandtranslation}
\end{figure*}

\section{Data Generation}
\label{sec:rdfdatamodel}
\label{sec:datagen}

{\bf The RDF Data Model.} From a logical point of view, RDF data bases are
collections of so-called {\rm triples of knowledge}. A triple
($\textit{subject}$,$\textit{predicate}$,$\textit{object}$)
models the binary relation $\textit{predicate}$ between $\textit{subject}$
and $\textit{object}$ and can be visualized in a directed graph by
an edge from the {\it subject} node to an {\it object} node under label
{\it predicate}. Figure~\ref{fig:rdfinstanceandtranslation}(b) shows a sample
RDF graph, where dashed lines represent edges that are labeled with
{\small \sf rdf:type}, and {\it sc} is an abbreviation for
{\small \sf rdfs:subClassOf}. For instance, the arc from node \textit{Proceeding1}
to node {\it \_:John\_Due} represents the triple
({\it Proceeding1},{\it swrc:editor},{\it \_:John\_Due}).

RDF graphs may contain three types of nodes. First, {\it URIs} (Uniform
Resource Identifiers) are strings that uniquely identify
abstract or physical resources, such as conferences or journals.
{\it Blank nodes} have an existential character, i.e.~are typically
used to denote resources that exist, but are not assigned a
fixed URI. We represent URIs and blank nodes by ellipses,
identifying blank nodes by the prefix ``\_:''. {\it Literals}
represent (possibly typed) values and usually describe URIs or 
blank nodes. Literals are represented by quoted strings. 

The RDF standard~\cite{rdfconceptsw3c} introduces a base vocabulary with
fixed semantics, e.g.~defines URI {\small \sf rdf:type} for type specifications.
This vocabulary also includes containers, such as bags or sequences.
RDFS~\cite{rdfschemaw3c} extends the RDF vocabulary and, among others,
provides URIs for subclass ({\small \sf rdfs:subClassOf}) and subproperty
({\small \sf rdf:subPropertyOf}) specifications. On top of RDF and RDFS, one
can easily create user-defined, domain-specific vocabularies. Our
data generator makes heavy use of such predefined vocabulary collections.

{\bf The DBLP RDF Scheme.} Our RDF scheme basically follows the approach
in~\cite{d2rdblp}, which presents an XML-to-RDF mapping of the
original DBLP data. However, we want to generate arbitrarily-sized
documents and provide lists of first and last names, publishers, and
random words to our data generator. Conference and journal names are always
of the form ``{\it Conference \$i (\$year)}'' and
``{\it Journal \$i (\$year)}'', where $\$i$ is a unique conference
(resp.~journal) number in year $\textit{\$year}$.

Similar to~\cite{d2rdblp}, we use existing RDF vocabularies to
describe resources in a uniform way. We borrow
vocabulary from FOAF\footnote{http://www.foaf-project.org/} for describing
persons, and from SWRC\footnote{http://ontoware.org/projects/swrc/} and
DC\footnote{http://dublincore.org/} for describing scientific resources.
Additionally, we introduce a namespace {\small \sf bench}, which defines
DBLP-specific document classes, such as {\tt bench:Book} and
{\tt bench:Article}. Figure~\ref{fig:rdfinstanceandtranslation}(a) shows
the translation of attributes to RDF properties. For each attribute, we also
list its range restriction, i.e.~the type of elements it refers to. For
instance, attribute {\small \sf author} is mapped to
{\small \sf dc:creator}, and references objects of type {\tt foaf:Person}.

The original DBLP RDF scheme neither contains blank nodes nor
RDF containers. As we want to test our queries on top
of such RDF-specific constructs, we use (unique) blank nodes 
``{\it \_:givenname\_lastname}'' for persons (instead of URIs) and
model outgoing citations of documents using standard
{\tt rdf:Bag} containers. We also enriched a small fraction
of \textsc{Article} and \textsc{Inproceedings} documents with
the new property {\small \sf bench:abstract} (about $1\%$, keeping
the modification low), which constitutes comparably large strings
(using a Gaussian distribution with $\mu=150$ expected words and
$\sigma=30$).

Figure~\ref{fig:rdfinstanceandtranslation}(b) shows a sample DBLP instance.
On the logical level, we distinguish between the {\it schema} layer (gray)
and the {\it instance} layer (white). Reference lists are modeled as blank
nodes of type {\tt rdf:Bag}, i.e.~using standard RDF containers
(see node {\it \_:references1}). Authors and editors are represented
by blank nodes of type {\tt foaf:Person}. Class {\tt foaf:Document}
splits up into the individual document classes {\tt bench:Journal},
{\tt bench:Article}, and so on. Our graph defines three
persons, one proceeding, two inproceedings, one journal, and one article.
For readability reasons, we plot only selected predicates. As also
illustrated, property {\small \sf dcterms:partOf} links inproceedings and proceedings
together, while {\small \sf swrc:journal} connects articles to their journals.

In order to provide an entry point for queries that access authors and
to provide a person with fixed characteristics, we created a special
author, named after the famous mathematician Paul Erd\"os. Per year,
we assign $10$ publications and $2$ editor activities to this prominent
person, starting from year 1940 up to 1996. For the ease of access,
Paul Erd\"os is modeled by a fixed URI. As an example query consider
$Q8$, which extracts all persons 
with {\it Erd\"os Number}\footnote{See http://www.oakland.edu/enp/.}~$1$ or~$2$.

\begin{figure}[t]
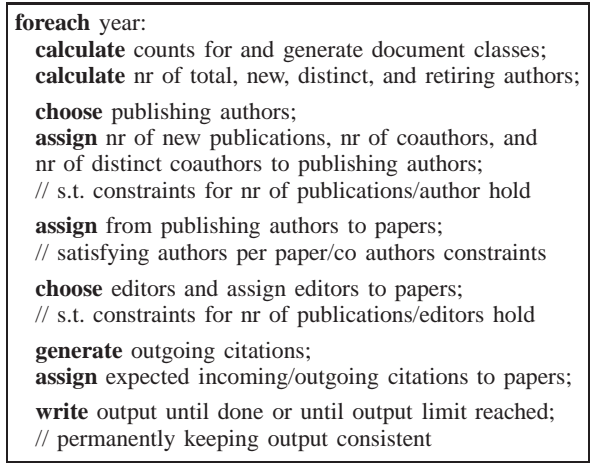

{\small
\begin{center}
\begin{boxedminipage}{6cm}
\begin{tabbing}
x \= xx \= x \= xl \= xx \= xx \= xx \= xx \= \kill
{\bf foreach} year:\\
\>{\bf calculate} counts for and generate document classes;\\
\>{\bf calculate} nr of total, new, distinct, and retiring authors;\\
\\[-0.23cm]
\>{\bf choose} publishing authors;\\
\>{\bf assign} nr of new publications, nr of coauthors, and\\
\>nr of distinct coauthors to publishing authors;\\
\>// s.t. constraints for nr of publications/author hold\\
\\[-0.23cm]
\>{\bf assign} from publishing authors to papers;\\
\>// satisfying authors per paper/co authors constraints\\
\\[-0.23cm]
\>{\bf choose} editors and assign editors to papers;\\
\>// s.t. constraints for nr of publications/editors hold\\
\\[-0.23cm]
\>{\bf generate} outgoing citations;\\
\>{\bf assign} expected incoming/outgoing citations to papers;\\
\\[-0.23cm]
\>{\bf write} output until done or until output limit reached;\\
\>// permanently keeping output consistent
\end{tabbing}
\end{boxedminipage}
\end{center}
}
\vspace{-0.2cm}
\caption{Data generation algorithm}
\label{fig:datageneration}
\vspace{-0.3cm}
\end{figure}

{\bf Data Generation.}
Our data generator was implemented in $C$++. It takes into account all
relationships and characteristics that have been studied in
Section~\ref{sec:dblp}. Figure~\ref{fig:datageneration}
shows the key steps in data generation. We simulate year by year and
generate data according to the structural constraints in a carefully
selected order. As a consequence, data generation is incremental,
i.e.~small documents are always contained in larger documents.

The generator offers two parameters, to fix either a triple count limit
or the year
up to which data will be generated. When the triple count limit is
set, we make sure to end up in a ``consistent'' state, e.g.~whenever
proceedings are written, the corresponding conference will be included.

The generation process is simulation-based. Among others, this means
that we assign life times to authors, and individually estimate their
future behavior, taking into account global publication and coauthor
characteristics, as well as the fraction of distinct and new authors
(cf.~Section~\ref{subsec:persons}).

All random functions (which, for example, are used to assign the attributes
according to Table~\ref{tbl:attributes}) base on a fixed seed. This makes
data generation deterministic, i.e.~the parameter setting uniquely identifies
the outcome. As data generation is also platform-independent,
we ensure that experimental results from different machines are comparable.

\begin{table*}
\caption{Selected properties of the benchmark queries; shortcuts are indicated by bold font}
\vspace{-0.23cm}
\hspace{-0.35cm}
{\scriptsize
\begin{tabular}{llcccccccccccc}
\toprule
\rowcolor{darkgray}
& {\bf Query} & 1 & 2 & 3abc & 4 & 5ab & 6 & 7 & 8 & 9 & 10 & 11 & 12c\\
\midrule
1 & Operators: \textsc{{\bf A}nd},\textsc{{\bf F}ilter},\textsc{{\bf U}nion},\textsc{{\bf O}ptional}	& \verb!A!	& \verb!A,O!	& \verb!A,F!	& \verb!A,F!	& \verb!A,F!	& \verb!A,F,O!	& \verb!A,F,O!	& \verb!A,F,U!	& \verb!A,U!	& - & - & -\\
\rowcolor{lightgray}
2 & Modifiers: \textsc{{\bf D}istinct},\textsc{{\bf L}imit},\textsc{{\bf Of}fset},\textsc{{\bf O}rder {\bf b}y}	& - & \verb!Ob!	& - & \verb!D! & \verb!D! & & \verb!D!	& \verb!D!	& \verb!D!	& - & \verb!L,Ob,Of! & -\\
4 & Filter Pushing Possible?	& - & - & $\checkmark$ & - & $\checkmark$/- & $\checkmark$ & $\checkmark$ & $\checkmark$ & - & - & - & -\\
\rowcolor{lightgray}
5 & Reusing of Graph Pattern Possible?	& - & - & - & $\checkmark$ & - & $\checkmark$ & $\checkmark$ & $\checkmark$ & $\checkmark$ & - & -\\
6 & Data Access: \textsc{{\bf B}lank nodes},\textsc{{\bf L}iterals},\textsc{{\bf U}RIs},	& \verb!L,U! & \verb!L,U,La! & \verb!L,U! & \verb!B,L,U! & \verb!B,L,U! & \verb!B,L,U! & \verb!L,U,C! & \verb!B,L,U! & \verb!B,L,U! &\verb!U! & \verb!L,U! & \verb!U!\\
 & \ \ \ \ \ \ \ \ \ \ \ \ \ \ \ \ \textsc{{\bf La}rge Literals},\textsc{{\bf C}ontainers}	& & & & & & & & & & & &\\
\bottomrule
\end{tabular}
}
\label{tbl:querycharacteristics}
\vspace{-0.3cm}
\end{table*}


\section{Benchmark Queries}
\label{sec:queries}

{\bf The SPARQL Query Language.} SPARQL is a declarative
language and bases upon a powerful graph matching facility,
allowing to match query subexpressions against the RDF input graph.
The very basic SPARQL constructs are triple
patterns $({\it subject}, {\it predicate}, {\it object})$,
where variables might be used in place of fixed values for each
of the three components. In evaluating SPARQL, these patterns
are mapped against one or more input graphs, thereby binding
variables to matching nodes or edges in the graph(s).
Since we are primarily interested in database aspects, such as
operator constellations and access patterns, we focus on 
queries that access a single graph.

The SPARQL standard~\cite{sparqlw3c} defines four distinct query
forms. \textsc{Select} queries retrieve all possible
variable-to-graph mappings, while \textsc{Ask} queries return {\it yes}
if at least one such mapping exists, and {\it no} otherwise.
The \textsc{Describe} form extracts additional information related
to the result mappings (e.g.~adjacent nodes), while \textsc{Construct}
transforms the result mapping into a new RDF graph. The most appropriate
for our purpose is \textsc{Select}, which best reflects SPARQL
core evaluation. \textsc{Ask} queries are also interesting,
as they might affect the choice of the query execution plan (QEP).
In contrast, \textsc{Construct} and \textsc{Describe}
build upon the core evaluation of \textsc{Select}, i.e.~transform
its result in a post-processing step. This step is not very challenging
from a database perspective, so we focus on \textsc{Select}
and \textsc{Ask} queries (though, on demand, these queries could easily
be translated into the other forms).

The most important SPARQL operator is~\textsc{And} (denoted as ``.''). If
two SPARQL expressions $A$ and $B$ are connected by~\textsc{And}, the
result is computed by joining the result mappings of $A$ and $B$ on
their shared variables~\cite{pag2006}.
Let us consider~$Q1$ from the Appendix,
which defines three triple patterns interconnected 
through~\textsc{And}. When first evaluating the patterns individually,
variable $\textit{?journal}$ is bound to nodes with
(1)~edge {\small \sf rdf:type} pointing to the URI {\tt bench:Journal},
(2)~edge {\small \sf dc:title} pointing to the Literal
``\textit{Journal 1 (1940)}'' of type string,
and (3)~edge {\small \sf dcterms:issued}, respectively. The next step 
is to join the individual mapping sets on variable
$\textit{?journal}$. The result then contains all mappings from $\textit{?journal}$
to nodes that satisfy all three patterns. Finally \textsc{Select}
projects for variable $\textit{?yr}$, which has been bound in
the third pattern.

Other SPARQL operators are \textsc{Union}, \textsc{Optional},
and \textsc{Filter}, akin to relational unions, left outer joins,
and selections, respectively. For space limitations, we omit
an explanation of these constructs and refer the reader to
the SPARQL semantics~\cite{sparqlw3c}. Beyond all these operators,
SPARQL provides functions to be used in~\textsc{Filter}
expressions, e.g.~for regular expression testing. We expect these
functions to only marginally affect engine performance, since
their implementation is mostly straightforward (or might be
realized through efficient libraries). They are unlikely to
bring insights into the core evaluation capabilities,
so we omit them intentionally. This decision also facilitates
benchmarking of research prototypes, which typically do not
implement the full standard.

The SP$^2$Bench queries also cover SPARQL solution modifiers, such
as~\textsc{Distinct}, \textsc{Order By}, \textsc{Limit}, and~\textsc{Offset}.
Like their SQL counterparts, they might heavily affect the choice of an
efficient QEP, so they are relevant for our benchmark. We point out
that the previous discussion captures virtually all key features of
the SPARQL query language. In particular, SPARQL does (currently) not
support aggregation, nesting, or recursion.

{\bf SPARQL Characteristics.}
Rows~$1$ and~$2$ in Table~\ref{tbl:querycharacteristics} survey the
operators used in the \textsc{Select} benchmark queries (the \textsc{Ask}-queries
$Q12a$ and $Q12b$ share the characteristics of their \textsc{Select} counterparts
$Q5a$ and $Q8$, respectively, and are not shown). The queries 
cover various operator constellations, combined with selected solution modifiers
combinations.

One very characteristic SPARQL feature is operator \textsc{Optional}.
An expression $A~\textsc{Optional}~\{ B \}$ joins result
mappings from $A$ with mappings from $B$, but -- unlike
\textsc{And} -- retains mappings from $A$
for which no join partner in $B$ is present. In the latter case,
variables that occur only inside $B$ might be unbound. 
By combining~\textsc{Optional} with \textsc{Filter} and function 
\textsc{bound}, which checks if a variable is bound
or not, one can simulate {\it closed world negation} in
SPARQL. Many interesting queries involve such an encoding
(c.f.~$Q6$ and $Q7$).

SPARQL operates on graph-structured data, thus engines should
perform well on different kinds of graph patterns. Unfortunately,
up to the present there exist only few real-world SPARQL scenarios.
It would be necessary to analyze a large set of such scenarios,
to extract graph patterns that frequently occur in practice. In the
absence of this possibility, we distinguish between {\it long path
chains}, i.e.~nodes linked to each other node via a long path, {\it bushy
patterns}, i.e.~single nodes that are linked to a
multitude of other nodes, and {\it combinations of these two
patterns}. Since it is impossible to give a precise definition of
``long'' and ``bushy'', we designed meaningful queries that contain
{\it comparably} long chains (i.e.~$Q4$, $Q6$) and {\it comparably}
bushy patterns (i.e.~$Q2$) w.r.t.~our scenario. 
These patterns contribute to the variety of characteristics that we cover.

{\bf SPARQL Optimization.}
Our objective is to design queries that are amenable to a
variety of SPARQL optimization approaches. To this end, we discuss
possible optimization techniques before presenting the benchmark
queries.

A promising approach to SPARQL optimization is the
{\it reordering of triple patterns} based on selectivity
estimation~\cite{ssbkr2008}, akin to relational join reordering. 
Closely related to triple reordering is {\it \textsc{Filter} pushing},
which aims at an early evaluation of filter conditions, similar to projection
pushing in Relational Algebra. Both techniques might speed up
evaluation by decreasing the size of intermediate results. An efficient
join order depends on selectivity estimations for triple patterns, but
might also be affected by available data access paths.
Join reordering might apply to most of our queries. Row~$4$ in
Table~\ref{tbl:querycharacteristics} lists the queries that support
\textsc{Filter} pushing. 

Another idea is to {\it reuse evaluation results of triple patterns}
(or even combinations thereof). This might be possible
whenever the same pattern is used multiple times. As an example
consider $Q4$. Here, $\textit{?article1}$ and $\textit{?article2}$
in the first and second triple pattern will be bound to
the same nodes. We survey the applicability of this technique in
Table~\ref{tbl:querycharacteristics}, row~$5$.

{\bf RDF Characteristics and Storage.}
SPARQL has been specifically designed to operate on top of
RDF~\cite{rdfconceptsw3c} rather than RDFS~\cite{rdfschemaw3c} data.
Although it is possible to access RDFS vocabulary with SPARQL, the
semantics of RDFS~\cite{rdfsemanticsw3c} is ignored when evaluating such 
queries. Consider for example the {\small \sf rdfs:subClassOf} property, which is
used to model subclass relationships between entities, and assume that
class {\tt Student} is a subclass of {\tt Person}. A SPARQL query like
``{\it Select all objects of type {\tt Person}}'' then does {\it not}
return students, although according to~\cite{rdfsemanticsw3c} each student
is also a person. Hence, queries that cover RDFS inference make no
sense unless the SPARQL standard is changed accordingly.

Recalling that persons are modeled as blank nodes, all queries
that deal with persons access blank nodes. Moreover, one of our
queries operates on top of the RDF bag container for reference
lists ($Q7$), and one accesses the comparably large abstract literals ($Q2$).
Row~$6$ in Table~\ref{tbl:querycharacteristics} provides a survey.

A comparison of RDF storage strategies is provided in~\cite{ammh2007}.
Storage scheme and indices finally imply a selection of efficient
{\it data access paths}. Our queries impose varying challenges to the
storage scheme, e.g.~test data access
through RDF subjects, predicates, objects, and combinations thereof. In
most cases, predicates are fixed and subject and/or object vary, but
we also test more uncommon access patterns. We will resume this
discussion when describing~$Q9$ and~$Q10$.

\subsection{Benchmark Queries}
The benchmark queries also vary in general characteristics like
{\it selectivity}, {\it query and output size}, and {\it different
types of joins}. We will point out such characteristics in the 
subsequent individual discussion of the benchmark queries.

In the following, we distinguish between
{\it in-memory} engines, which load the document from file and process
queries in main memory, and {\it native} engines, which rely on a
physical database system. When discussing challenges to and evaluation
strategies for native engines, we always assume that the document has
already been loaded in the database before.

We finally emphasize that in this paper we focus on the SPARQL versions
of our queries, which can be processed directly by real SPARQL engines.
One might also be interested in the SQL-translations of these
queries available at the SP$^2$Bench project page. We refer the
interested reader to~\cite{shklp2008} for an elaborate discussion
of these translations.

\bigskip

\noindent
{\small
\begin{tabularx}{8.4cm}{X}
\toprule[.11em]
\rowcolor{darkgray}
{\bf Q1.} {\it Return the year of publication of ``Journal 1 (1940)''.}\\
\midrule
This simple query returns exactly one result (for arbitrarily large documents).
Native engines might use index lookups in order to answer this query in (almost)
constant time, i.e.~execution time should be independent from document size.
\end{tabularx}
}

\medskip

\noindent
{\small
\begin{tabularx}{8.4cm}{X}
\toprule[.11em]
\rowcolor{darkgray}
{\bf Q2.} {\it Extract all inproceedings with properties
{\footnotesize \sf dc:creator}, {\footnotesize \sf bench:booktitle},
{\footnotesize \sf dcterms:issued},
{\footnotesize \sf dcterms:partOf},
{\footnotesize \sf rdfs:seeAlso},
{\footnotesize \sf dc:title},
{\footnotesize \sf swrc:pages},
{\footnotesize \sf foaf:homepage}, 
and optionally {\footnotesize \sf bench:abstract}, including their values.}\\
\midrule
This query implements a bushy graph pattern. It contains a single,
simple \textsc{Optional} expression, and accesses large strings
(i.e.~the abstracts). Result size grows with database size,
and a final result ordering is necessary due to operator \textsc{Order By}.
Both native and in-memory engines might reach evaluation times that are
almost linear to the document size.
\end{tabularx}
}

\medskip

\noindent
{\small
\begin{tabularx}{8.4cm}{X}
\toprule[.11em]
\rowcolor{darkgray}
{\bf Q3abc.} {\it Select all articles with property 
(a)~{\footnotesize \sf swrc:pages}, (b)~{\footnotesize \sf swrc:month},
or (c) {\footnotesize \sf swrc:isbn}.}\\
\midrule
This query tests \textsc{Filter} expressions with varying
selectivity. According to Table~\ref{tbl:attributes}, the \textsc{Filter}
expression in $Q3a$ is not very selective (i.e.~retains about $92.61\%$ of
all articles). Data access through a secondary index for $Q3a$ 
is probably not very efficient, but might work well for $Q3b$,
which selects only $0.65\%$ of all articles. The filter condition in
$Q3c$ is never satisfied, as no articles have {\small \sf swrc:isbn}
predicates. Schema statistics might be used to answer $Q3c$ in constant
time.
\end{tabularx}
}

\medskip

\noindent
{\small
\begin{tabularx}{8.4cm}{X}
\toprule[.11em]
\rowcolor{darkgray}
{\bf Q4.} {\it Select all distinct pairs of article author names for
authors that have published in the same journal.}\\
\midrule
$Q4$ contains a rather long graph chain, i.e.~variables $\textit{?name1}$
and $\textit{?name2}$ are linked through the articles the authors
have published, and a common journal. The result is very large,
basically quadratic in number and size of journals. Instead of evaluating
the outer pattern block and applying the \textsc{Filter} afterwards,
engines might embed the \textsc{Filter} expression in the computation
of the block, e.g.~by exploiting indices on author names.
The \textsc{Distinct} modifier further complicates the query. 
We expect superlinear behavior, even for native engines.
\end{tabularx}
}

\medskip

\noindent
{\small
\begin{tabularx}{8.4cm}{X}
\toprule[.11em]
\rowcolor{darkgray}
{\bf Q5ab.} {\it Return the names of all persons that occur as author
of at least one inproceeding and at least one article.}\\
\midrule
Queries $Q5a$ and $Q5b$ test different variants of joins.
$Q5a$ implements an implicit join on author names, which is encoded in the
\textsc{Filter} condition, while $Q5b$ explicitly joins the authors on
variable $\textit{name}$. Although in general the queries are not
equivalent, the one-to-one mapping between authors and their names 
(i.e.~author names constitute primary keys) in our scenario implies
equivalence. In~\cite{lms2008}, semantic optimization on top of such
keys for RDF has been proposed. Such an approach might detect
the equivalence of both queries in this scenario and select the more efficient
variant.
\end{tabularx}
}

\medskip

\noindent
{\small
\begin{tabularx}{8.4cm}{X}
\toprule[.11em]
\rowcolor{darkgray}
{\bf Q6.} {\it Return, for each year, the set of all publications authored by
persons that have not published in years before.}\\
\midrule
$Q6$ implements closed world negation (CWN), expressed through a combination
of operators \textsc{Optional}, \textsc{Filter}, and \textsc{bound}. The idea of
the construction is that the block outside the \textsc{Optional} expression
computes all publications, while the inner one constitutes earlier publications
from authors that appear outside. The outer \textsc{Filter} expression then
retains publications for which $\textit{?author2}$ is unbound, i.e.~exactly the
publications of authors without publications in earlier years. 
\end{tabularx}
}

\medskip

\noindent
{\small
\begin{tabularx}{8.4cm}{X}
\toprule[.11em]
\rowcolor{darkgray}
{\bf Q7.} {\it Return the titles of all papers that have been cited at 
least once, but not by any paper that has not been cited itself.}\\
\midrule
This query tests double negation, which requires nested CWN.
Recalling that the citation system of DBLP is rather incomplete
(cf.~Section~\ref{subsec:citations}), we expect only few results.
Though, the query is challenging due to the double negation.
Engines might reuse graph pattern results, for instance, the block 
{\sf ?class[i] rdf:type foaf:Document. ?doc[i] rdf:type ?class[i].}
occurs three times, for empty $[i]$, $[i]$=$3$, and $[i]$=$4$. 
\end{tabularx}
}

\medskip

\noindent
{\small
\begin{tabularx}{8.4cm}{X}
\toprule[.11em]
\rowcolor{darkgray}
{\bf Q8.} {\it Compute authors that have published with
Paul Erd\"os or with an author that has published
with Paul Erd\"os.}\\
\midrule
Here, the evaluation of the second \textsc{Union} part is basically
``contained'' in the evaluation of the first part. Hence, techniques like
graph pattern (or subexpression) reusing might apply. Another very promising
optimization approach is to decompose the filter expressions and push down
its components, in order to decrease the size of intermediate results.
\end{tabularx}
}

\medskip

\noindent
{\small
\begin{tabularx}{8.4cm}{X}
\toprule[.11em]
\rowcolor{darkgray}
{\bf Q9.} {\it Return incoming and outgoing properties of persons.}\\
\midrule
$Q9$ has been designed to test non-standard data access
patterns. Naive implementations would compute the triple patterns
of the \textsc{Union} subexpressions separately, thus evaluate
patterns where no component is bound. Then, pattern
{\tt ?subject ?predicate ?person}
would select all graph triples, which is rather inefficient. Another idea
is to evaluate the first triple in each \textsc{Union} subexpression,
afterwards using the bindings for variable $\textit{?person}$ to evaluate
the second patterns more efficiently. In this case, we observe patterns
where only the $\textit{subject}$ (resp.~the $\textit{object}$) is bound.
Also observe that this query extracts schema information. The result size
is exactly $4$ (for sufficiently large documents). Statistics about
incoming/outgoing properties of $\textit{Person}$-typed objects in
native engines might be used to answer this query in constant time,
even without data access. In-memory engines always must load the
document, hence might scale linearly to document size.
\end{tabularx}
}

\medskip

\noindent
{\small
\begin{tabularx}{8.4cm}{X}
\toprule[.11em]
\rowcolor{darkgray}
{\bf Q10.} {\it Return all subjects that stand in any relation
to person ``{\footnotesize \sf Paul Erd\"os}''.} In our scenario the query can be
reformulated as {\it Return publications and venues in
which ``{\footnotesize \sf Paul Erd\"os}'' is involved either as author or as editor.}\\
\midrule
$Q10$ implements an object bound-only access pattern. In contrast to
$Q9$, statistics are not immediately useful, since the result includes
subjects. Recall that ``Paul Erd\"os'' is active only between 1940 and 1996,
so result size stabilizes for sufficiently large documents. Native engines
might exploit indices and reach (almost) constant execution time.
\end{tabularx}
}

\medskip

\noindent
{\small
\begin{tabularx}{8.4cm}{X}
\toprule[.11em]
\rowcolor{darkgray}
{\bf Q11.} {\it Return (up to) 10 electronic edition URLs starting
from the 51$^{th}$ publication, in lexicographical order.}\\
\midrule
This query focuses on the combination of solution modifiers
\textsc{Order By}, \textsc{Limit}, and \textsc{Offset}. In-memory
engines have to read, process, and sort electronic editions
prior to processing \textsc{Limit} and \textsc{Offset}.
In contrast, native engines might exploit indices to access only a
fraction of all electronic editions and, as the result is
limited to $10$, reach constant runtimes.
\end{tabularx}
}

\medskip

\noindent
{\small
\begin{tabularx}{8.4cm}{X}
\toprule[.11em]
\rowcolor{darkgray}
{\bf Q12.} {\it (a) Return {\it yes} if a person occurs as author
of at least one inproceeding and article, {\it no} otherwise; (b)
Return {\it yes} if an author has published with Paul Erd\"os
or with an author that has published with ``{\footnotesize \sf Paul Erd\"os}'',
and {\it no} otherwise.}; (c) Return {\it yes} if person
``{\footnotesize \sf John Q. Public}'' is present in the database.\\
\midrule
$Q12a$ and $Q12b$ share the properties of their
\textsc{Select} counterparts $Q5a$ and $Q8$, respectively. 
They always return {\it yes} for sufficiently large documents. 
When evaluating such \textsc{Ask} queries, engines should break as soon
a solution has been found. They might adapt the QEP, to efficiently locate
a witness. For instance, based on execution time estimations it might be
favorable to evaluate the second part of the \textsc{Union} in $Q12b$ first.
Both native and in-memory engines should answer these queries very fast,
independent from document size.

$Q12c$ asks for a single triple that is not present in the database. With
indices, native engines might execute this query in constant time.
Again, in-memory engines must scan (and hence, load) the whole document.
\end{tabularx}
}

\begin{table}
\caption{Document generation evaluation}
\vspace{-0.15cm}
\hspace{-0.3cm}
\begin{tabular}{lrrrrrrr}
\toprule[.11em]
\rowcolor{darkgray}
\#triples & $10^3$ & $10^4$ & $10^5$ & $10^6$ & $10^7$ & $10^8$ & $10^9$\\
\midrule
elapsed time [s] & 0.08 & 0.13 & 0.60 & 5.76 & 70 & 1011 & 13306\\
\bottomrule
\end{tabular}
\label{tbl:docgenperformance}
\end{table}

\section{Experiments}
\label{sec:experiments}
All experiments were conducted under Linux ubuntu v7.10 gutsy, on top of
an Intel Core2 Duo E6400 2.13GHz CPU and 3GB DDR2 667 MHz
nonECC physical memory. We used a 250GB Hitachi P7K500 SATA-II
hard drive with 8$\textit{MB}$ Cache. The Java engines were executed
with JRE v1.6.0\_04.

\subsection{Data Generator}

To prove the practicability of our data generator, we
measured data generation times for documents of different sizes.
Table~\ref{tbl:docgenperformance} shows the performance results
for documents containing up to one billion triples. The generator
scales almost linearly with document size and creates even large
documents very fast (the $10^9$ triples document has a physical
size of about $103\textit{GB}$). Moreover, it runs with constant main
memory consumption (i.e., gets by with about $1.2\textit{GB}$ RAM).

We verified the implementation of all characteristics from
Section~\ref{sec:dblp}. Table~\ref{tbl:docgen} shows selected data
generator and output document characteristics for documents up to
$25M$ triples. We list the size of the output file, the year in which
simulation ended, the number of total authors and distinct authors
contained in the data set (cf.~Section~\ref{subsec:persons}), and
the counts of the document class instances (cf.~Section~\ref{subsec:keycharacteristics}).
We observe superlinear growth for the number of authors (w.r.t.~the
number of triples).
This is primarily caused by the increasing average number of authors per paper 
(cf.~Section~\ref{subsec:structure}). The growth rate of
proceedings and inproceedings is also superlinear, while the rate
of journals and articles is sublinear. These observations reflect
the yearly document class counts in Figure~\ref{fig:dblpanalysis}(b).
We remark that -- like in the original DBLP database -- in the early
years instances of several document classes are missing, e.g.~there
are no \textsc{Book} and \textsc{Www} documents. Also note
that the counts of inproceedings and articles clearly dominate the
remaining document classes.

Table~\ref{tbl:resultsizes} surveys the result sizes for the queries on
documents up to $25M$ triples. We observe for example that the outcome
of $Q3a$, $Q3b$, and $Q3c$ reflects the selectivities of their
\textsc{Filter} attributes {\small \sf swrc:pages}, {\small \sf swrc:month},
and {\small \sf swrc:isbn} (cf.~Table~\ref{tbl:attributes} and~\ref{tbl:docgen}).
We will come back to the result size listing when discussing the benchmark results
later on.

\subsection{Benchmark Metrics}

Depending on the scenario, we will report on user time ({\tt usr}), system
time ({\tt sys}), and the high watermark of resident memory consumption
({\tt rmem}). These values were extracted from the {\it proc} file
system, whereas we measured elapsed time ({\tt tme}) through
timers. It is important to note that experiments were carried out
on a DuoCore CPU, where the linux kernel sums up the {\tt usr} and
{\tt sys} times of the individual processor units. As a consequence, 
in some scenarios the sum {\tt usr}+{\tt sys} might be greater than
the elapsed time {\tt tme}.

We propose several metrics that capture different aspects of the
evaluation. Reports of the benchmark results would, in the best case,
include all these metrics, but might also ignore metrics
that are irrelevant to the underlying scenario. We propose to 
perform three runs over documents comprising $10k$, $50k$, $250k$,
$1M$, $5M$, and $25M$ triples, using a fixed timeout of $30$min per
query and document, always reporting on the average value over all
three runs and, if significant, the errors within these runs.
We point out that this setting can be evaluated in reasonable time
(typically within few days). If the implementation is fast enough,
nothing prevents the user from adding larger documents. All reports
should, of course, include the hardware and software specifications.
Performance results should list {\tt tme}, and optionally {\tt usr}
and {\tt sys}. In the following, we shortly describe a set of 
interesting metrics.

\begin{enumerate}
\item \textsc{Success Rate}: We propose to separately report on the
success rates for the engine on top of all document sizes, distinguishing
between {\small \sf Success}, {\small \sf Timeout} (e.g.~an execution
time $>30min$ as used in our experiments here), {\small \sf Memory Exhaustion}
(if an additional memory limit was set), and general {\small \sf Errors}.
This metric gives a good survey over scaling properties and might give first
insights into the behavior of engines.
 
\item \textsc{Loading Time}: The user should report on the loading times
for the documents of different sizes. This metric primarily applies to
engines with a database backend and might be ignored for in-memory engines,
where loading is typically part of the evaluation process.

\item \textsc{Per-Query Performance}: The report should include the
individual performance results for all queries over all document sizes.
This metric is more detailed than the \textsc{Success Rate} report and
forms the basis for a deep study of the results, in order to identify
strengths and weaknesses of the tested implementation.

\item \textsc{Global Performance}: We propose to combine the per-query
results into a single performance measure. Here we recommend to list
for execution times the arithmetic as well as the geometric mean,
which is defined as the n$^{th}$ root of the product
over $n$ numbers. In the context of SP$^2$Bench, this means we multiply
the execution time of all $17$ queries (queries that failed should be
ranked with $3600s$, to penalize timeouts and other errors) and compute
the 17$^{th}$ root of this product (for each document size, accordingly).
This metric is well-suited to compare the performance of engines.

\item \textsc{Memory Consumption}: In particular for engines
with a physical backend, the user should also report on the high
watermark of main memory consumption and ideally also the 
average memory consumption over all queries (cf. Table \ref{tbl:means_in_memory}
and \ref{tbl:means_native}). \end{enumerate}

\subsection{Benchmark Results for Selected Engines}
It is beyond the scope of this paper to provide an in-depth comparison
of existing SPARQL engines. Rather than that, we use our metrics to
give first insights into the state-of-the art and exemplarily illustrate
that the benchmark indeed gives valuable hints on bottlenecks in
current implementations. In this line, we are not primarily interested
in concrete values (which, however, might be of great interest in the
general case), but focus on the principal behavior and properties of
engines, e.g.~discuss how they scale with document size. We will exemplarily
discuss some interesting cases and refer the interested reader to the
Appendix for the complete results.

\begin{table*}[t]
\caption{Success rates for queries on RDF documents up to $25M$ triples. Queries are encoded in hexadecimal (e.g., 'A' stands for $Q10$). We use the shortcuts +:=Success, T:=Timeout, M:=Memory Exhaustion, and E:=Error.}
\vspace{-0.3cm}
\begin{center}
\begin{tabular}{lllll}
\toprule[.11em]
\rowcolor{darkgray}
      & {\bf ARQ}                & {\bf Sesame}$_M$         & {\bf Sesame}$_{DB}$      & {\bf Virtuoso}\\
\midrule
Query & {\small \verb!123  45 6789ABC  !} & {\small \verb!123  45 6789ABC  !} & {\small \verb!123  45 6789ABC  !} & {\small \verb!123  45 6789ABC  !}\\
      & {\small \verb!  abc ab         !} & {\small \verb!  abc ab         !} & {\small \verb!  abc ab         !} & {\small \verb!  abc ab         !}\\
\midrule
\rowcolor{lightgray}
10k   & {\small \verb!+++++++++++++++++!} & {\small \verb!+++++++++++++++++!} & {\small \verb!+++++++++++++++++!} & {\small \verb!++++++++E++++++++!}\\
50k   & {\small \verb!+++++++++++++++++!} & {\small \verb!+++++++++++++++++!} & {\small \verb!+++++++++++++++++!} & {\small \verb!++++++++E++++++++!}\\
\rowcolor{lightgray}
250k  & {\small \verb!+++++T+++++++++++!} & {\small \verb!++++++T+T++++++++!} & {\small \verb!++++++T+TT+++++++!} & {\small \verb!+++++TT+E++++++++!}\\
1M    & {\small \verb!+++++TT+TT+++++++!} & {\small \verb!++++++T+TT+++++++!} & {\small \verb!++++++T+TT+++++++!} & {\small \verb!+++++TTTET+++++++!}\\
\rowcolor{lightgray}
5M    & {\small \verb!+++++TT+TT+++++++!} & {\small \verb!+++++TT+TT+++++++!} & {\small \verb!+++++MT+TT+++++++!} & {\small \verb!+++++TTTET+++++++!}\\
25M   & {\small \verb!TTTTTTTTTTTTTTTTT!} & {\small \verb!MMMMMMMTMMMMMTMMT!} & {\small \verb!+++++TT+TT+++++++!} & (loading of document failed)\\
\bottomrule[.11em]
\end{tabular}
\end{center}
\label{tbl:succqueries}
\vspace{-0.3cm}
\end{table*}

We conducted benchmarks for (1)~the Java engine
{\bf ARQ}\footnote{http://jena.sourceforge.net/ARQ/} v2.2 on top of
Jena~2.5.5, (2)~the {\bf Redland}\footnote{http://librdf.org/}
RDF Processor v1.0.7 (written in C), using the Raptor Parser
Toolkit v.1.4.16 and Rasqal Library v0.9.15,
(3)~{\bf SDB}\footnote{http://jena.sourceforge.net/SDB/},
which link ARQ to an SQL database back-end (i.e., we used mysql v5.0.34) , 
(4)~the Java implementation {\bf Sesame}\footnote{http://www.openrdf.org/}
v2.2beta2, and finally~(5)~OpenLink
{\bf Virtuoso}\footnote{http://www.openlinksw.com/virtuoso/}
v5.0.6 (written in C).

For Sesame we tested two configurations: {\it Sesame}$_M$, which
processes queries in memory, and {\it Sesame}$_{DB}$, which
stores data physically on disk, using the native {\it Mulgara}
SAIL (v1.3beta1). We thus distinguish between the in-memory engines
({\it ARQ}, {\it Sesame}$_M$) and engines with physical backend,
namely ({\it Redland}, {\it SBD}, {\it Sesame}$_{DB}$,
{\it Virtuoso}). The latter can further be divided into engines with
a native RDF store ({\it Redland}, {\it Sesame}$_{DB}$, {\it Virtuoso})
and a relational database backend ({\it SDB}). For all
physical-backend databases we created indices wherever possible
(immediately after loading the documents) and consider loading
and execution time separately (index creation time is included in
the reported loading times).

We performed three cold runs over all queries and documents 
of $10k$, $50k$, $250k$, $1M$, $5M$, and $25M$ triples, i.e.~in-between
each two runs we restarted the engines and cleared the
database. We set a timeout of $30$min ({\tt tme})
per query and a memory limit of $2.6\textit{GB}$, either using {\it ulimit} or
restricting the JVM (for higher limits, the initialization of the JVM failed).
Negative and positive variation of the average
(over the runs) was $<2\%$ in almost all cases, so we omit error bars.
For {\it SDB} and {\it Virtuoso}, which follow a client-server
architecture, we monitored both processes and sum up these values.

We verified all results by comparing the outputs,
observing that {\it SDB} and {\it Redland} returned wrong results
for a couple of queries, so we restrict ourselves on the discussion
of the remaining four engines. Table~\ref{tbl:succqueries} shows the
success rates. All queries that are not listed succeeded, except for
{\it ARQ} and {\it Sesame$_M$} on the $25M$ document (either due to
timeout or memory exhaustion) and Virtuoso on $Q6$ (due to missing
standard compliance). Hence, $Q4$, $Q5a$, $Q6$, and $Q7$ are the
most challenging queries, where we observe many timeouts even for
small documents. Note that we did not succeed in loading the $25M$
triples document into the {\it Virtuoso} database.

\begin{table*}[t]
\vspace{0.2cm}
\caption{Number of query results on documents up to $25$ million triples}
\vspace{-0.3cm}
\begin{center}
{\small
\begin{tabular}{l|rrrrrrrrrrrrrrr}
\toprule[.11em]
\rowcolor{darkgray}
{\bf Query}  & $Q1$ & $Q2$ & $Q3a$ & $Q3b$ & $Q3c$ & $Q4$ & $Q5a$ & $Q5b$ & $Q6$ & $Q7$ & $Q8$ & $Q9$ & $Q10$ & $Q11$\\
\midrule
$10k$  & 1 & 147     & 846    & 9    & 0 & 23226    & 155    & 155    & 229     & 0    & 184 & 4 & 166 & 10\\
\rowcolor{lightgray}
$50k$  & 1 & 965     & 3647   & 25   & 0 & 104746   & 1085   & 1085   & 1769    & 2    & 264 & 4 & 307 & 10\\
$250k$ & 1 & 6197    & 15853  & 127  & 0 & 542801   & 6904   & 6904   & 12093   & 62   & 332 & 4 & 452 & 10\\
\rowcolor{lightgray}
$1M$   & 1 & 32770   & 52676  & 379  & 0 & 2586733  & 35241  & 35241  & 62795   & 292  & 400 & 4 & 572 & 10\\
$5M$   & 1 & 248738  & 192373 & 1317 & 0 & 18362955 & 210662 & 210662 & 417625  & 1200 & 493 & 4 & 656 & 10\\
\rowcolor{lightgray}
$25M$  & 1 & 1876999 & 594890 & 4075 & 0 & n/a      & 696681 & 696681 & 1945167 & 5099 & 493 & 4 & 656  & 10\\
\bottomrule[.11em]
\end{tabular}
}
\end{center}
\label{tbl:resultsizes}
\vspace{-0.4cm}
\end{table*}


\subsection{Discussion of Benchmark Results}
\label{subsec:bench}

{\bf Main Memory.} For the in-memory engines we observe that the high
watermark of main memory consumption during query evaluation increases
sublinearly to document size (cf.~Table \ref{tbl:means_in_memory}),
e.g.~for ARQ we measured an average (over runs and queries) of
$85\textit{MB}$ on $10k$, $166\textit{MB}$ on $50k$, $318\textit{MB}$ on
$250k$, $526\textit{MB}$ on $1M$,
and $1.3\textit{GB}$ on $5M$ triples. Somewhat surprisingly, also the
memory consumption of the native engines {\it Virtuoso} and {\it Sesame}$_{DB}$
increased with document size.

{\bf Arithmetic and Geometric Mean.} For the in-memory engines
we observe that {\it Sesame$_M$} is superior to {\it ARQ} regarding 
both means (see Table~\ref{tbl:means_in_memory}). For instance,
the arithmetic ($T_a$) and geometric ($T_g$)
mean for the engines on the $1M$
document over all queries\footnote{We always penalize failure queries
with $3600s$.} are $T_a^{SesM}=683.16s$,
$T_g^{SesM}=106.84s$, $T_a^{ARQ}=901.73s$, and $T_g^{ARQ}=179.42s$.

For the native engines on $1M$ triples (cf.~Table~\ref{tbl:means_native})
we have $T_a^{SesDB}=653.17s$, $T_g^{SesDB}=10.17s$, $T_a^{Virt}=850.06s$,
and $T_g^{Virt}=3.03s$. The arithmetic mean of {\it Sesame$_{DB}$}
is superior, which is mainly due to the fact that it failed only on $4$
(vs.~$5$) queries. The geometric mean moderates the impact of these
outliers. {\it Virtuoso} shows a better overall performance for the
success queries, so its geometric mean is superior.

\begin{figure*}[t]
\mbox{
\hspace{-0.9cm}
\subfigure{\rotatebox{270}{\includegraphics[scale=0.20]{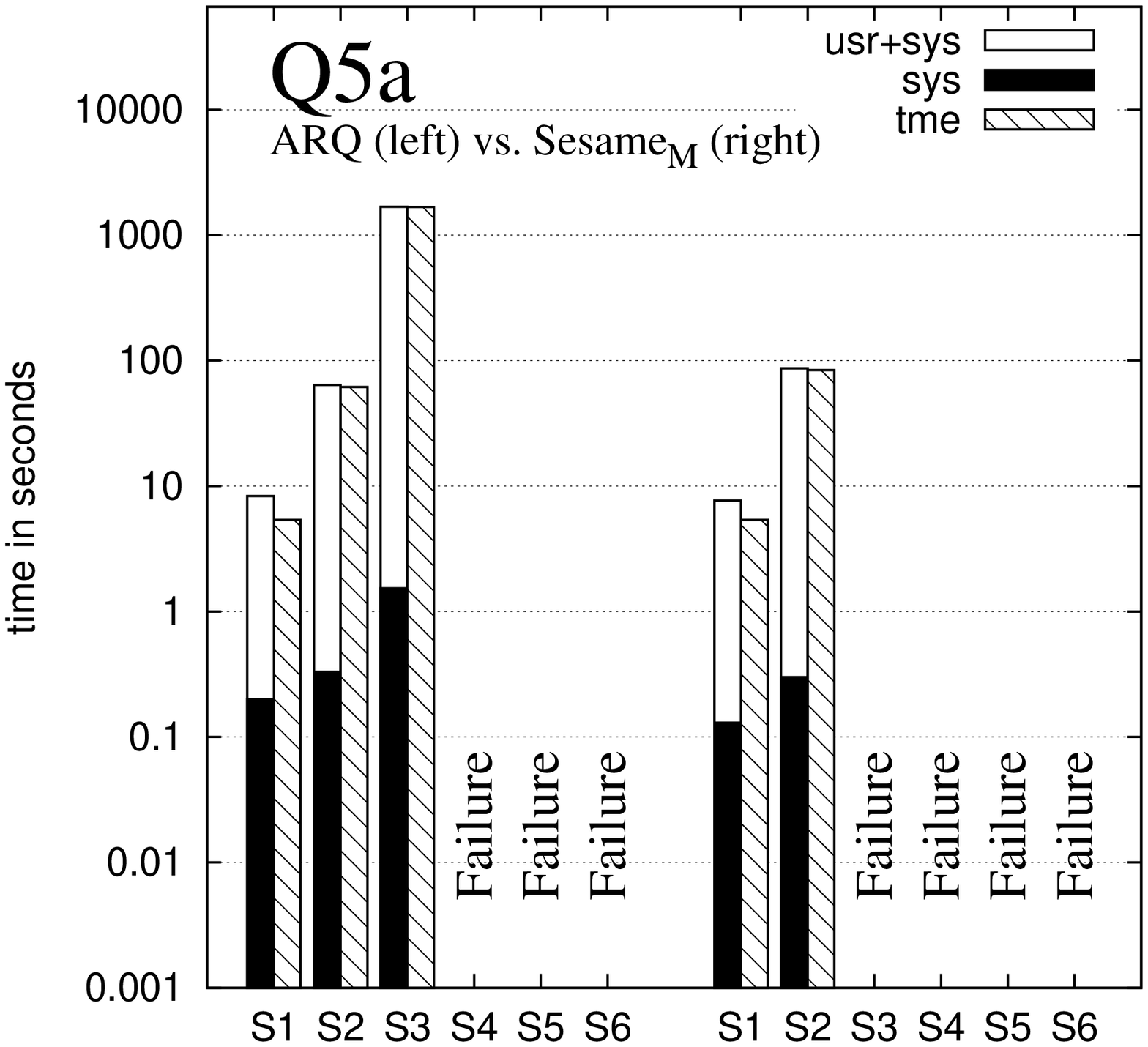}}}
\hspace{-1.7cm}
\subfigure{\rotatebox{270}{\includegraphics[scale=0.20]{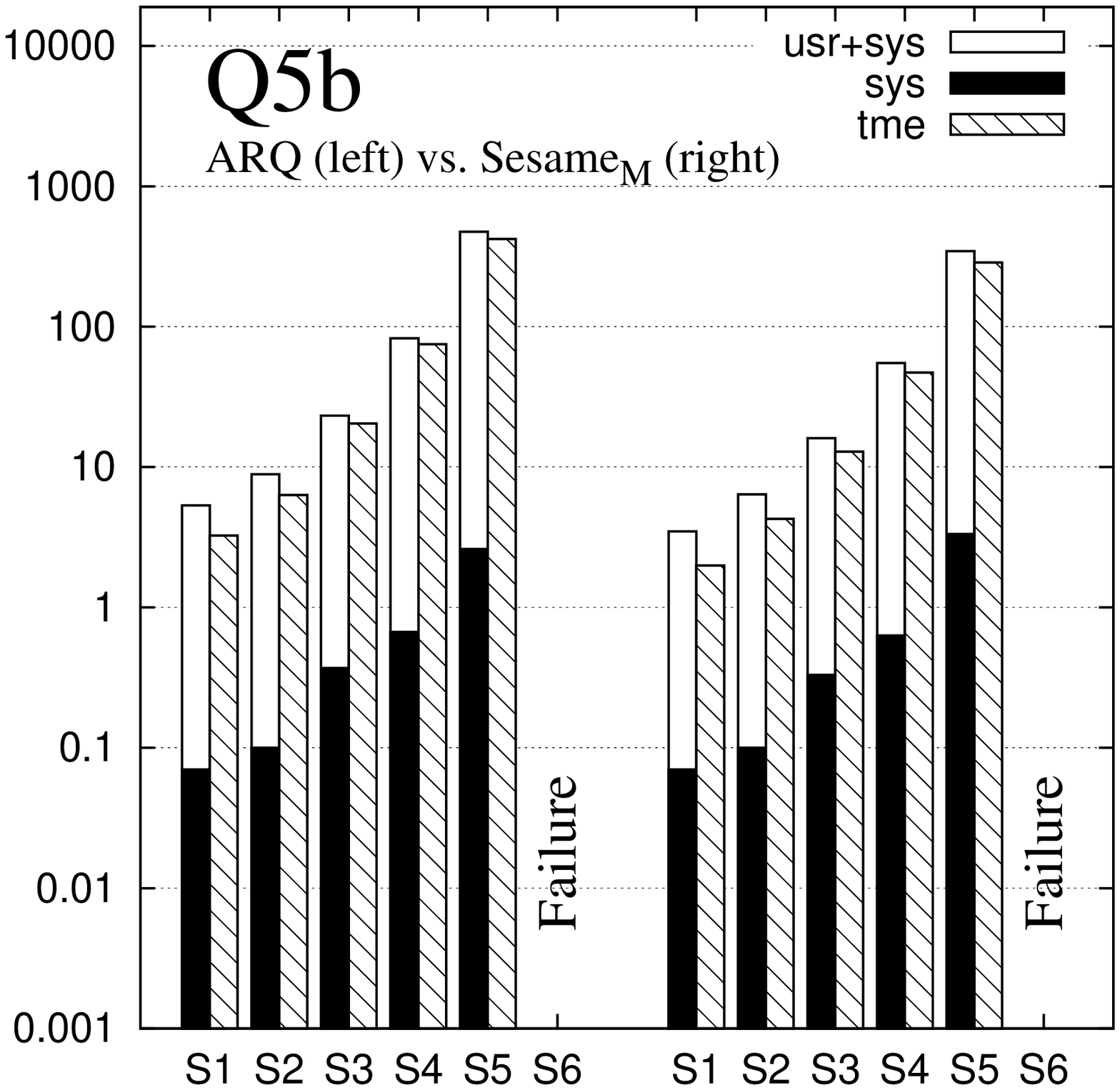}}}
\hspace{-1.7cm}
\subfigure{\rotatebox{270}{\includegraphics[scale=0.20]{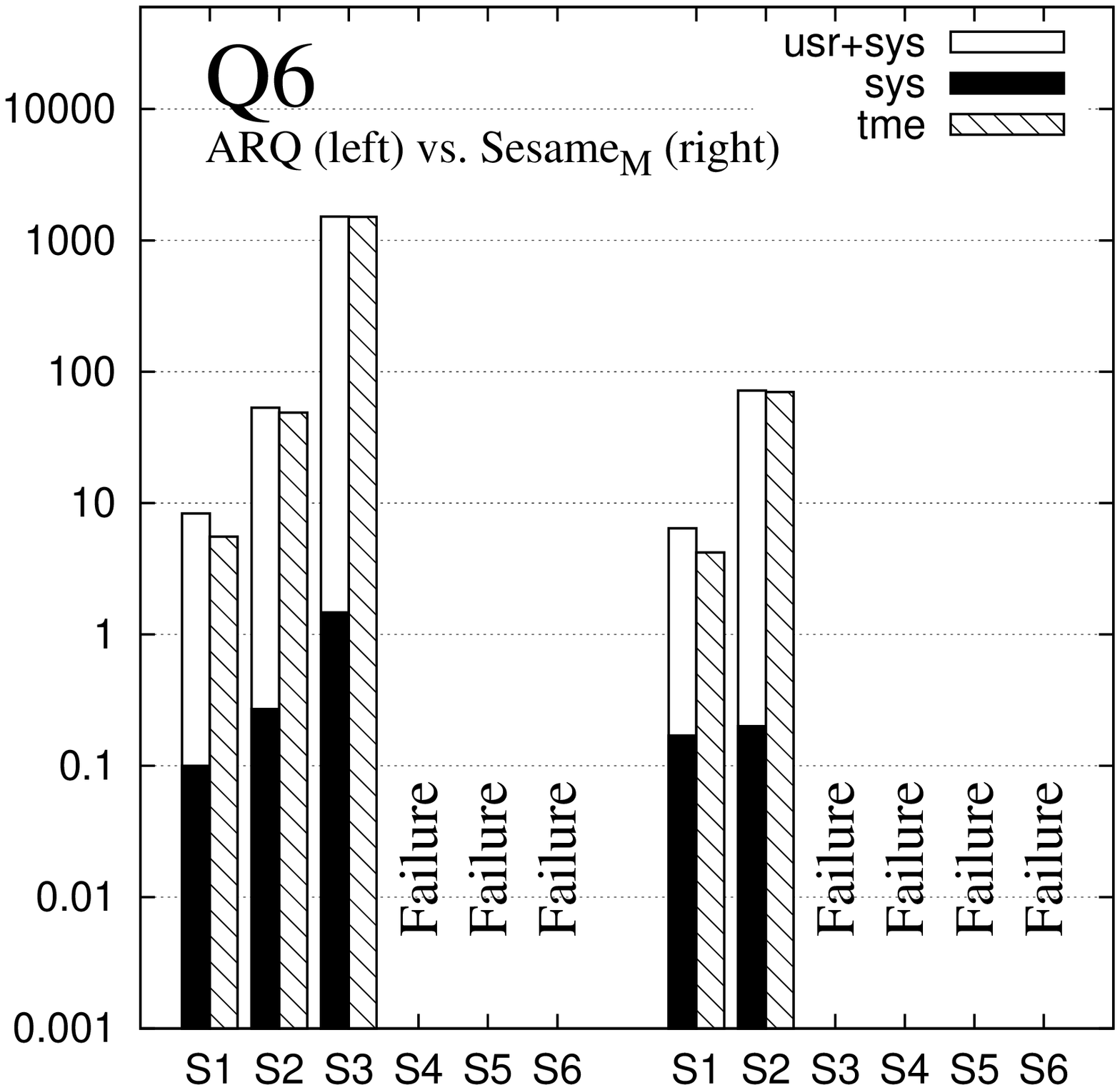}}}
\hspace{-1.7cm}
\subfigure{\rotatebox{270}{\includegraphics[scale=0.20]{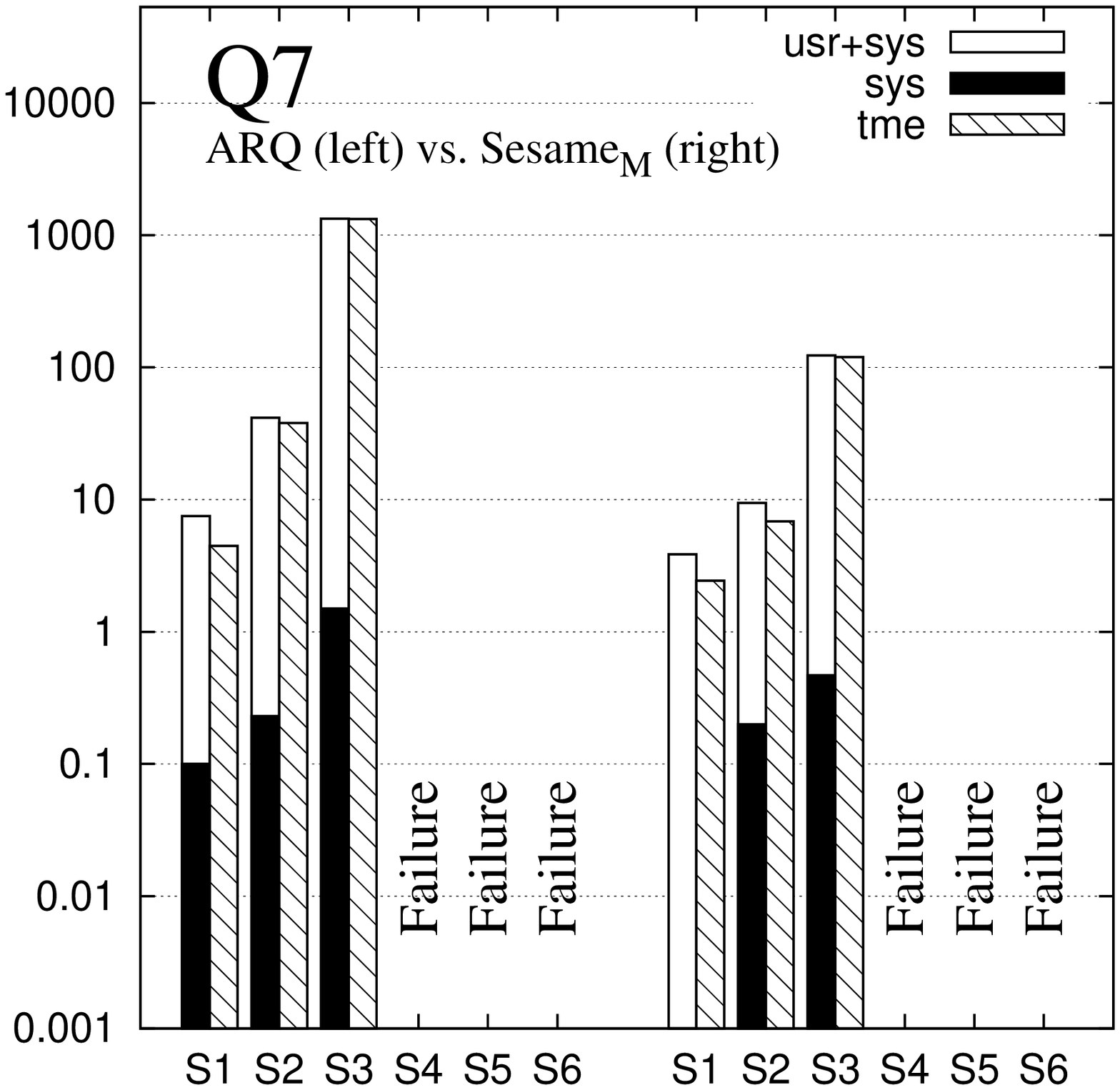}}}
\hspace{-1.7cm}
\subfigure{\rotatebox{270}{\includegraphics[scale=0.20]{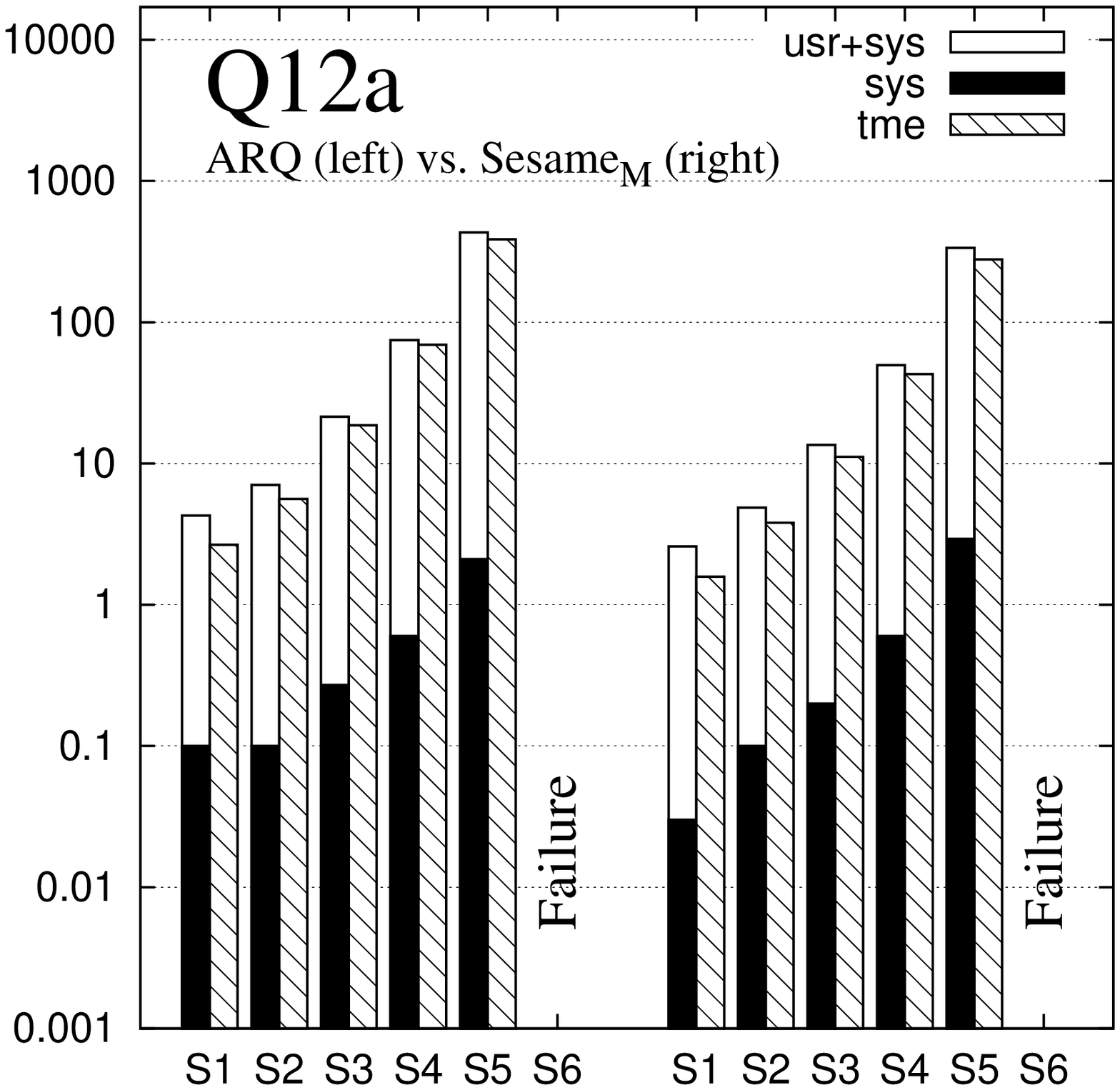}}}
}
\\[-0.2cm]
\mbox{
\hspace{-0.9cm}
\subfigure{\rotatebox{270}{\includegraphics[scale=0.20]{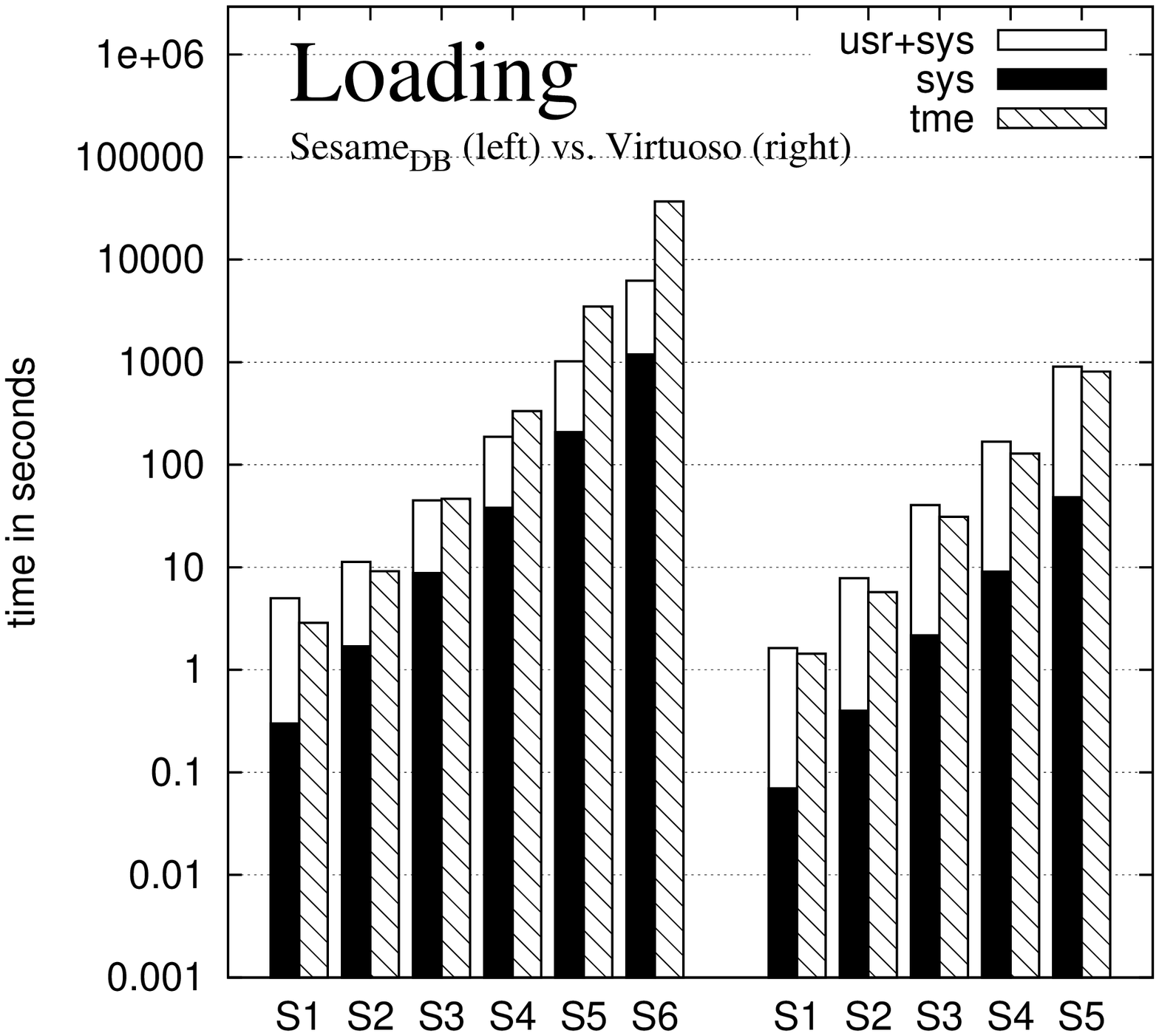}}}
\hspace{-1.7cm}
\subfigure{\rotatebox{270}{\includegraphics[scale=0.20]{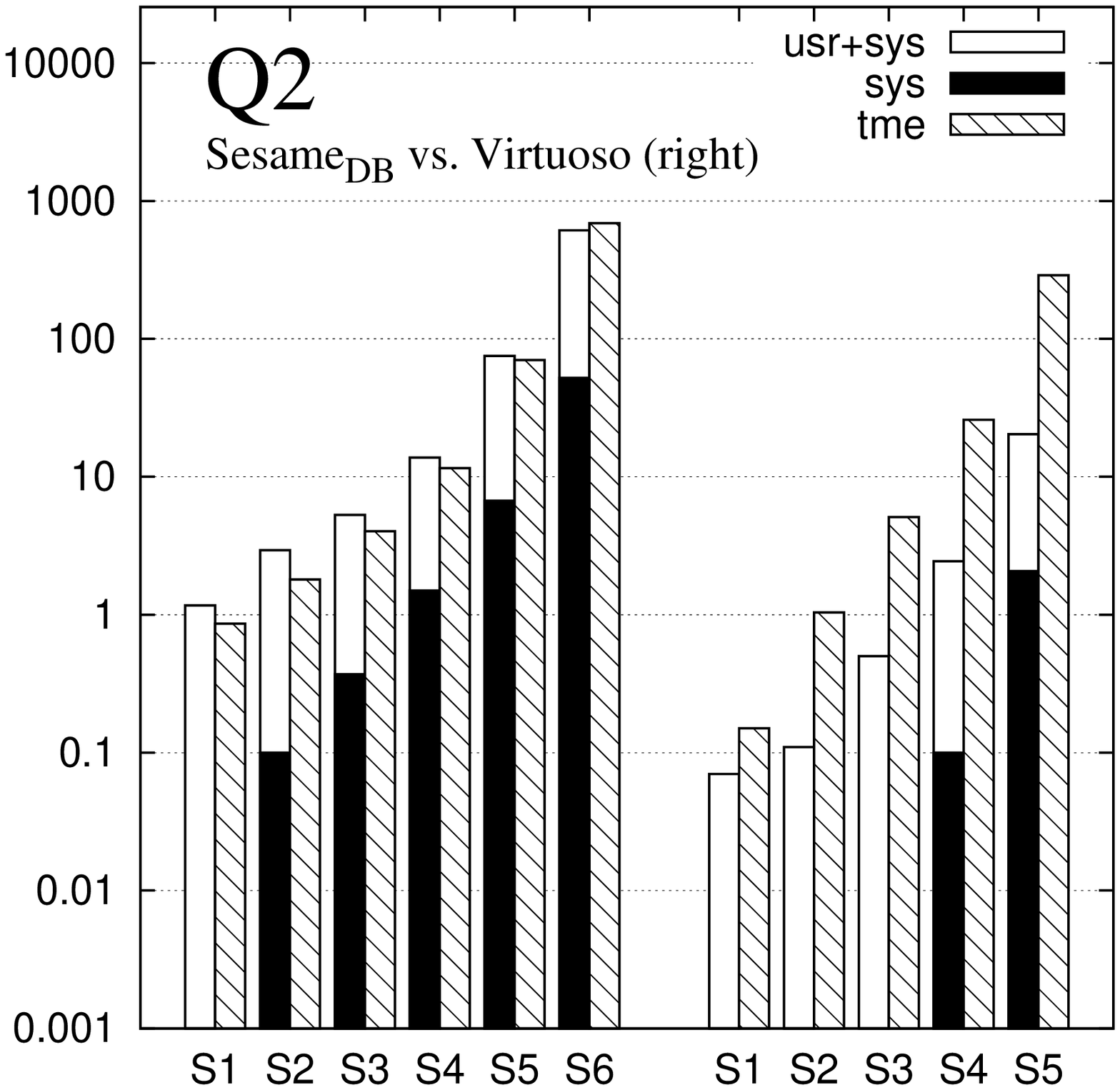}}}
\hspace{-1.7cm}
\subfigure{\rotatebox{270}{\includegraphics[scale=0.20]{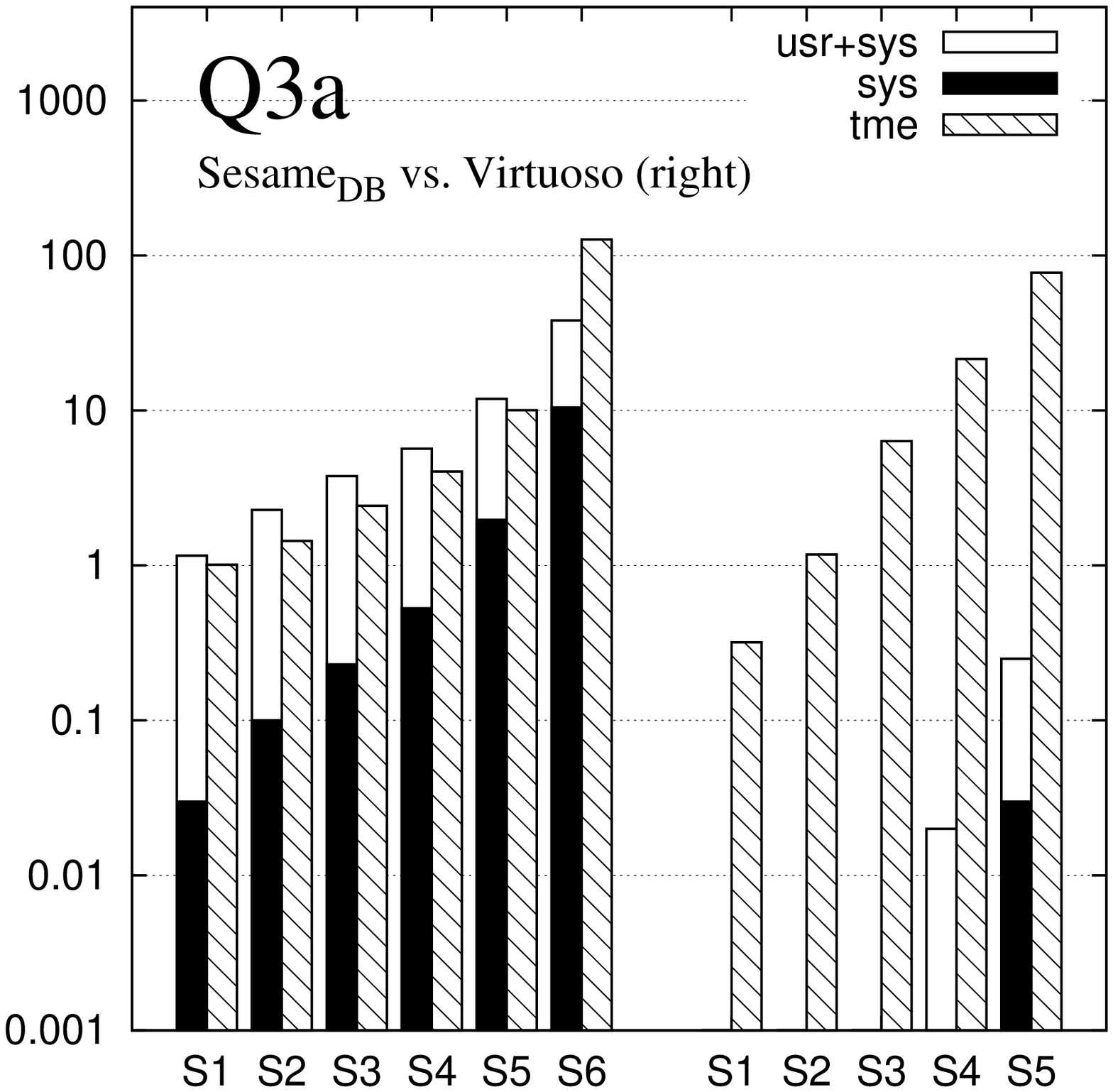}}}
\hspace{-1.7cm}
\subfigure{\rotatebox{270}{\includegraphics[scale=0.20]{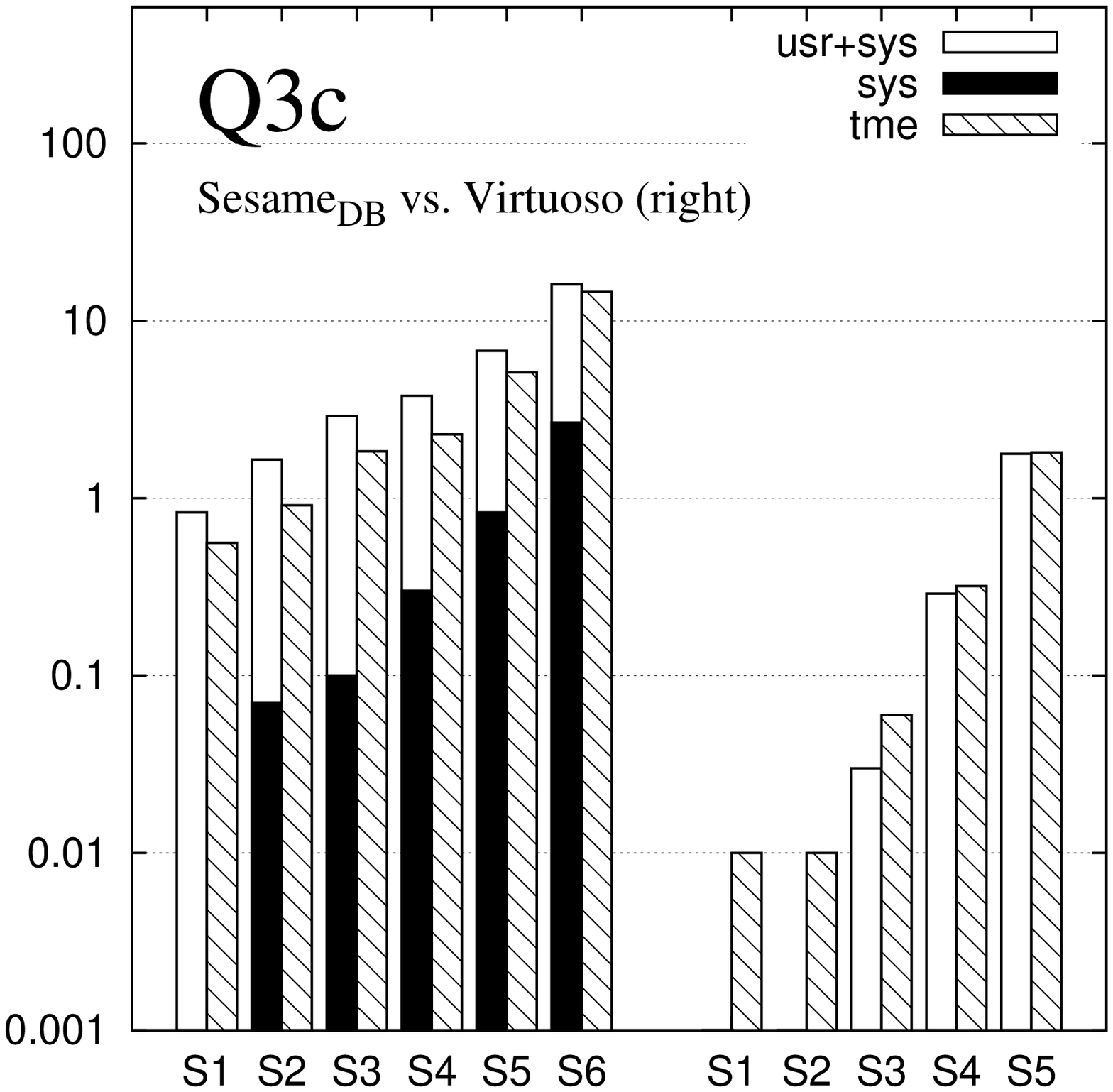}}}
\hspace{-1.7cm}
\subfigure{\rotatebox{270}{\includegraphics[scale=0.20]{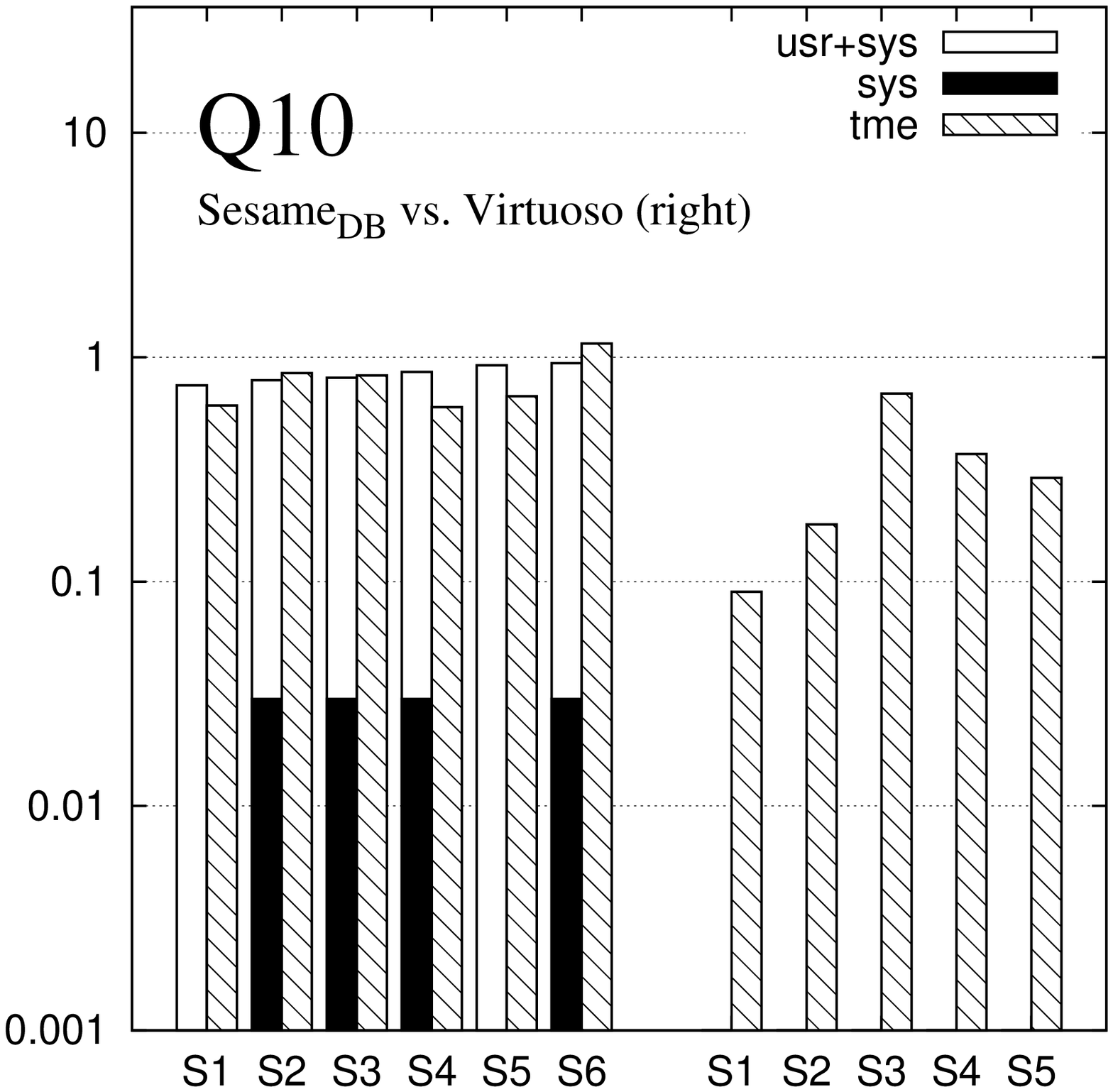}}}
}
\vspace{-0.45cm}
\caption{Results for in-memory engines (top) and native engines (bottom) on S1=10k, S2=50k, S3=250k, S4=1M, S5=5M, and S6=25M triples}
\label{fig:combined}
\vspace{-0.4cm}
\end{figure*}

\begin{table}[t]
\caption{Arithmetic and geometric means of execution time ($\mbox{T}_{a}$/$\mbox{T}_g$) and arithmetic mean of memory consumption ($\mbox{M}_a$) for the in-memory engines}
\vspace{-0.5cm}
\begin{center}
\begin{tabular}{r|rrr|rrr}
\toprule[.11em]
\rowcolor{darkgray}
      & \multicolumn{3}{>{\columncolor{darkgray}}c|}{{\bf ARQ}} & \multicolumn{3}{>{\columncolor{darkgray}}c}{{\bf Sesame}$_M$}\\
\midrule
\rowcolor{darkgray}
 & \multicolumn{1}{>{\columncolor{darkgray}}c}{$\mbox{T}_a$[s]} & \multicolumn{1}{>{\columncolor{darkgray}}c}{$\mbox{T}_g$[s]} & \multicolumn{1}{>{\columncolor{darkgray}}c|}{$\mbox{M}_a$[MB]} & \multicolumn{1}{>{\columncolor{darkgray}}c}{$\mbox{T}_a$[s]} & \multicolumn{1}{>{\columncolor{darkgray}}c}{$\mbox{T}_g$[s]} & \multicolumn{1}{>{\columncolor{darkgray}}c}{$\mbox{M}_a$[MB]}\\
\midrule
250k  & 491.87  & 56.35  & 318.25  & 442.47  & 28.64  & 272.27\\
\rowcolor{lightgray}
1M    & 901.73  & 179.42 & 525.61  & 683.16  & 106.38 & 561.79\\
5M    & 1154.80 & 671.41 & 1347.55 & 1059.03 & 506.14 & 1437.38\\
\bottomrule[.11em]
\end{tabular}
\end{center}
\label{tbl:means_in_memory}
\vspace{-0.3cm}
\end{table}

\begin{table}[t]
\caption{Arithmetic and geometric means of execution time ($\mbox{T}_a$/$\mbox{T}_g$) and arithmetic mean of memory consumption ($\mbox{M}_{a}$) for the native engines}
\vspace{-0.3cm}
\begin{center}
\begin{tabular}{r|rrr|rrr}
\toprule[.11em]
\rowcolor{darkgray}
      & \multicolumn{3}{>{\columncolor{darkgray}}c|}{{\bf Sesame}$_{DB}$}  & \multicolumn{3}{>{\columncolor{darkgray}}c}{\bf Virtuoso}\\
\midrule
\rowcolor{darkgray}
 & \multicolumn{1}{>{\columncolor{darkgray}}c}{$\mbox{T}_{a}$[s]} & \multicolumn{1}{>{\columncolor{darkgray}}c}{$\mbox{T}_{g}$[s]} & \multicolumn{1}{>{\columncolor{darkgray}}c|}{$\mbox{M}_{a}$[MB]} & \multicolumn{1}{>{\columncolor{darkgray}}c}{$\mbox{T}_{a}$[s]} & \multicolumn{1}{>{\columncolor{darkgray}}c}{$\mbox{T}_{g}$[s]} & \multicolumn{1}{>{\columncolor{darkgray}}c}{$\mbox{M}_{a}$[MB]} \\
\midrule
250k  & 639.86 & 6.79  & 73.92  & 546.31 & 1.31 & 377.60 \\
\rowcolor{lightgray}
1M    & 653.17 & 10.17 & 145.97 & 850.06 & 3.03 & 888.72 \\
5M    & 860.33 & 22.91 & 196.33 & 870.16 & 8.96 & 1072.84 \\
\bottomrule[.11em]
\end{tabular}
\end{center}
\label{tbl:means_native}
\vspace{-0.3cm}
\end{table}


{\bf In-memory Engines.}
Figure~\ref{fig:combined} (top) plot selected results for in-memory
engines. We start with $Q5a$ and $Q5b$. Although both compute the same
result, the engines perform much better for the explicit join in $Q5b$.
We may suspect that the implicit join in $Q5a$ is not recognized,
i.e.~that both engines compute the cartesian product and apply the
filter afterwards.

$Q6$ and $Q7$ implement simple and double negation, respectively.
Both engines show insufficient behavior. At the first glance, we might
expect that $Q7$ (which involves double negation) is more complicated
to evaluate, but we observe that {\it Sesame$_M$} scales even worse for $Q6$.
We identify two possible explanations. First, $Q7$ ``negates'' documents
with incoming citations, but -- according to Section~\ref{subsec:citations}
-- only a small fraction of papers has incoming citations at all. In
contrast, $Q6$ negates arbitrary documents, i.e.~a much larger set.
Another reasonable cause might be the non-equality filter
subexpression \verb!?yr2 < ?yr! inside the inner \textsc{Filter} of $Q6$.

For \textsc{Ask} query $Q12a$ both engines scale linearly with document
size. However, from Table~\ref{tbl:resultsizes} and the fact that our data
generator is incremental and deterministic, we know that a
``witness'' is already contained in the first $10k$ triples of the
document. It might be located even without reading the whole document, so both
evaluation strategies are suboptimal.

{\bf Native Engines.} The leftmost plot at the bottom of
Figure~\ref{fig:combined} shows the loading times for the native engines
{\it Sesame}$_{DB}$ and {\it Virtuoso}. Both engines scale well concerning
{\tt usr} and {\tt sys}, essentially linear to document size.
For {\it Sesame}$_{DB}$, however, {\tt tme} grows superlinearly
(e.g., loading of the $25M$ document is about ten times slower than
loading of the $5M$ document). This might cause problems for larger documents.

The running times for $Q2$ increase superlinear for both engines (in
particular for larger documents). This reflects the superlinear growth
of inproceedings and the growing result size
(cf.~Tables~\ref{tbl:docgen} and~\ref{tbl:resultsizes}).
What is interesting here is the significant difference between
{\tt usr+sys} and {\tt tme} for {\it Virtuoso}, which indicates
disproportional disk I/O. Since {\it Sesame} does not exhibit
this peculiar behavior, it might be an interesting starting point
for further optimizations in the {\it Virtuoso} engine.

Queries $Q3a$ and $Q3c$ have been designed to test the intelligent
choice of indices in the context of \textsc{Filter} expressions with
varying selectivity. {\it Virtuoso} gets by with an economic
consumption of {\tt usr} and {\tt sys} time for both queries,
which suggests that it makes heavy use of indices. While this strategy
pays off for $Q3c$, the elapsed time for $Q3a$ is unreasonably high
and we observe that {\it Sesame$_M$} scales better for this query.

$Q10$ extracts subjects and predicates that are associated with
{\it Paul Erd\"os}. First recall that, for each year up to $1996$,
{\it Paul Erd\"os} has exactly $10$ publications and occurs twice as
editor (cf.~Section~\ref{sec:datagen}). Both engines answer this query
in about constant time, which is possible due to the upper result size
bound (cf.~Table~\ref{tbl:resultsizes}). Regarding {\tt usr+sys},
{\it Virtuoso} is even more efficient: These times are
diminishing in all cases.
Hence, this query constitutes an example for desired engine behavior.
\begin{table}
\caption{Characteristics of generated documents}
\vspace{-0.2cm}
{\footnotesize
\begin{tabular}{lrrrrrrr}
\toprule[.11em]
\rowcolor{darkgray}
\#{\bf Triples} & {\bf 10k} & {\bf 50k} & {\bf 250k} & {\bf 1M} & {\bf 5M} & {\bf 25M}\\
\midrule
file size [MB]        & 1.0  & 5.1  & 26    & 106    & 533    & 2694\\
\rowcolor{lightgray}
data up to            & 1955 & 1967 & 1979  & 1989   & 2001   & 2015\\
\midrule
\#{\tt Tot.Auth.}     & 1.5k  & 6.8k & 34.5k & 151.0k & 898.0k & 5.4M\\
\rowcolor{lightgray}
\#{\tt Dist.Auth.}    & 0.9k  & 4.1k & 20.0k & 82.1k  & 429.6k & 2.1M\\
\midrule
\#{\tt Journals}	  &  25  &  104 &  439  & 1.4k   & 4.6k   & 11.7k\\
\rowcolor{lightgray}
\#{\tt Articles}	  & 916  & 4.0k & 17.1k & 56.9k  & 207.8k & 642.8k\\
\#{\tt Proc.}		  &   6  &   37 &   213 &   903  &   4.7k & 24.4k\\
\rowcolor{lightgray}
\#{\tt Inproc.}	      & 169  & 1.4k &  9.2k & 43.5k  & 255.2k & 1.5M\\
\#{\tt Incoll.}       &  18  &   56 &  173  &   442  & 1.4k   & 4.5k\\
\rowcolor{lightgray}
\#{\tt Books}         &   0  &   0  &    39 &    356 &    973 & 1.7k\\
\#{\tt PhD Th.}       &   0  &   0  &     0 &    101 &    237 & 365\\
\rowcolor{lightgray}
\#{\tt Mast.Th.}      &   0  &   0  &     0 &     50 &     95 & 169\\
\#{\tt WWWs}          &   0  &   0  &     0 &     35 &     92 & 168\\ 
\bottomrule[.11em]
\end{tabular}
}
\label{tbl:docgen}
\end{table}


\section{Conclusion}
\label{sec:conclusion}

We have presented the SP$^2$Bench performance benchmark for SPARQL,
which constitutes the first methodical approach for testing the
performance of SPARQL engines w.r.t.~different operator
constellations, RDF access paths, typical RDF constructs, and a
variety of possible optimization approaches.

Our data generator relies on a deep study of DBLP. Although it is
not possible to mirror {\it all} correlations found in the original
DBLP data (e.g., we simplified when assuming independence between
attributes in Section~\ref{subsec:structure}), many aspects are modeled
in faithful detail and the queries are designed in such a way
that they build on exactly those aspects, which makes them
realistic, understandable, and predictable.

Even without knowledge about the internals of engines, we identified
deficiencies and reasoned about suspected causes.
We expect the benefit of our benchmark to be even higher
for developers that are familiar with the engine internals.

To give another proof of concept, in~\cite{shklp2008} we have
successfully used SP$^2$Bench to identify previously unknown
limitations of RDF storage schemes: Among others, we
identified scenarios where the advanced vertical storage
scheme from~\cite{ammh2007} was slower than a simple triple store approach.

With the understandable DBLP scenario we clear the way for
coming language modifications. For instance, SPARQL update and aggregation
support are currently discussed as possible
extensions.\footnote{See http://esw.w3.org/topic/SPARQL/Extensions.}
Updates, for instance, could be realized by minor extensions to our
data generator. Concerning aggregations, the detailed knowledge of
the document class counts and distributions (cf.~Section~\ref{sec:dblp})
facilitates the design of challenging aggregate
queries with fixed characteristics.

\begin{small}
\bibliographystyle{IEEEtran}
\bibliography{sparqlbench}
\end{small}

\newpage
\begin{tabbing}
xxx \= \kill
\>
\end{tabbing}
\newpage

\begin{appendix}
\section{Benchmark Queries}
\label{sec:benchmarkqueries}

\vspace{0.1cm}

\noindent
\begin{boxedminipage}{8.7cm}
{\footnotesize
\begin{tabbing}
xxxxxxxxxxxxxxxxxxxxxxxxxxxxxxxxxxxxxxxxxxxxxxxxxxxxxxxl \= \kill
\verb!SELECT ?yr!\>\fbox{\bf Q1}\\
\verb!WHERE {!\\
\verb! ?journal rdf:type bench:Journal.!\\
\verb! ?journal dc:title "Journal 1 (1940)"^^xsd:string.!\\
\verb! ?journal dcterms:issued ?yr }!
\end{tabbing}
}
\end{boxedminipage}

\vspace{0.05cm}

\noindent
\begin{boxedminipage}{8.7cm}
{\footnotesize
\begin{tabbing}
xxxxxxxxxxxxxxxxxxxxxxxxxxxxxxxxxxxxxxxxxxxxxxxxxxxxxxxl \= \kill
\verb!SELECT ?inproc ?author ?booktitle ?title !\>\fbox{\bf Q2}\\
\verb!       ?proc ?ee ?page ?url ?yr ?abstract!\\
\verb!WHERE {!\\
\verb!  ?inproc rdf:type bench:Inproceedings.!\\
\verb!  ?inproc dc:creator ?author.!\\
\verb!  ?inproc bench:booktitle ?booktitle.!\\
\verb!  ?inproc dc:title ?title.!\\
\verb!  ?inproc dcterms:partOf ?proc.!\\
\verb!  ?inproc rdfs:seeAlso ?ee.!\\
\verb!  ?inproc swrc:pages ?page.!\\
\verb!  ?inproc foaf:homepage ?url.!\\
\verb!  ?inproc dcterms:issued ?yr!\\
\verb!  OPTIONAL { ?inproc bench:abstract ?abstract }!\\
\verb!} ORDER BY ?yr!
\end{tabbing}
}
\end{boxedminipage}

\vspace{0.05cm}

\noindent
\begin{boxedminipage}{8.7cm}
{\footnotesize
\begin{tabbing}
xxxl\=xxxxxxxxxxxxxxxxxxxxxxxxxxxxxxxxxxxxxxxxxxxxxxxxxxxx \= \kill
(a)\>\verb!SELECT ?article!\>\fbox{\bf Q3}\\
\>\verb!WHERE { ?article rdf:type bench:Article.!\\
\>\verb!        ?article ?property ?value!\\
\>\verb!        FILTER (?property=swrc:pages) }!\\
\\[-0.3cm]
(b)\>\verb!Q3a, but "swrc:month" instead of "swrc:pages"!\\
\\[-0.3cm]
(c)\>\verb!Q3a, but "swrc:isbn" instead of "swrc:pages"!
\end{tabbing}
}
\end{boxedminipage}

\vspace{0.04cm}

\noindent
\begin{boxedminipage}{8.7cm}
{\footnotesize
\begin{tabbing}
xxxxxxxxxxxxxxxxxxxxxxxxxxxxxxxxxxxxxxxxxxxxxxxxxxxxxxxl \= \kill
\verb!SELECT DISTINCT ?name1 ?name2!\>\fbox{\bf Q4}\\
\verb!WHERE { ?article1 rdf:type bench:Article.!\\
\verb!        ?article2 rdf:type bench:Article.!\\
\verb!        ?article1 dc:creator ?author1.!\\
\verb!        ?author1 foaf:name ?name1.!\\
\verb!        ?article2 dc:creator ?author2.!\\
\verb!        ?author2 foaf:name ?name2.!\\
\verb!        ?article1 swrc:journal ?journal.!\\
\verb!        ?article2 swrc:journal ?journal!\\
\verb!        FILTER (?name1<?name2) }!
\end{tabbing}
}
\end{boxedminipage}

\vspace{0.05cm}

\noindent
\begin{boxedminipage}{8.7cm}
{\footnotesize
\begin{tabbing}
xxxl\=xxxxxxxxxxxxxxxxxxxxxxxxxxxxxxxxxxxxxxxxxxxxxxxxxxxx \= \kill
(a)\>\verb!SELECT DISTINCT ?person ?name!\>\fbox{\bf Q5}\\
\>\verb!WHERE { ?article rdf:type bench:Article.!\\
\>\verb!        ?article dc:creator ?person.!\\
\>\verb!        ?inproc rdf:type bench:Inproceedings.!\\
\>\verb!        ?inproc dc:creator ?person2.!\\
\>\verb!        ?person foaf:name ?name.!\\
\>\verb!        ?person2 foaf:name ?name2!\\
\>\verb!        FILTER(?name=?name2) }!\\
\\[-0.2cm]
(b)\>\verb!SELECT DISTINCT ?person ?name!\\
\>\verb!WHERE { ?article rdf:type bench:Article.!\\
\>\verb!        ?article dc:creator ?person.!\\
\>\verb!        ?inproc rdf:type bench:Inproceedings.!\\
\>\verb!        ?inproc dc:creator ?person.!\\
\>\verb!        ?person foaf:name ?name }!
\end{tabbing}
}
\end{boxedminipage}

\vspace{0.05cm}

\noindent
\begin{boxedminipage}{8.7cm}
{\footnotesize
\begin{tabbing}
xxxxxxxxxxxxxxxxxxxxxxxxxxxxxxxxxxxxxxxxxxxxxxxxxxxxxxxl \= \kill
\verb!SELECT ?yr ?name ?doc!\>\fbox{\bf Q6}\\
\verb!WHERE {!\\
\verb!  ?class rdfs:subClassOf foaf:Document.!\\
\verb!  ?doc rdf:type ?class.!\\
\verb!  ?doc dcterms:issued ?yr.!\\
\verb!  ?doc dc:creator ?author.!\\
\verb!  ?author foaf:name ?name!\\
\verb!  OPTIONAL {!\\
\verb!    ?class2 rdfs:subClassOf foaf:Document.!\\
\verb!    ?doc2 rdf:type ?class2.!\\
\verb!    ?doc2 dcterms:issued ?yr2.!\\
\verb!    ?doc2 dc:creator ?author2!\\
\verb!    FILTER (?author=?author2 && ?yr2<?yr) }!\\
\verb;  FILTER (!bound(?author2)) };
\end{tabbing}
}
\end{boxedminipage}

\vspace{0.05cm}

\noindent
\begin{boxedminipage}{8.7cm}
{\footnotesize
\begin{tabbing}
xxxxxxxxxxxxxxxxxxxxxxxxxxxxxxxxxxxxxxxxxxxxxxxxxxxxxxxl \= \kill
\verb!SELECT DISTINCT ?title!\>\fbox{\bf Q7}\\
\verb!WHERE {!\\
\verb!  ?class rdfs:subClassOf foaf:Document.!\\
\verb!  ?doc rdf:type ?class.!\\
\verb!  ?doc dc:title ?title.!\\
\verb!  ?bag2 ?member2 ?doc.!\\
\verb!  ?doc2 dcterms:references ?bag2!\\
\verb!  OPTIONAL {!\\
\verb!    ?class3 rdfs:subClassOf foaf:Document.!\\
\verb!    ?doc3 rdf:type ?class3.!\\
\verb!    ?doc3 dcterms:references ?bag3.!\\
\verb!    ?bag3 ?member3 ?doc!\\
\verb!    OPTIONAL {!\\
\verb!      ?class4 rdfs:subClassOf foaf:Document.!\\
\verb!      ?doc4 rdf:type ?class4.!\\
\verb!      ?doc4 dcterms:references ?bag4.!\\
\verb!      ?bag4 ?member4 ?doc3 }!\\
\verb;    FILTER (!bound(?doc4)) };\\
\verb;  FILTER (!bound(?doc3)) };
\end{tabbing}
}
\end{boxedminipage}

\vspace{0.05cm}

\noindent
\begin{boxedminipage}{8.7cm}
{\footnotesize
\begin{tabbing}
xxxxxxxxxxxxxxxxxxxxxxxxxxxxxxxxxxxxxxxxxxxxxxxxxxxxxxxl \= \kill
\verb!SELECT DISTINCT ?name!\>\fbox{\bf Q8}\\
\verb!WHERE {!\\
\verb!  ?erdoes rdf:type foaf:Person.!\\
\verb!  ?erdoes foaf:name "Paul Erdoes"^^xsd:string.!\\
\verb!  { ?doc dc:creator ?erdoes.!\\
\verb!    ?doc dc:creator ?author.!\\
\verb!    ?doc2 dc:creator ?author.!\\
\verb!    ?doc2 dc:creator ?author2.!\\
\verb!    ?author2 foaf:name ?name!\\
\verb;    FILTER (?author!=?erdoes &&;\\
\verb;            ?doc2!=?doc &&;\\
\verb;            ?author2!=?erdoes &&;\\
\verb;            ?author2!=?author);\\
\verb!  } UNION {!\\
\verb!    ?doc dc:creator ?erdoes.!\\
\verb!    ?doc dc:creator ?author.!\\
\verb!    ?author foaf:name ?name!\\
\verb;    FILTER (?author!=?erdoes) } };
\end{tabbing}
}
\end{boxedminipage}

\vspace{0.05cm}

\noindent
\begin{boxedminipage}{8.7cm}
{\footnotesize
\begin{tabbing}
xxxxxxxxxxxxxxxxxxxxxxxxxxxxxxxxxxxxxxxxxxxxxxxxxxxxxxxl \= \kill
\verb!SELECT DISTINCT ?predicate!\>\fbox{\bf Q9}\\
\verb!WHERE {!\\
\verb!  { ?person rdf:type foaf:Person.!\\
\verb!    ?subject ?predicate ?person } UNION!\\
\verb!  { ?person rdf:type foaf:Person.!\\
\verb!    ?person ?predicate ?object } }!
\end{tabbing}
}
\end{boxedminipage}

\vspace{0.05cm}

\noindent
\begin{boxedminipage}{8.7cm}
{\footnotesize
\begin{tabbing}
xxxxxxxxxxxxxxxxxxxxxxxxxxxxxxxxxxxxxxxxxxxxxxxxxxxxxxl \= \kill
\verb!SELECT ?subj ?pred!\>\fbox{\bf Q10}\\
\verb!WHERE { ?subj ?pred person:Paul_Erdoes }!
\end{tabbing}
}
\end{boxedminipage}

\vspace{0.05cm}

\noindent
\begin{boxedminipage}{8.7cm}
{\footnotesize
\begin{tabbing}
xxxxxxxxxxxxxxxxxxxxxxxxxxxxxxxxxxxxxxxxxxxxxxxxxxxxxxl \= \kill
\verb!SELECT ?ee!\>\fbox{\bf Q11}\\
\verb!WHERE { ?publication rdfs:seeAlso ?ee }!\\
\verb!ORDER BY ?ee LIMIT 10 OFFSET 50!
\end{tabbing}
}
\end{boxedminipage}

\vspace{0.05cm}

\noindent
\begin{boxedminipage}{8.7cm}
{\footnotesize
\begin{tabbing}
xxxl\=xxxxxxxxxxxxxxxxxxxxxxxxxxxxxxxxxxxxxxxxxxxxxxxxxxx \= \kill
(a)\>\verb!Q5a as ASK query!\>\fbox{\bf Q12}\\
\\[-0.2cm]
(b)\>\verb!Q8 as ASK query!\\
\\[-0.2cm]
(c)\>\verb!ASK {person:John_Q_Public rfd:type foaf:Person}!
\end{tabbing}
}
\end{boxedminipage}

\begin{figure*}[t]
\hspace{-1.1cm}
\begin{tabular}{cccc}
\rotatebox{270}{\includegraphics[scale=0.182]{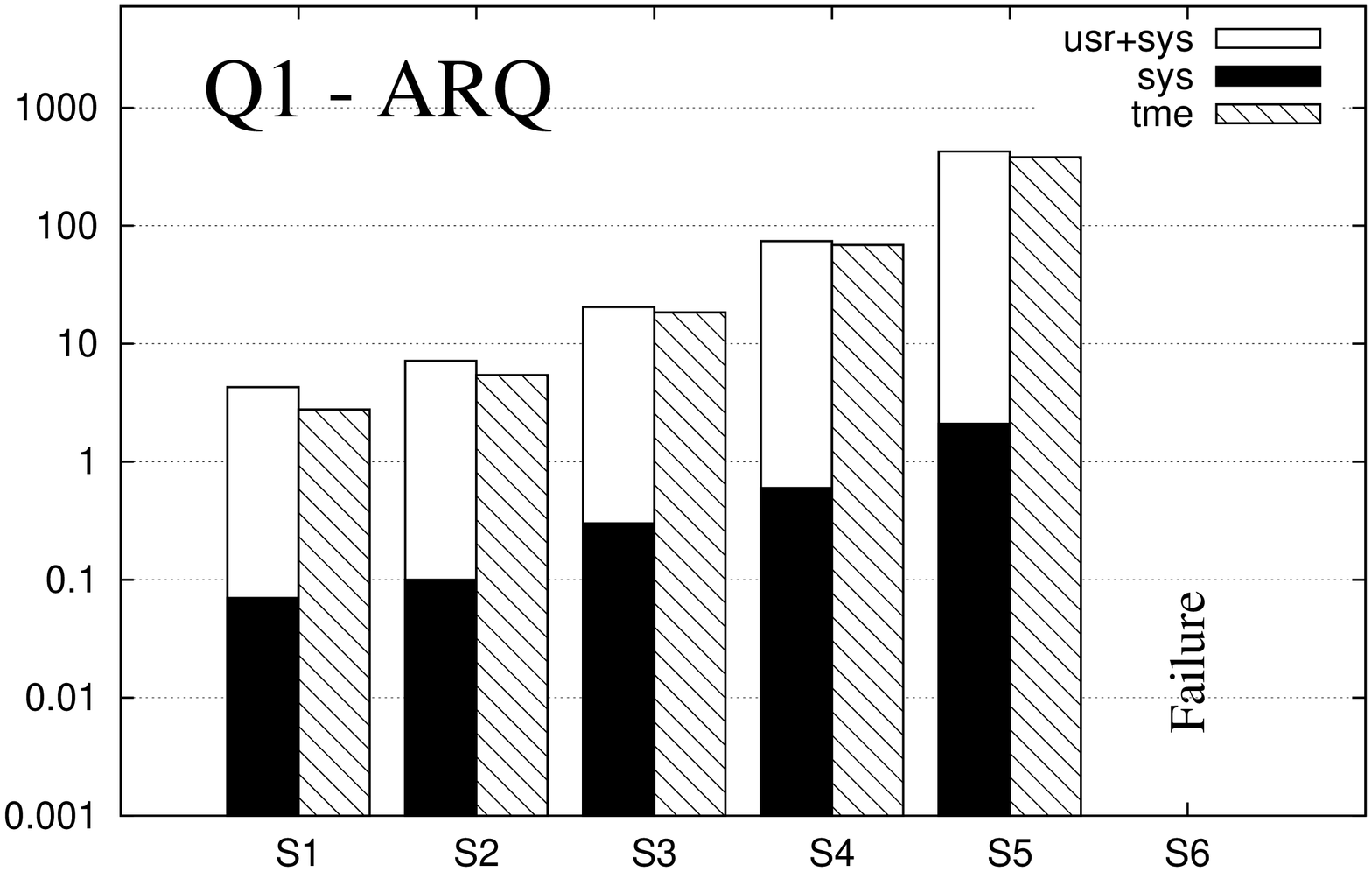}}
&
\hspace{-0.5cm}
\rotatebox{270}{\includegraphics[scale=0.182]{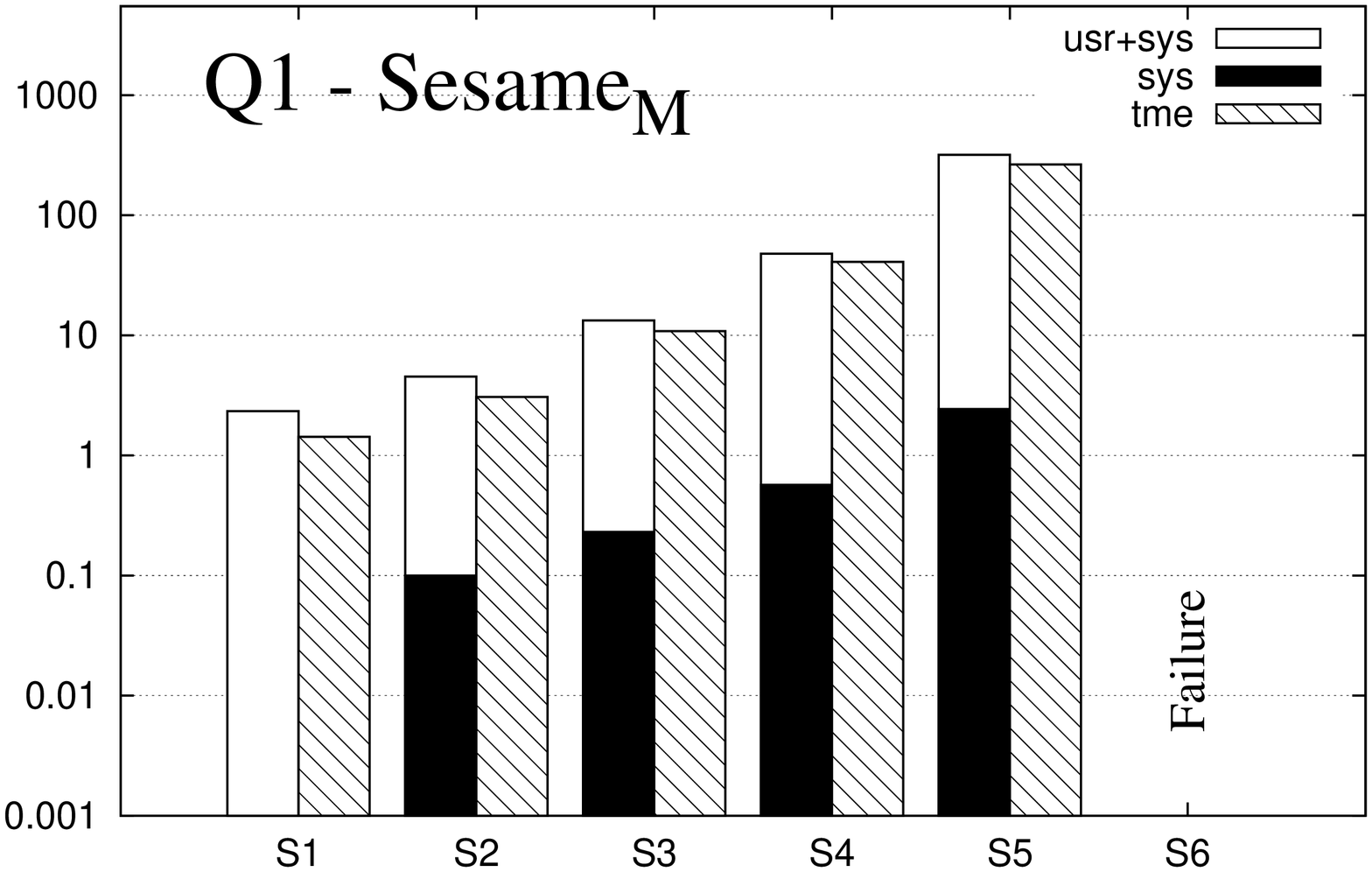}}
&
\hspace{-0.5cm}
\rotatebox{270}{\includegraphics[scale=0.182]{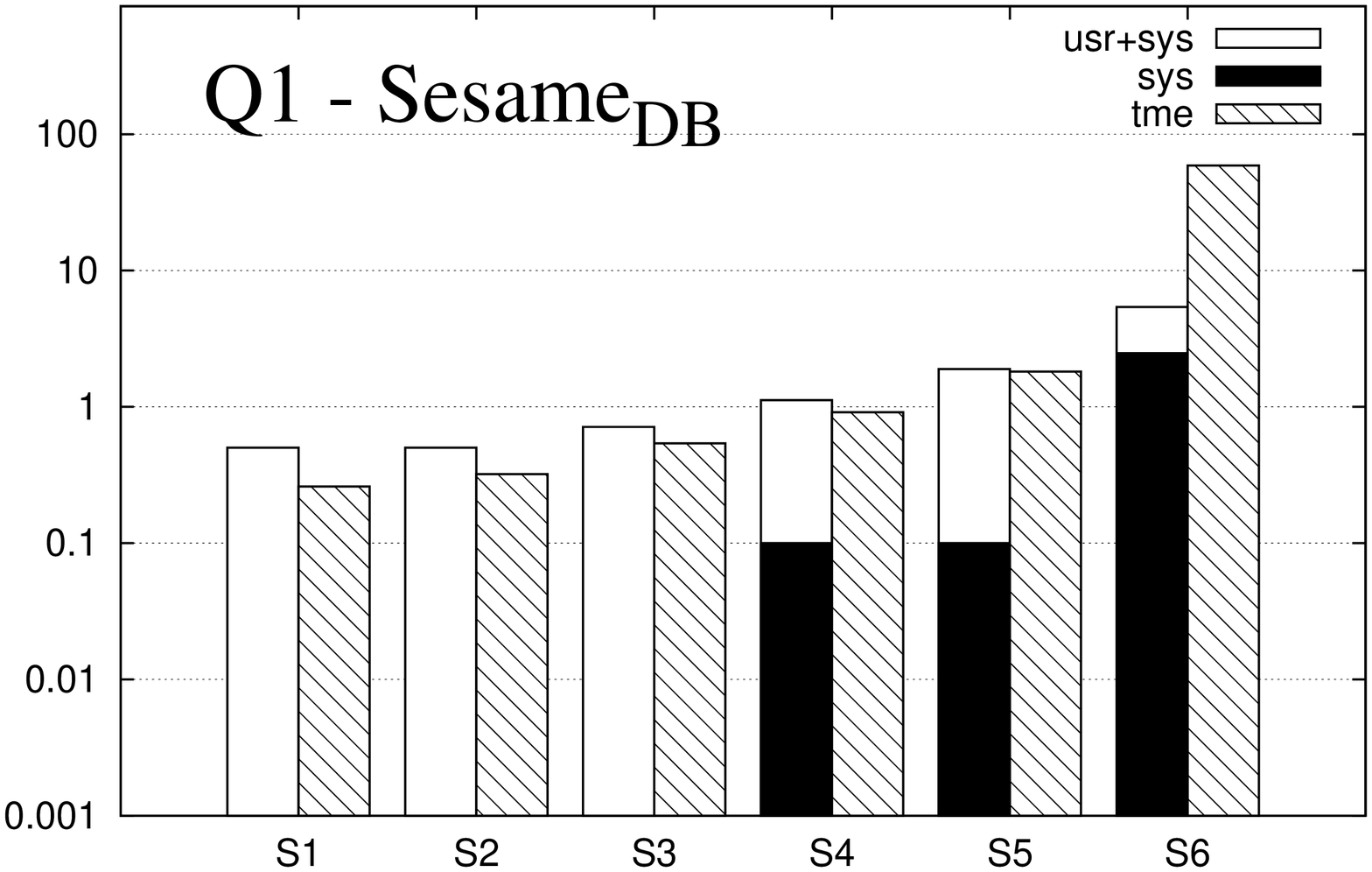}}
&
\hspace{-0.5cm}
\rotatebox{270}{\includegraphics[scale=0.182]{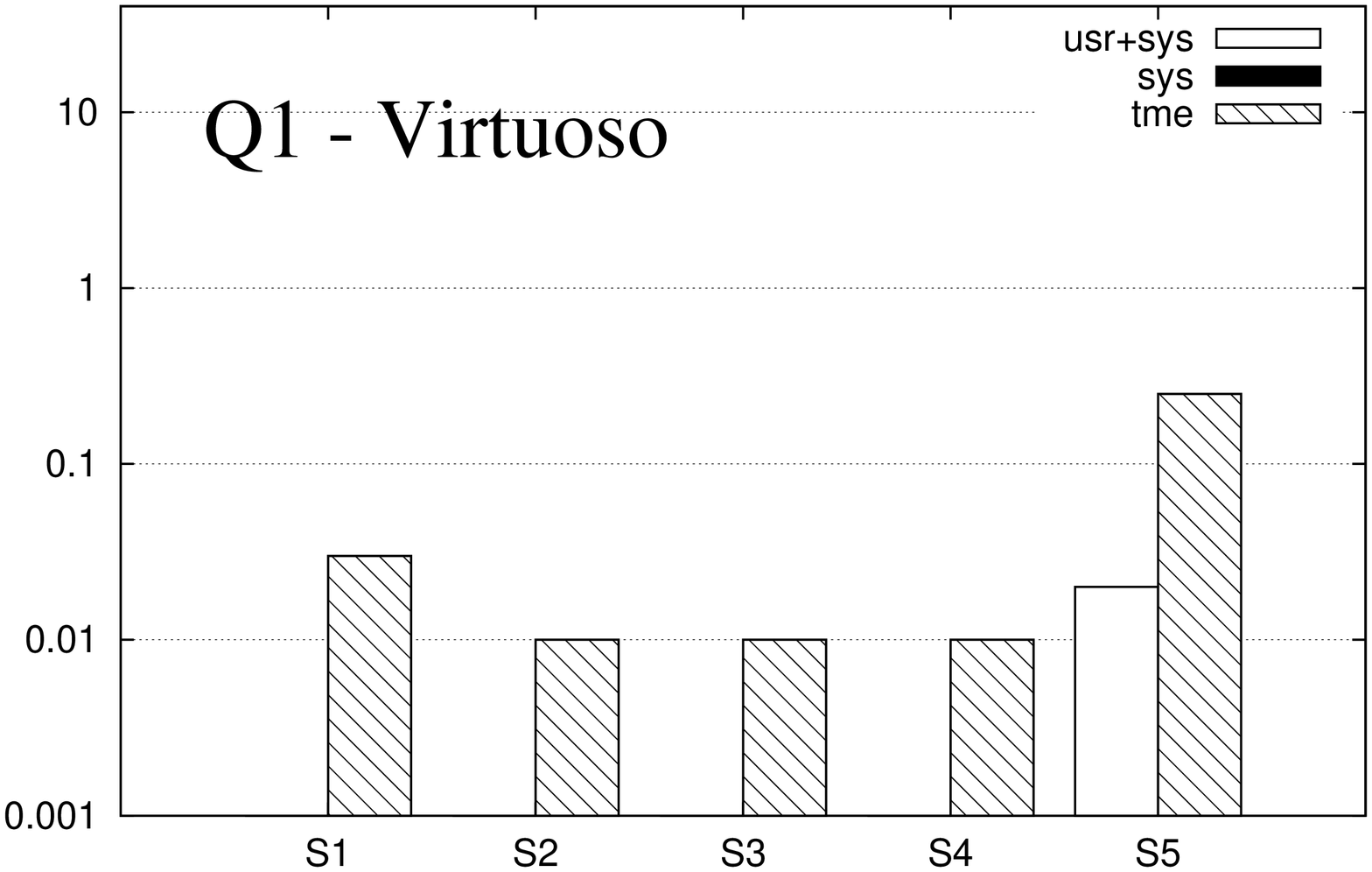}}
\\[-0.65cm]
\rotatebox{270}{\includegraphics[scale=0.18]{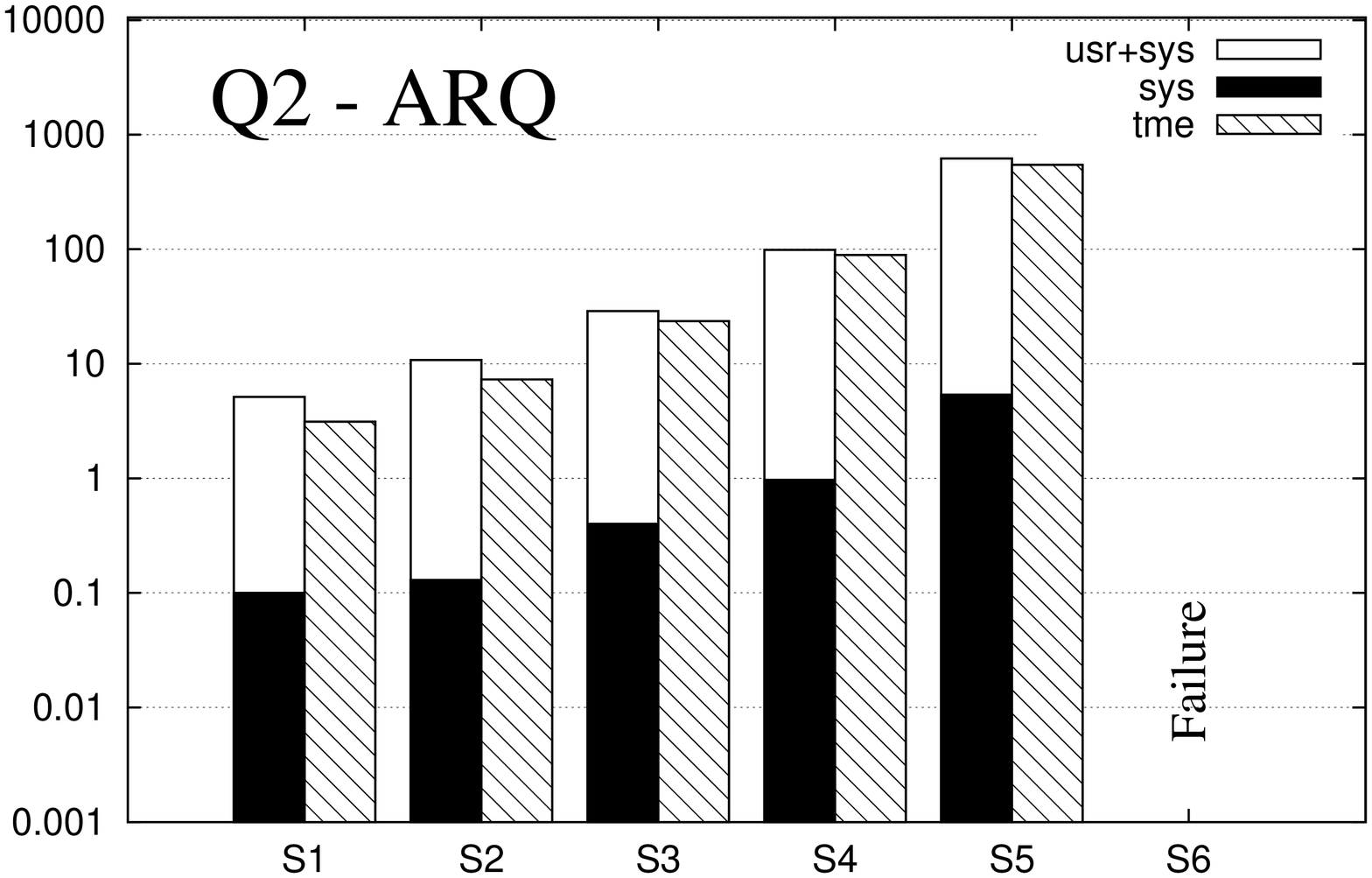}}
&
\hspace{-0.5cm}
\rotatebox{270}{\includegraphics[scale=0.18]{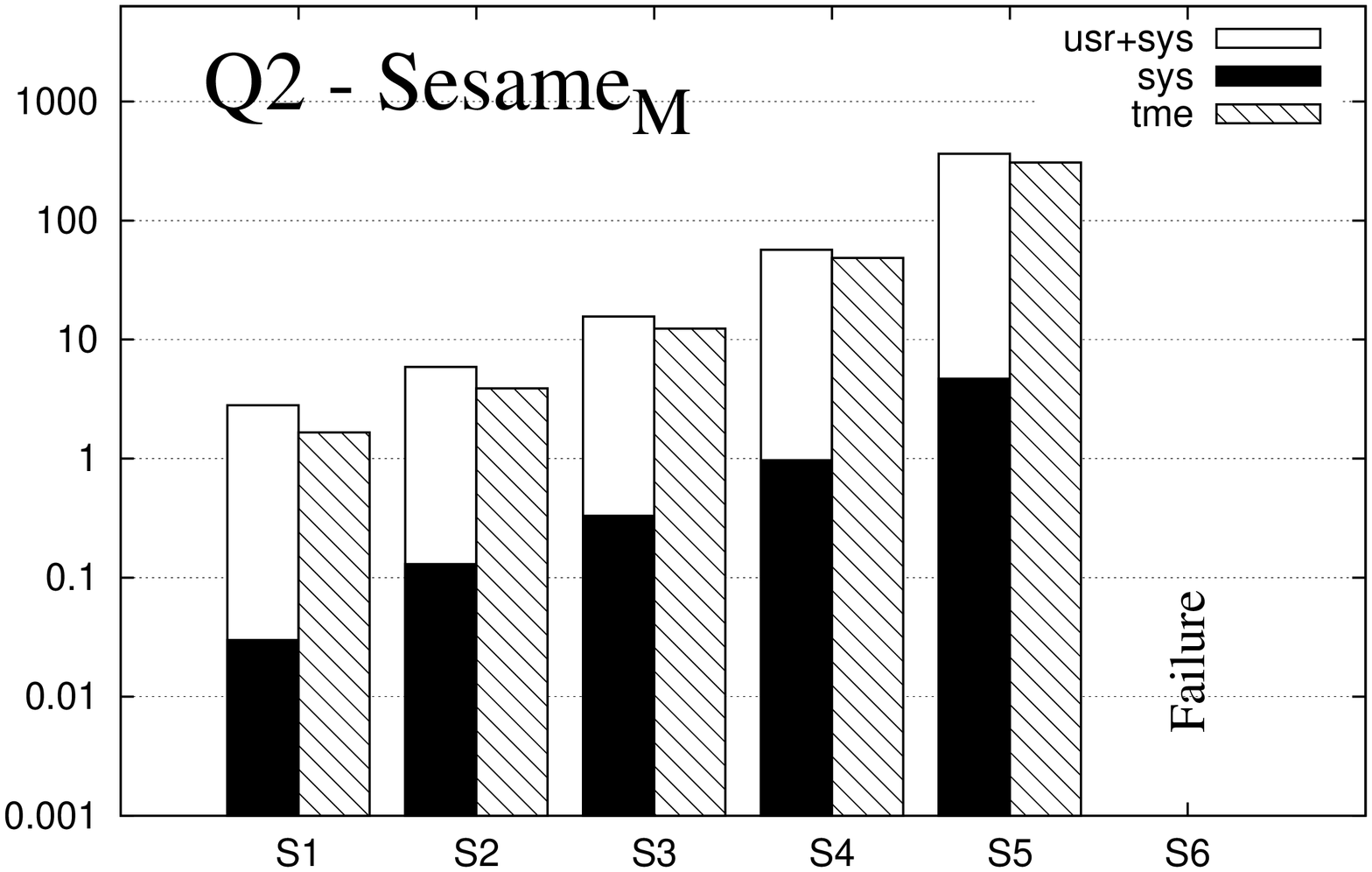}}
&
\hspace{-0.5cm}
\rotatebox{270}{\includegraphics[scale=0.18]{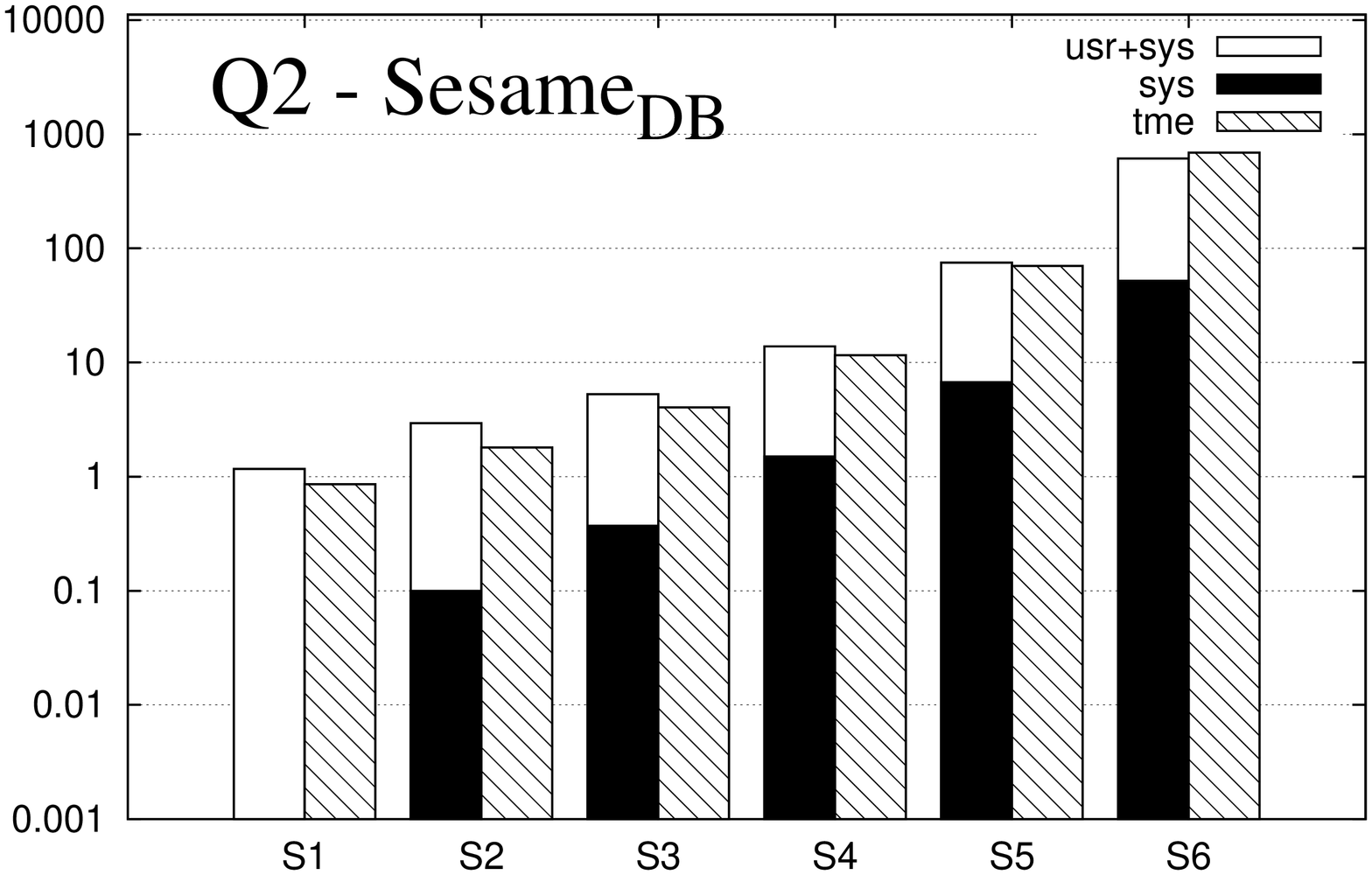}}
&
\hspace{-0.5cm}
\rotatebox{270}{\includegraphics[scale=0.18]{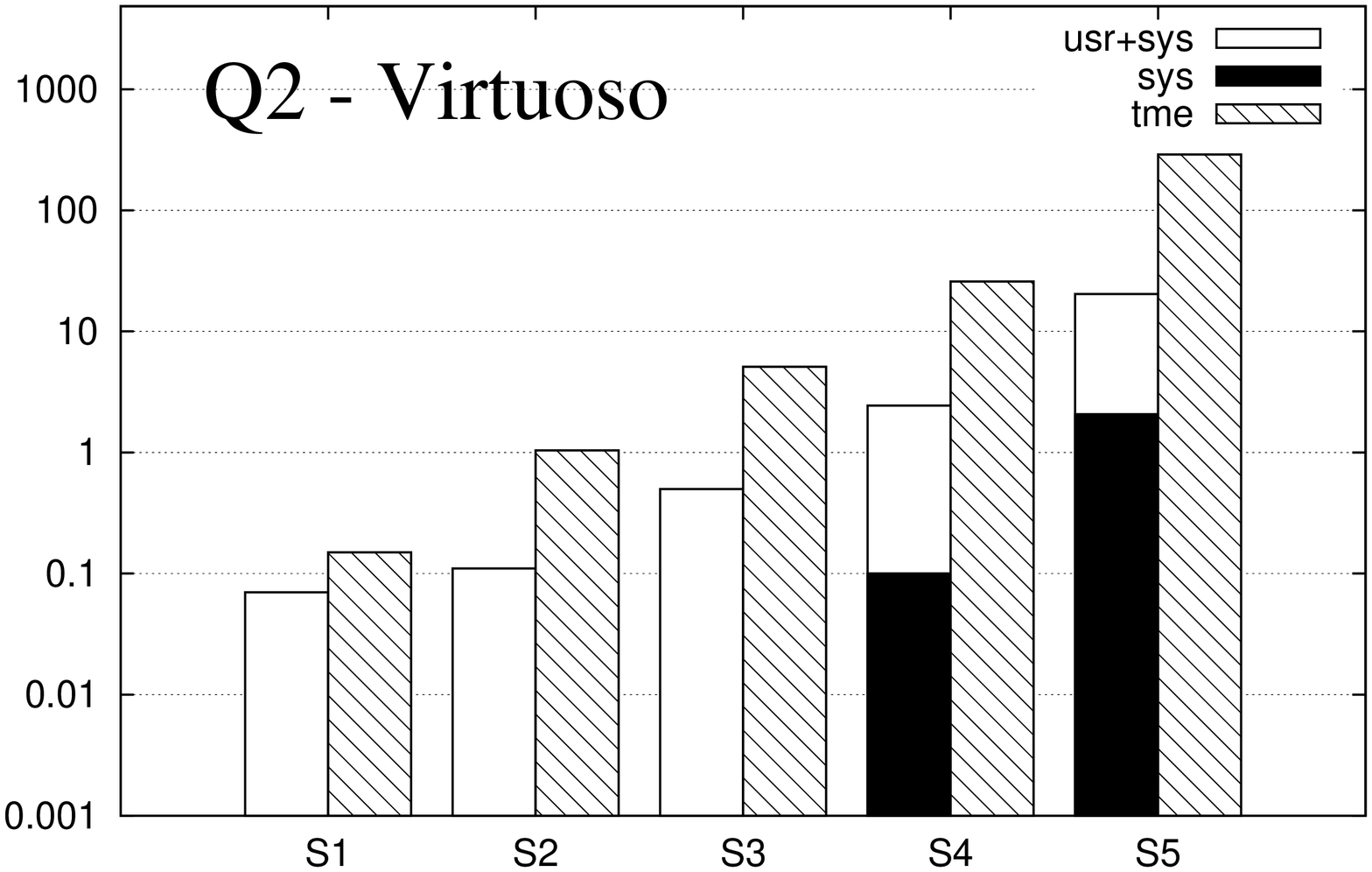}}
\\[-0.65cm]
\rotatebox{270}{\includegraphics[scale=0.18]{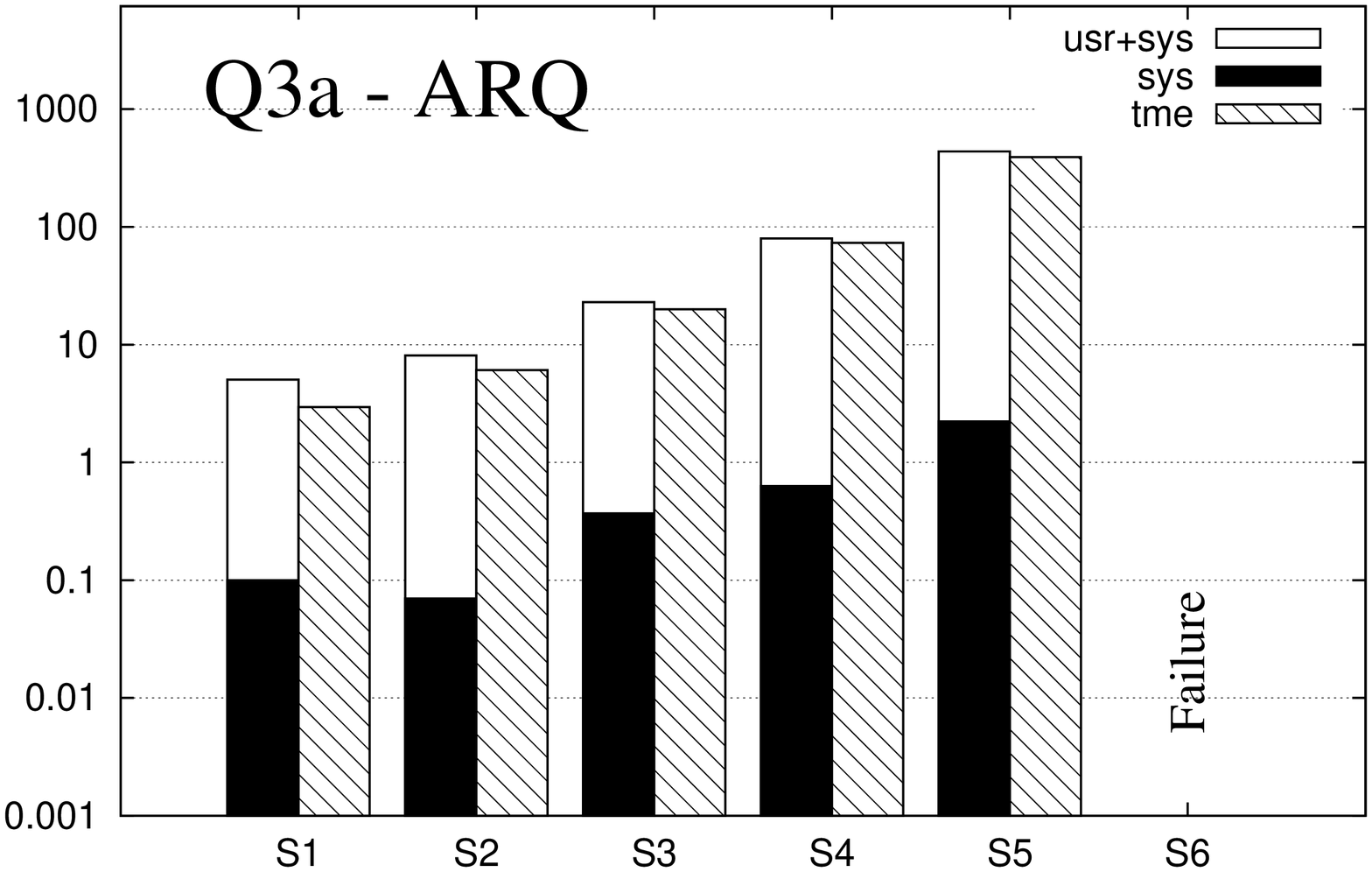}}
&
\hspace{-0.5cm}
\rotatebox{270}{\includegraphics[scale=0.18]{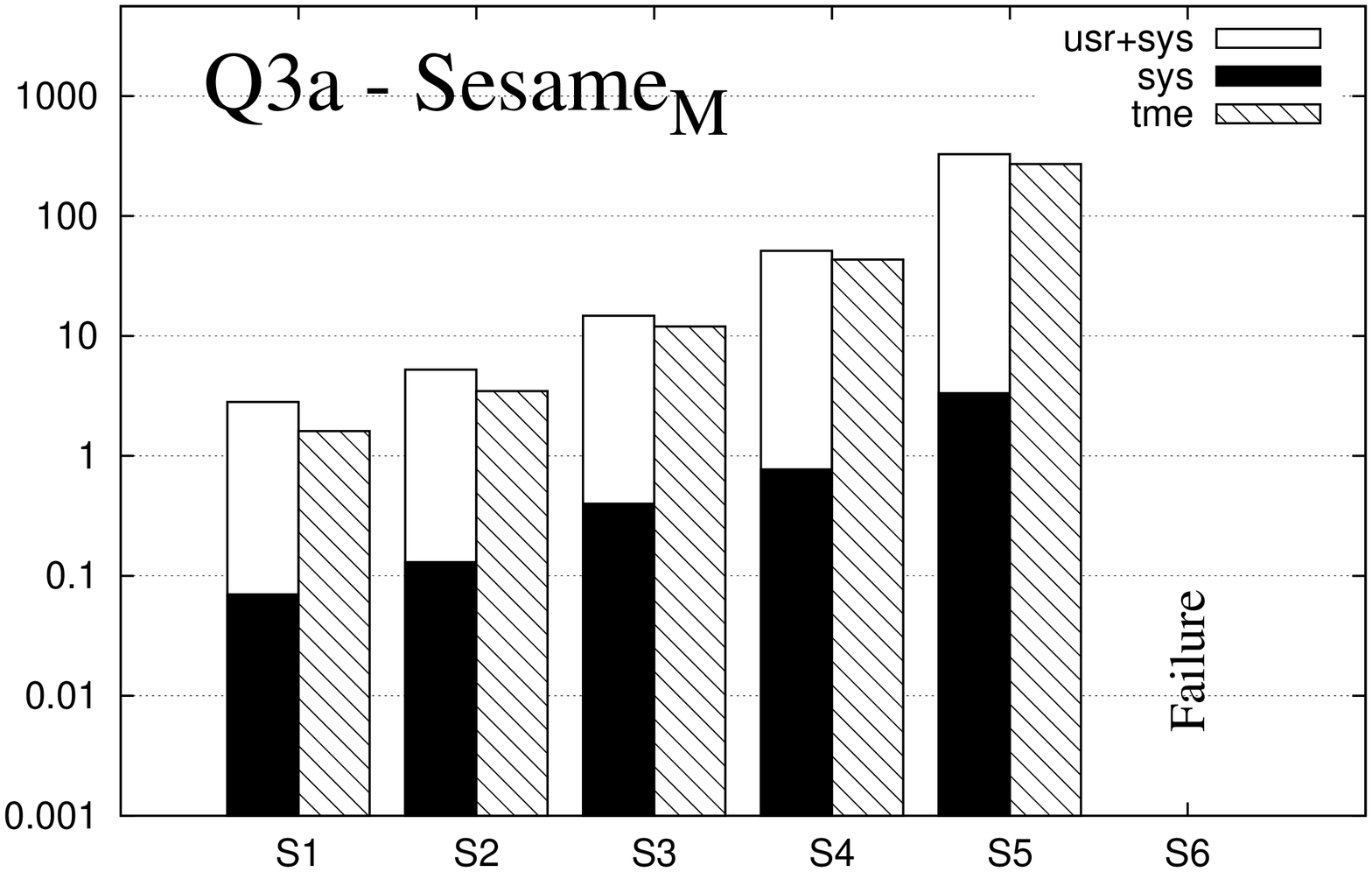}}
&
\hspace{-0.5cm}
\rotatebox{270}{\includegraphics[scale=0.18]{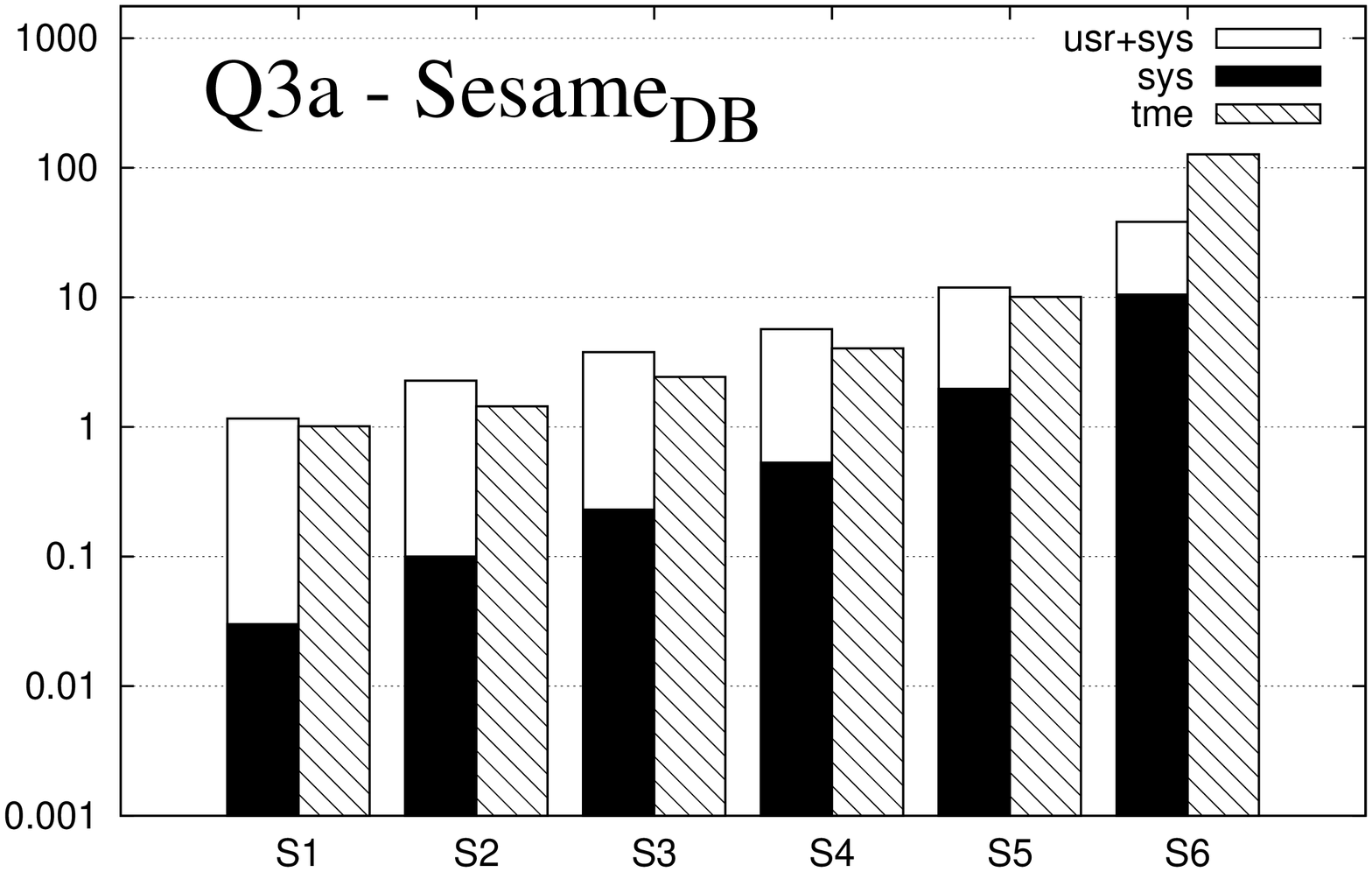}}
&
\hspace{-0.5cm}
\rotatebox{270}{\includegraphics[scale=0.18]{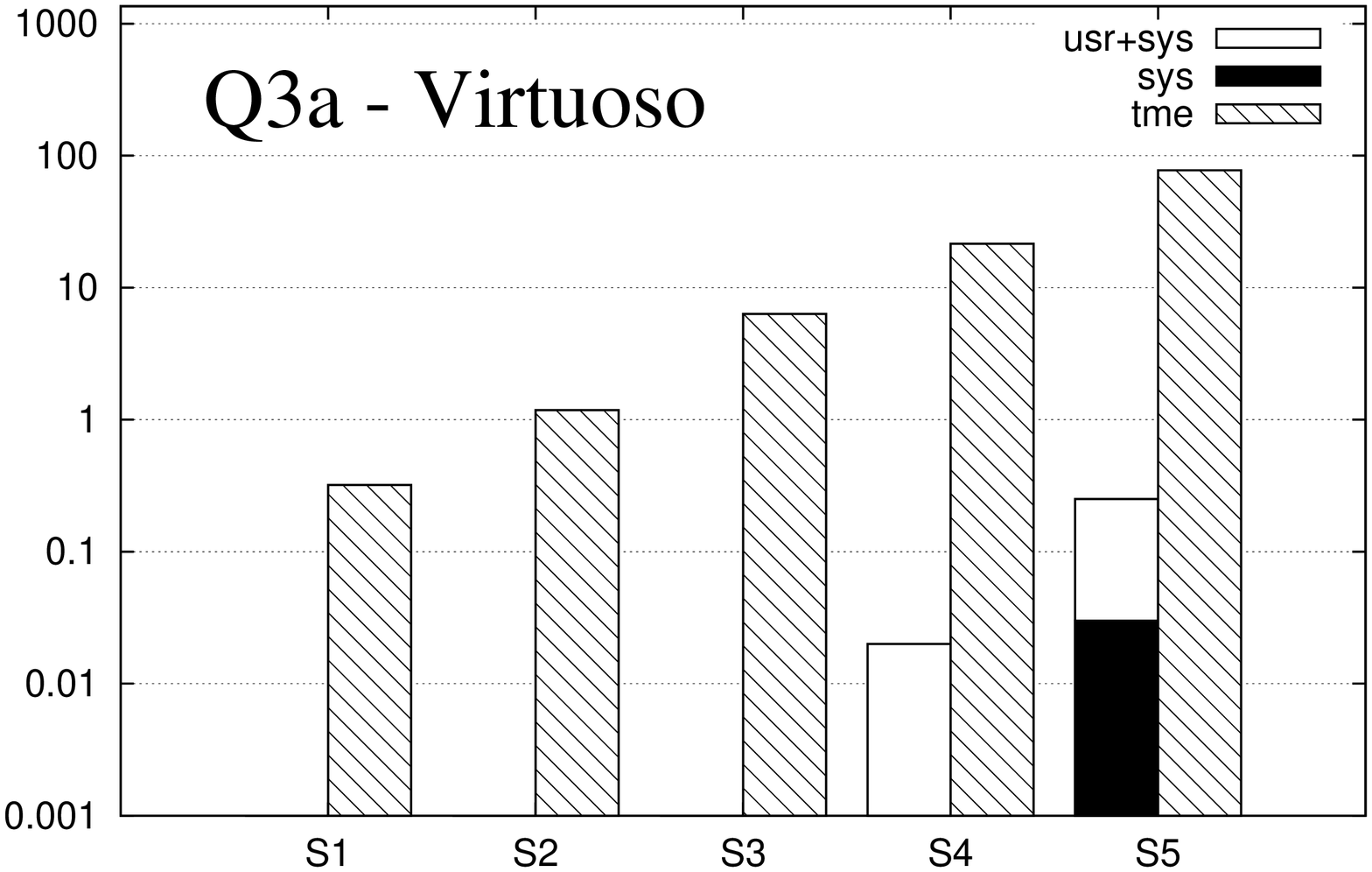}}
\\[-0.65cm]
\rotatebox{270}{\includegraphics[scale=0.18]{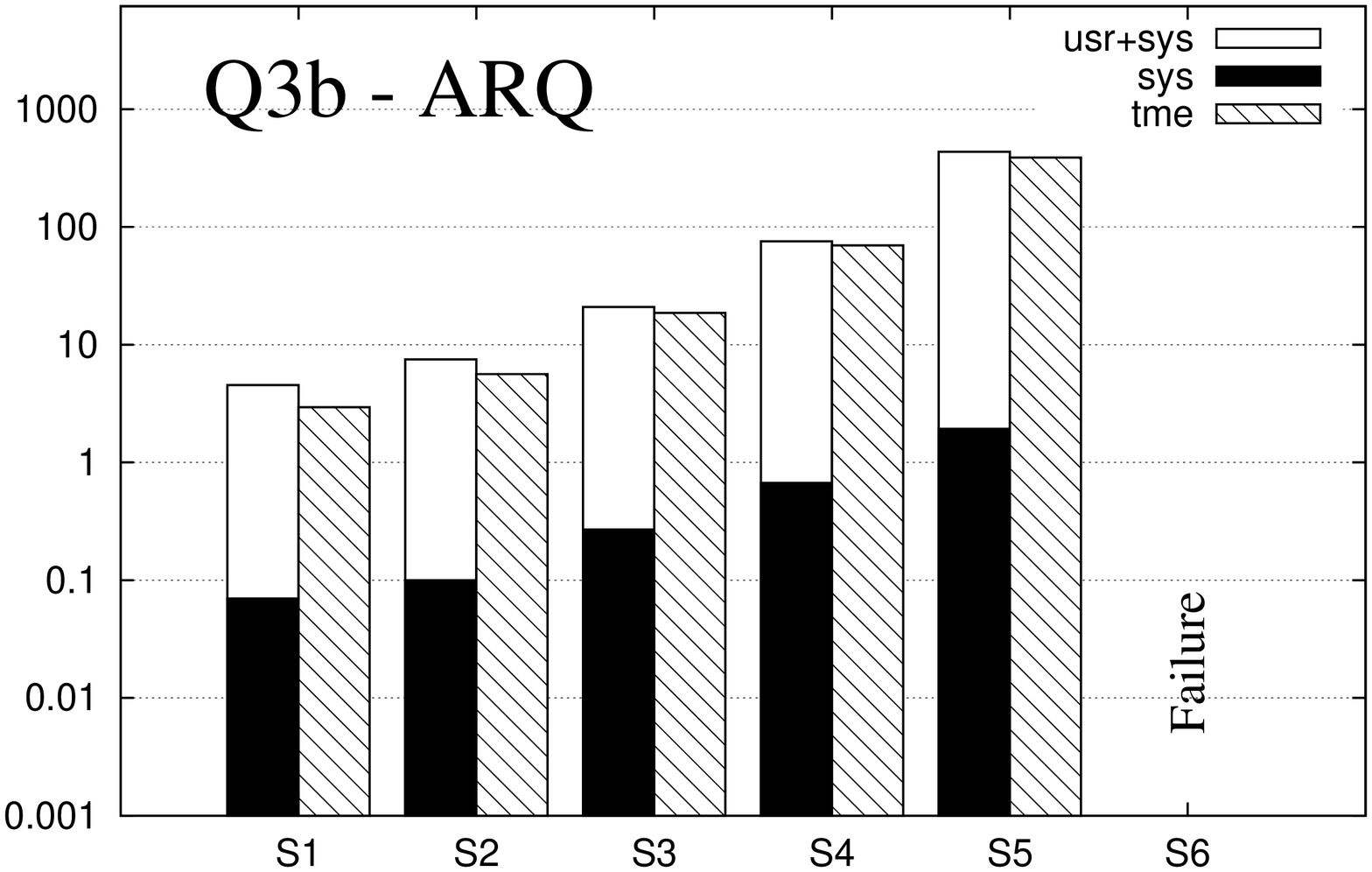}}
&
\hspace{-0.5cm}
\rotatebox{270}{\includegraphics[scale=0.18]{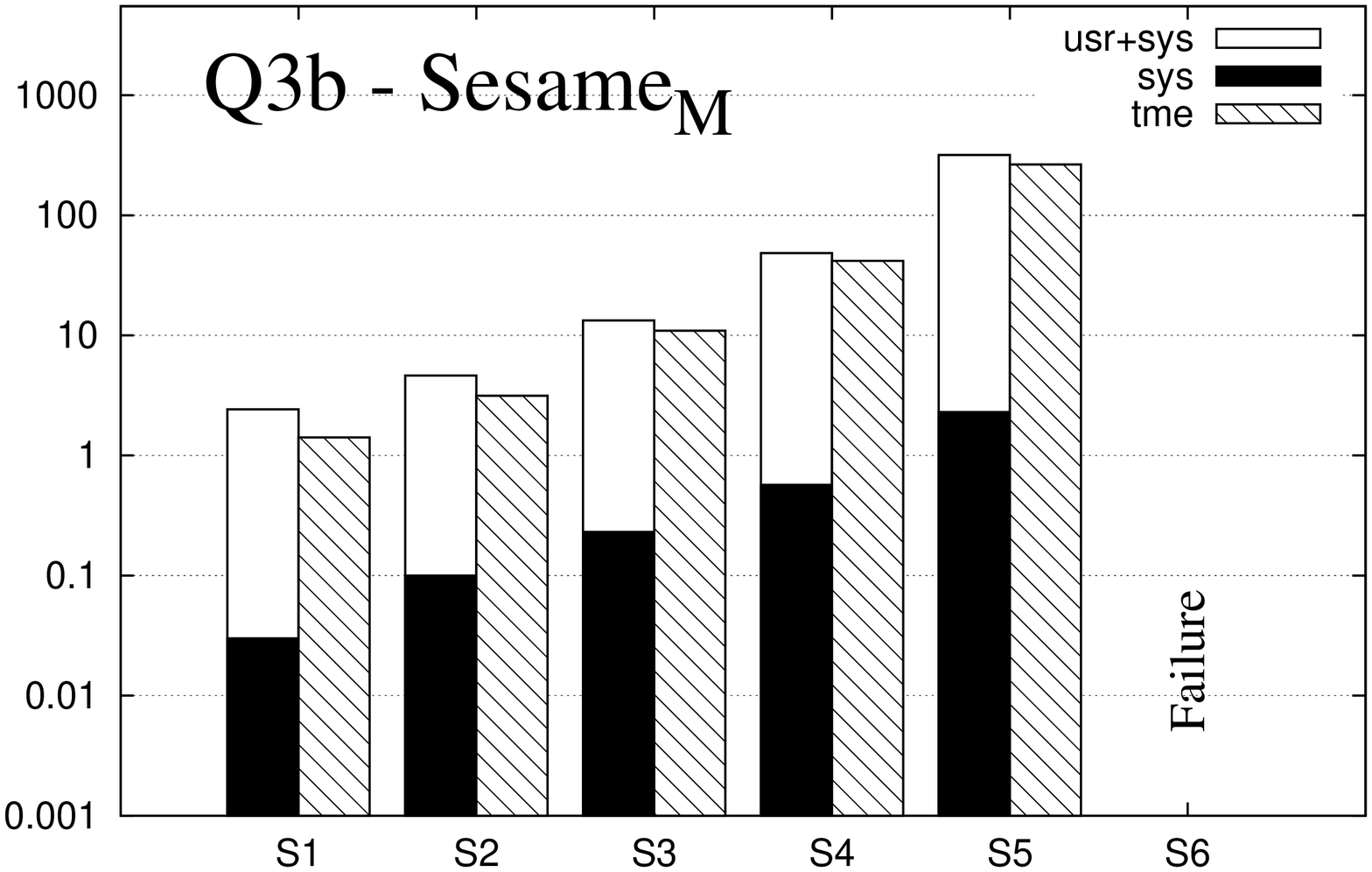}}
&
\hspace{-0.5cm}
\rotatebox{270}{\includegraphics[scale=0.18]{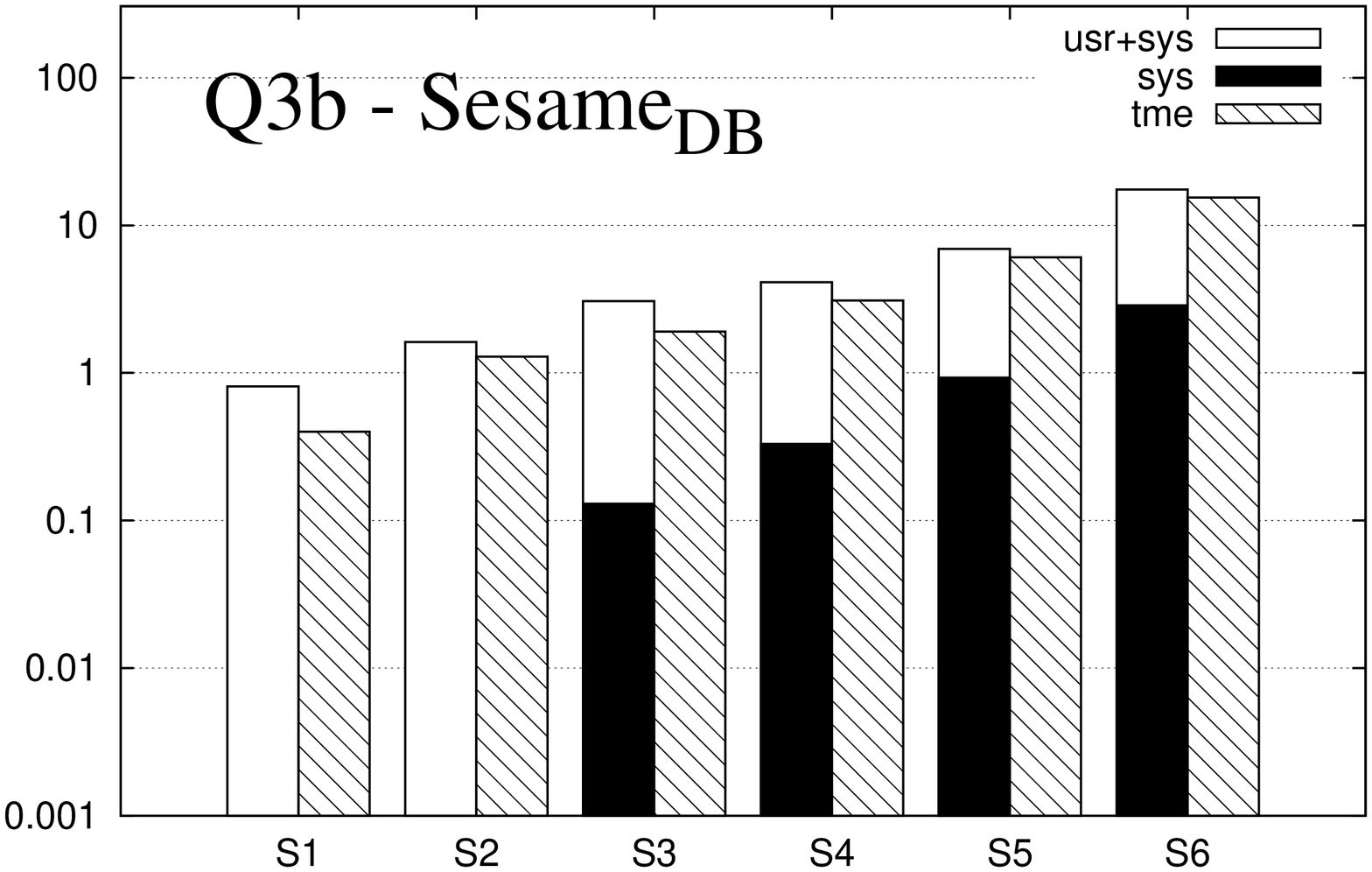}}
&
\hspace{-0.5cm}
\rotatebox{270}{\includegraphics[scale=0.18]{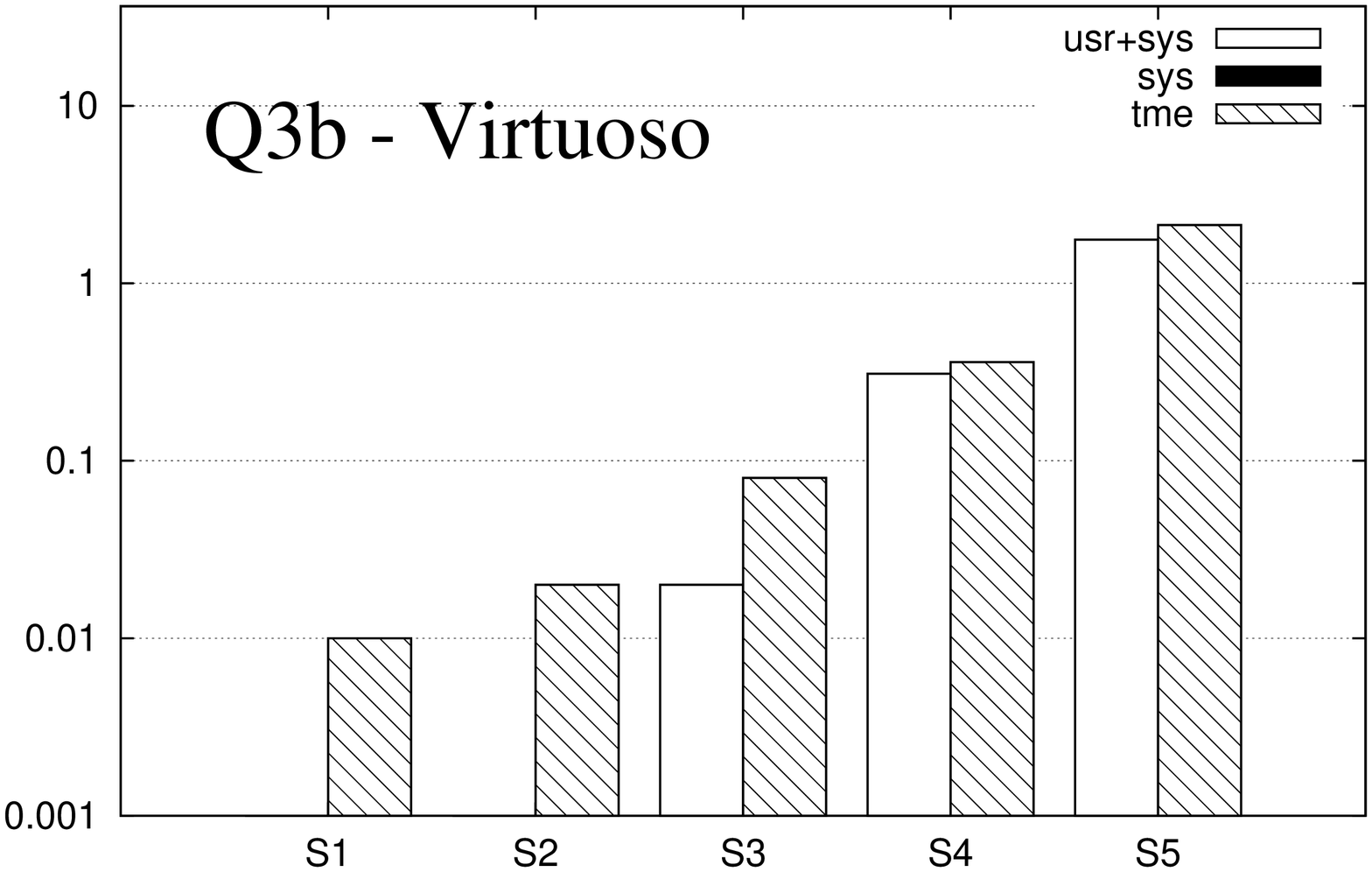}}
\\[-0.65cm]
\rotatebox{270}{\includegraphics[scale=0.18]{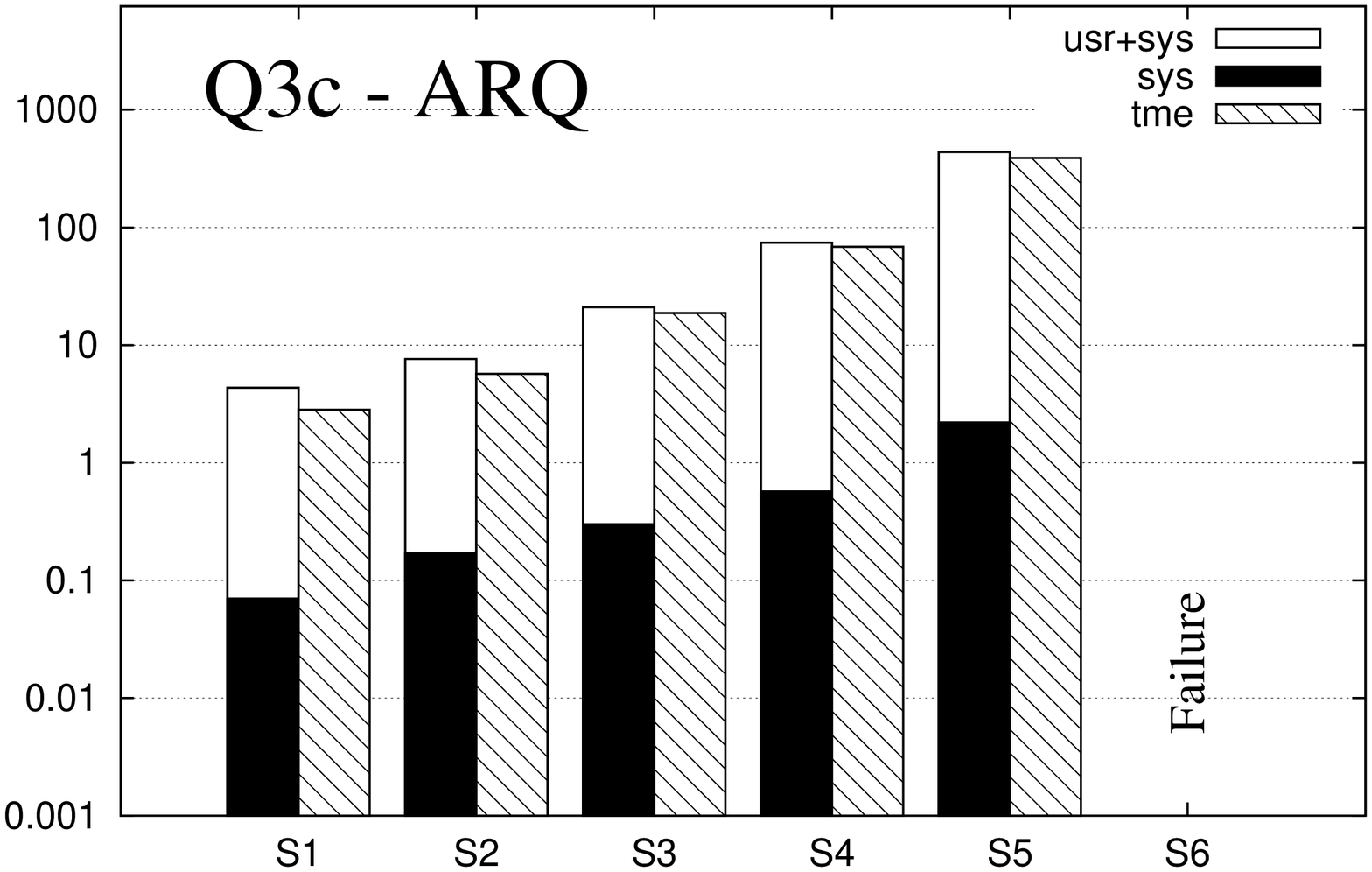}}
&
\hspace{-0.5cm}
\rotatebox{270}{\includegraphics[scale=0.18]{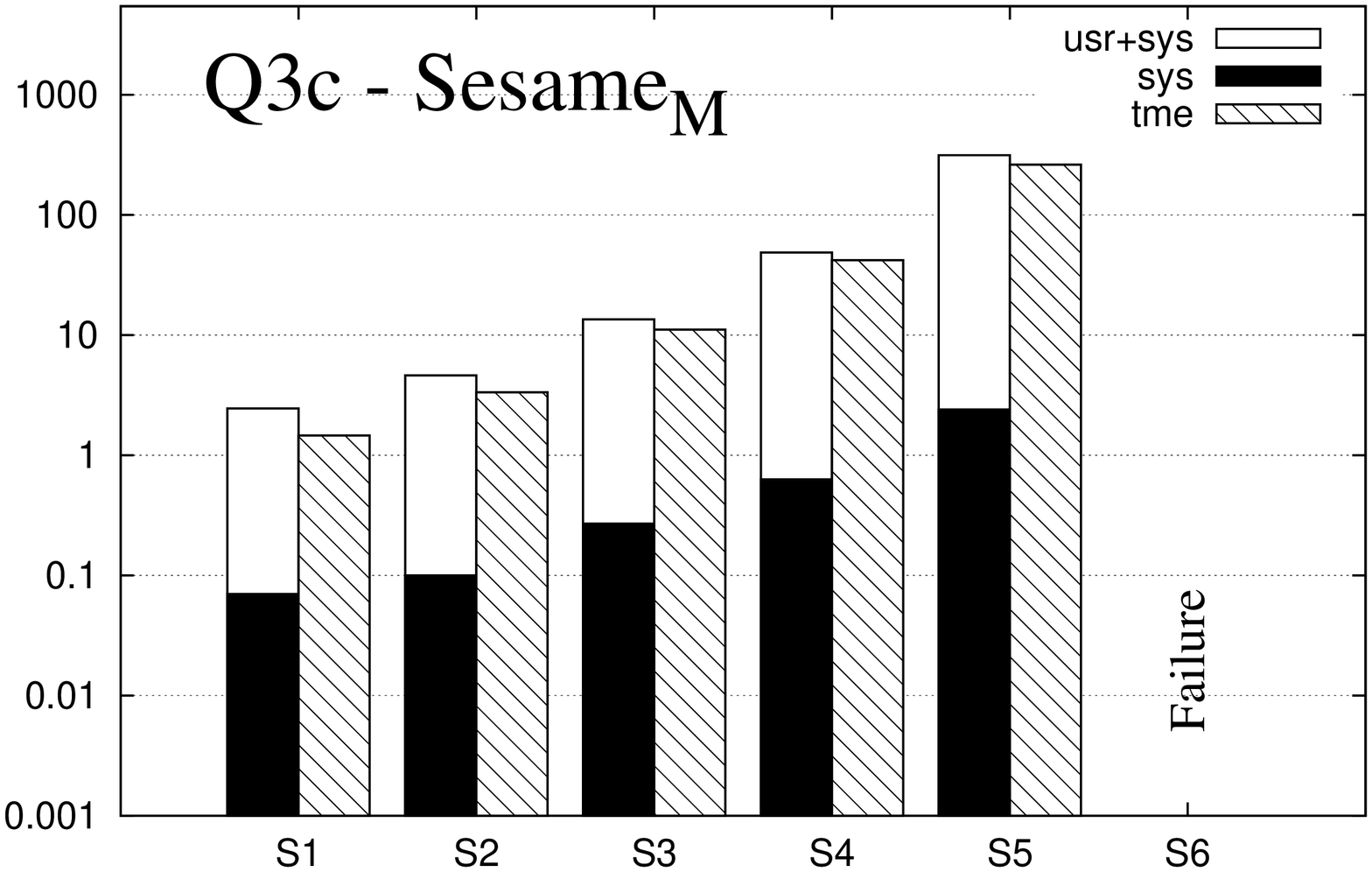}}
&
\hspace{-0.5cm}
\rotatebox{270}{\includegraphics[scale=0.18]{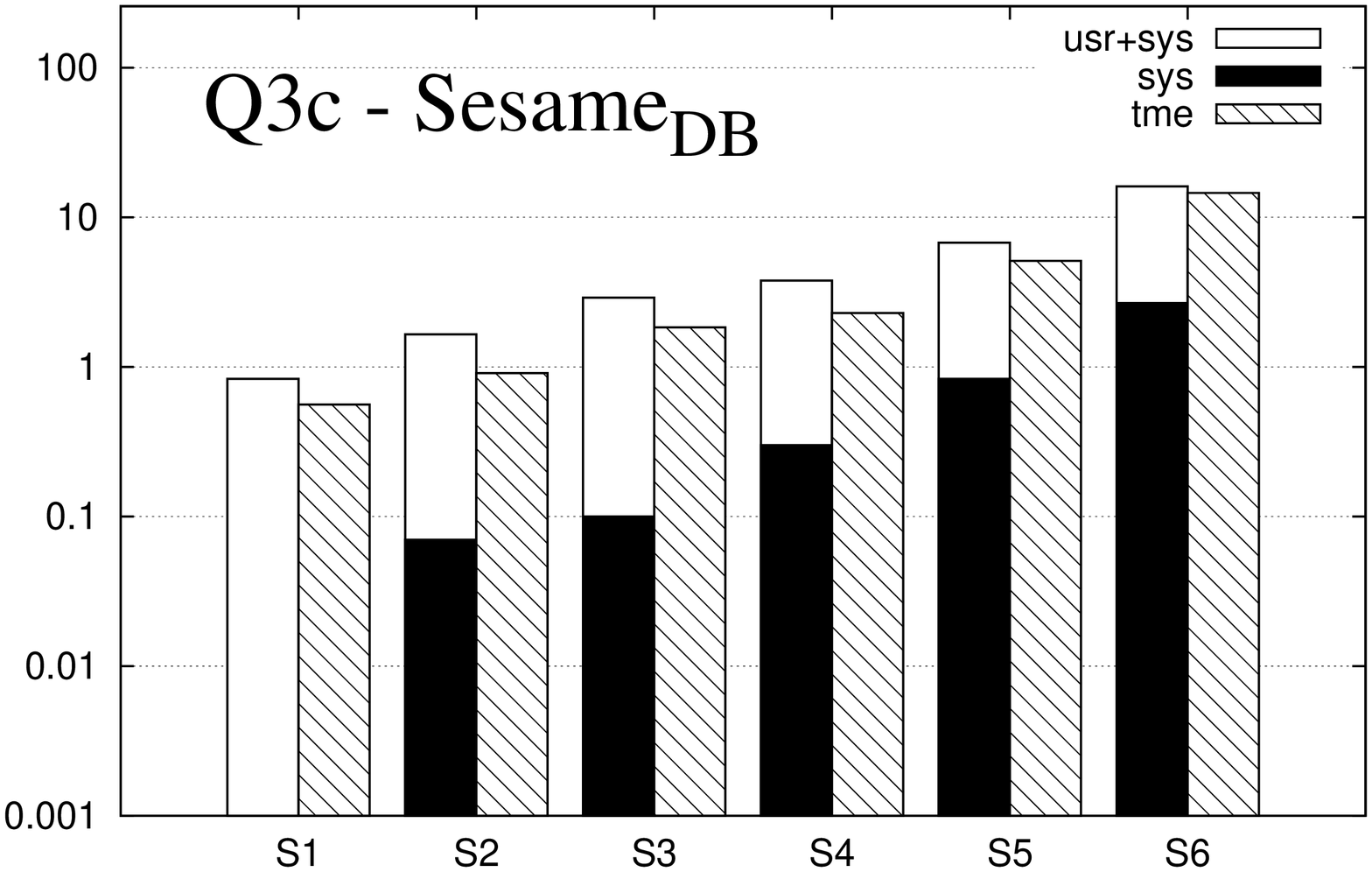}}
&
\hspace{-0.5cm}
\rotatebox{270}{\includegraphics[scale=0.18]{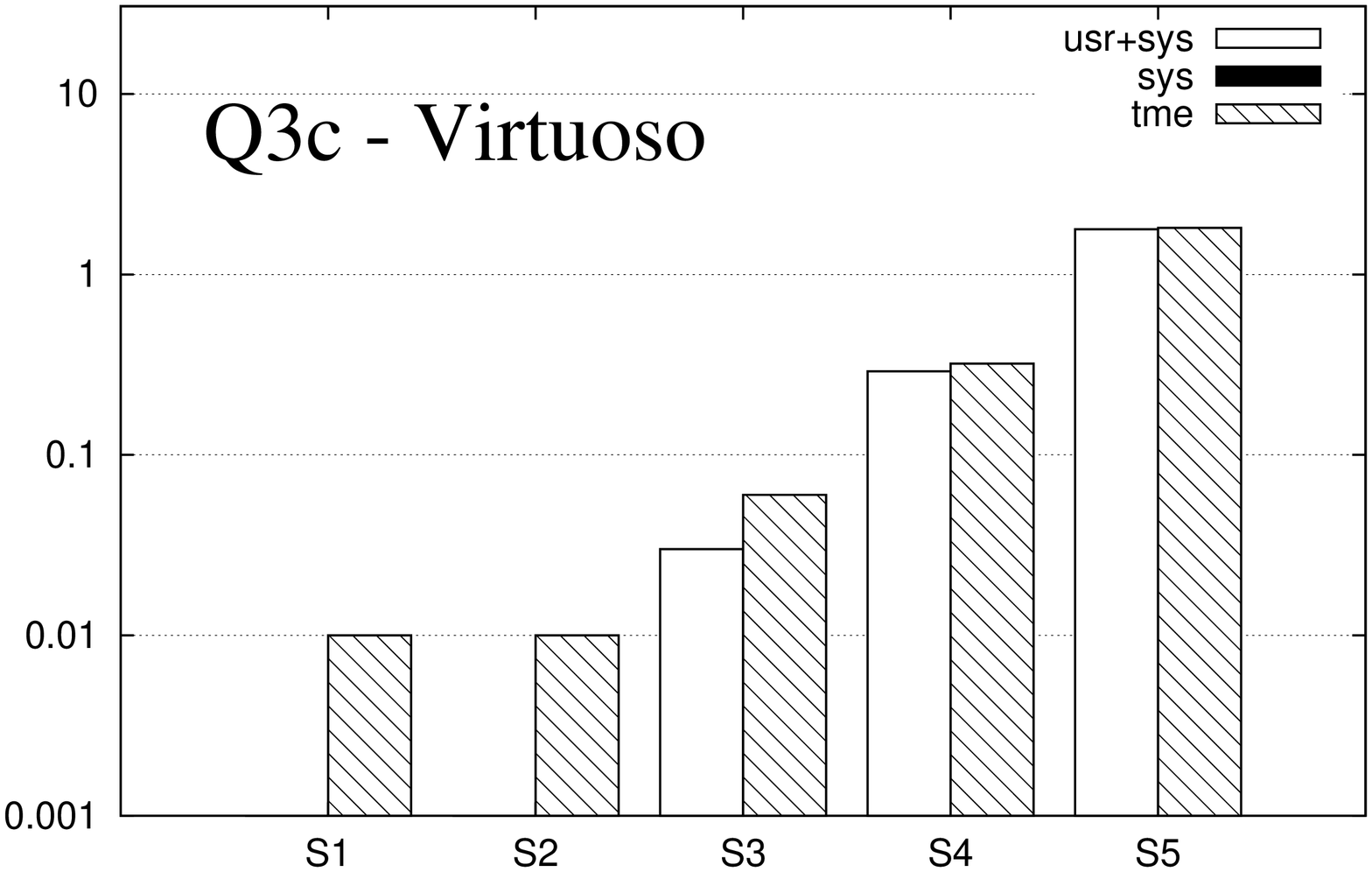}}
\\[-0.65cm]
\rotatebox{270}{\includegraphics[scale=0.18]{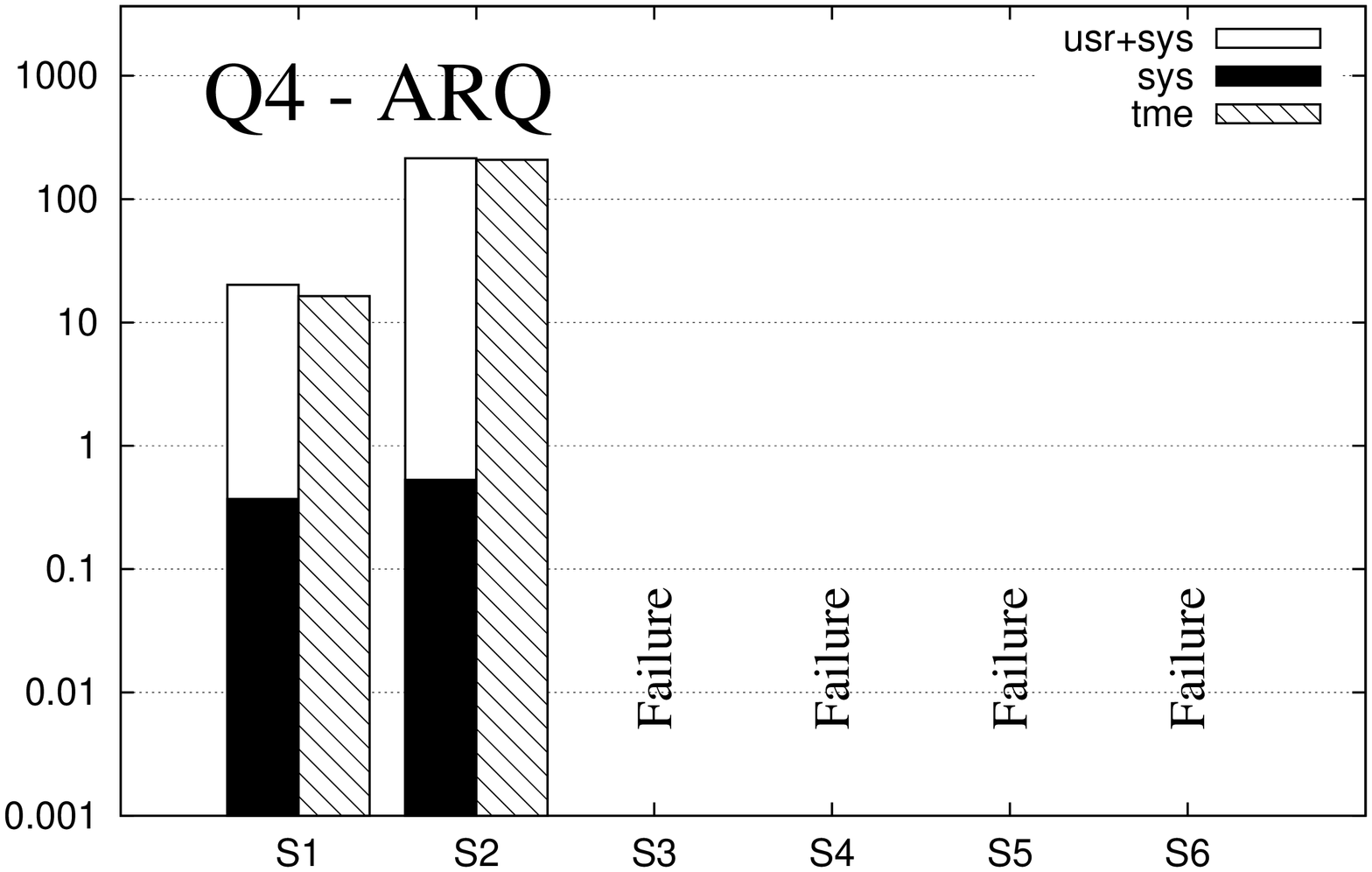}}
&
\hspace{-0.5cm}
\rotatebox{270}{\includegraphics[scale=0.18]{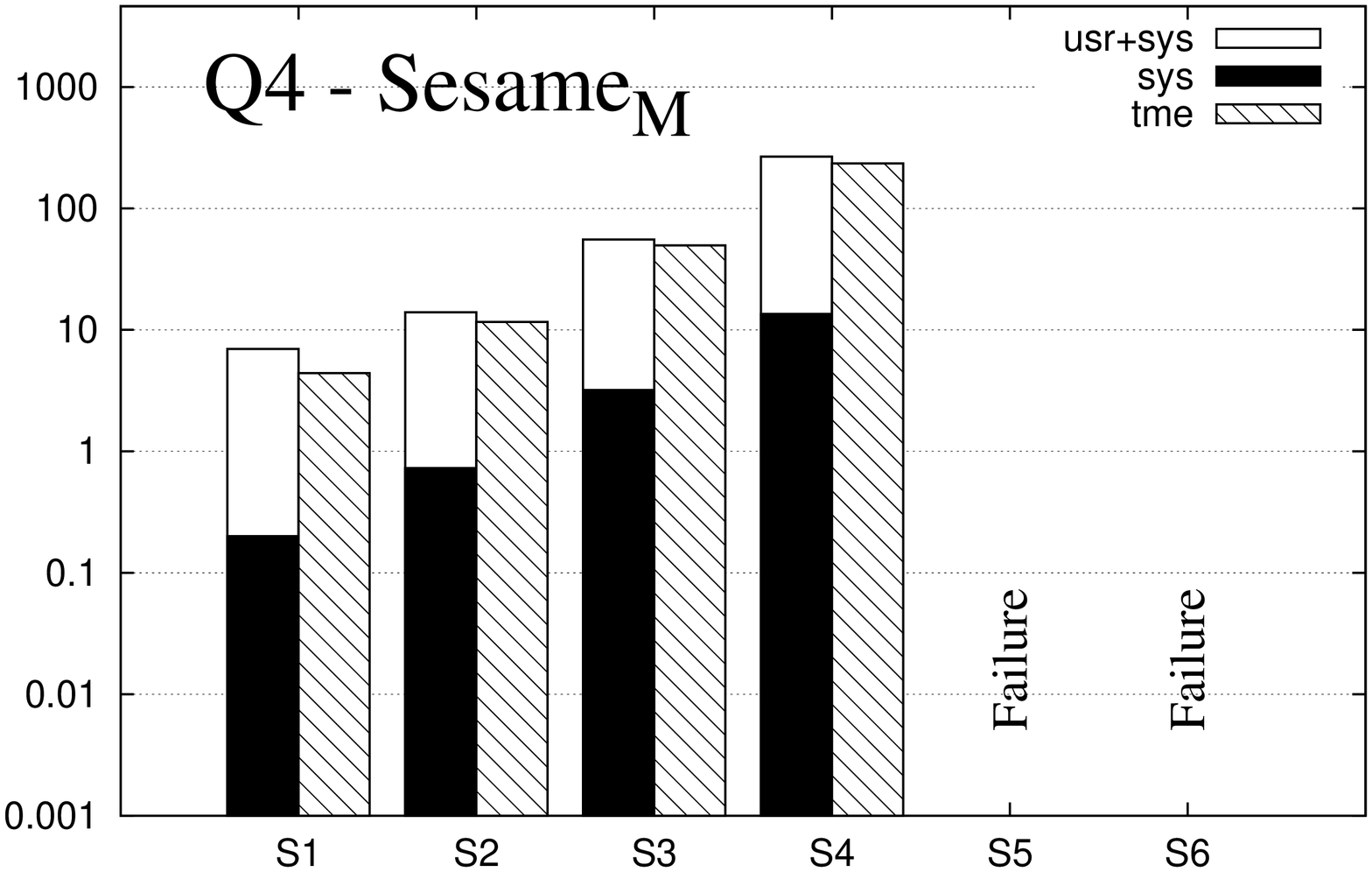}}
&
\hspace{-0.5cm}
\rotatebox{270}{\includegraphics[scale=0.18]{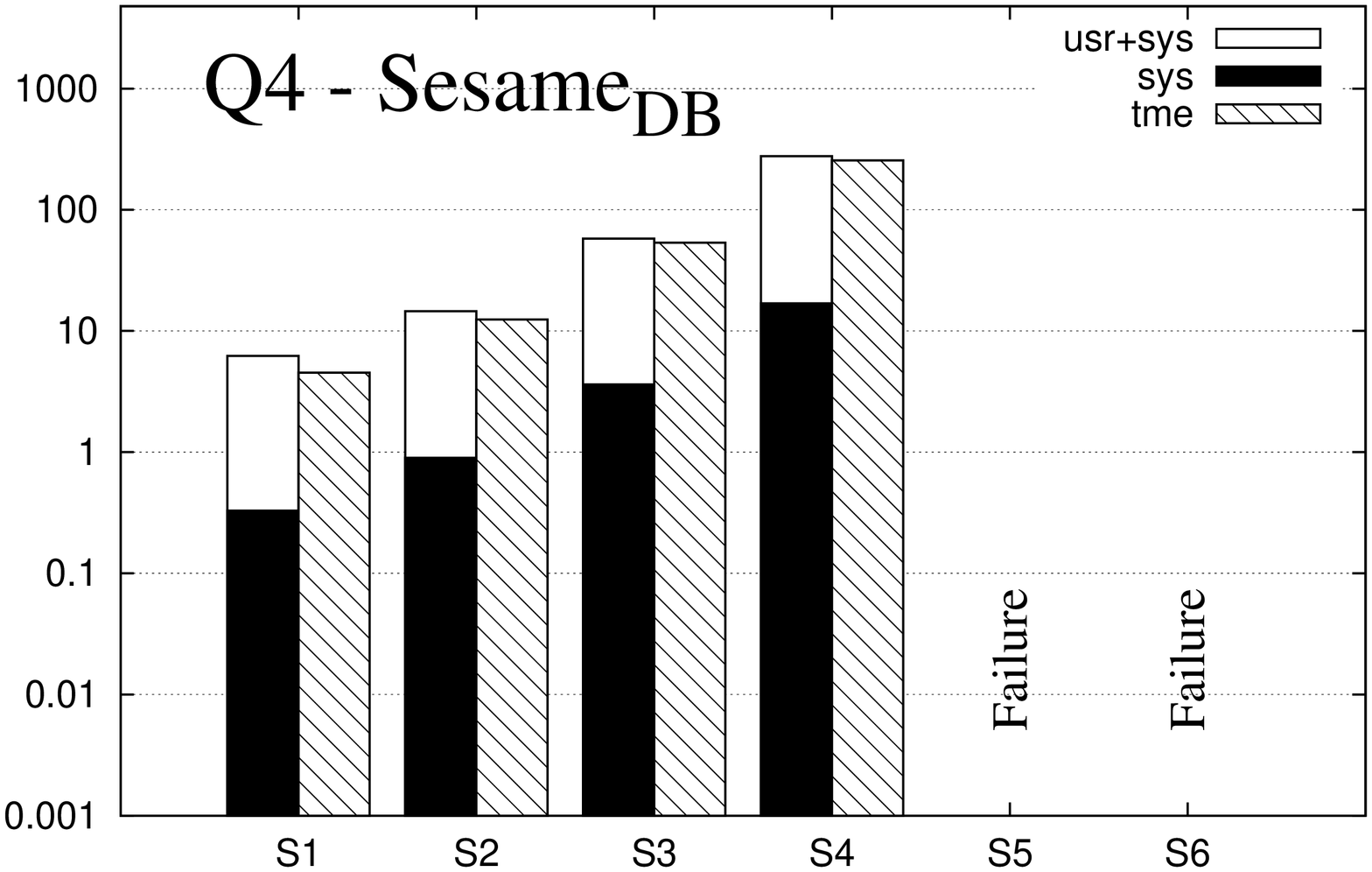}}
&
\hspace{-0.5cm}
\rotatebox{270}{\includegraphics[scale=0.18]{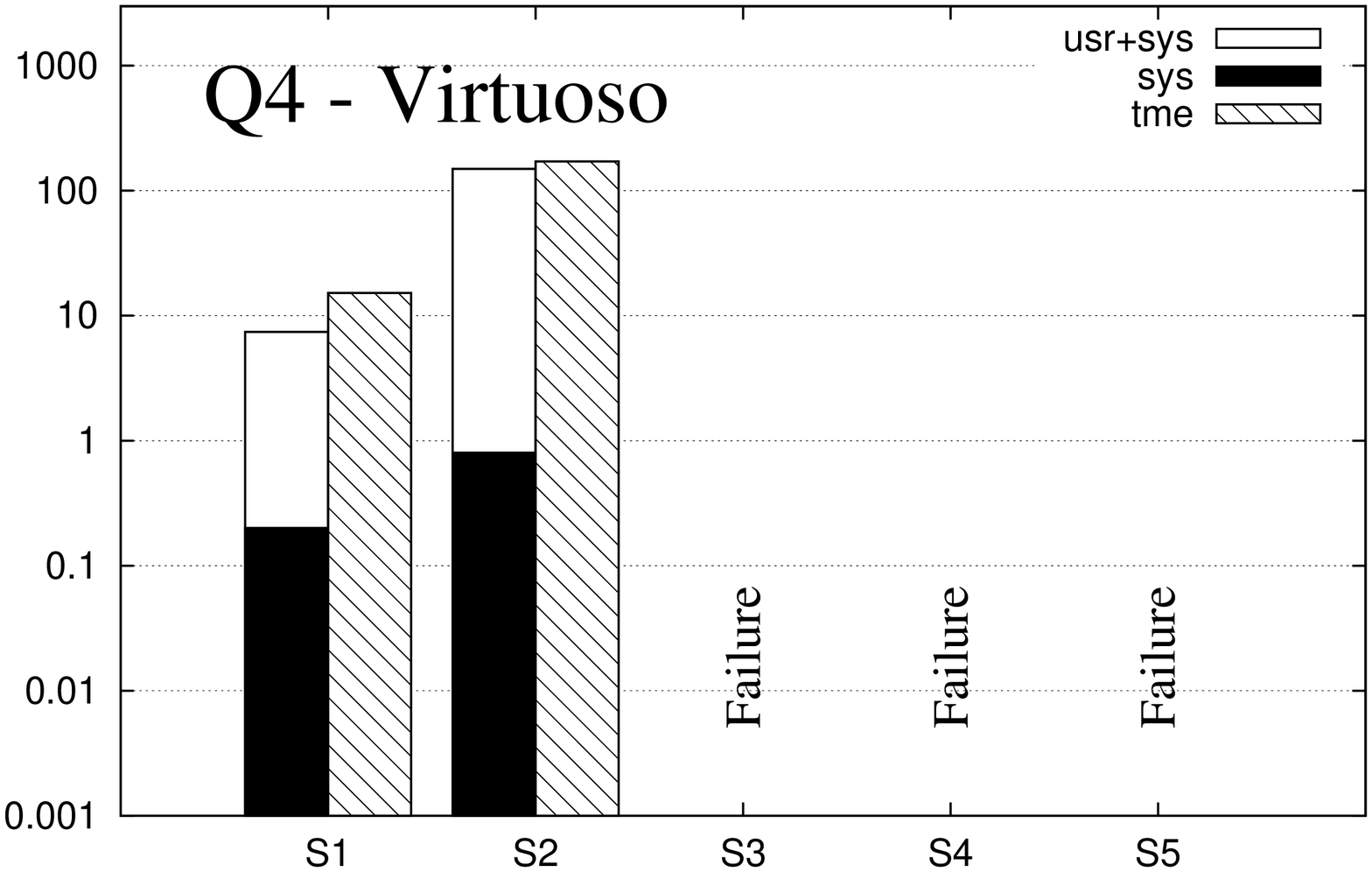}}
\end{tabular}
\caption{Query evaluation results on S1=10k, S2=50k, S3=250k, S4=1M, S5=5M, and S6=25M triples}
\label{fig:experiments1}
\end{figure*}

\begin{figure*}[t]
\hspace{-1.1cm}
\begin{tabular}{cccc}
\rotatebox{270}{\includegraphics[scale=0.182]{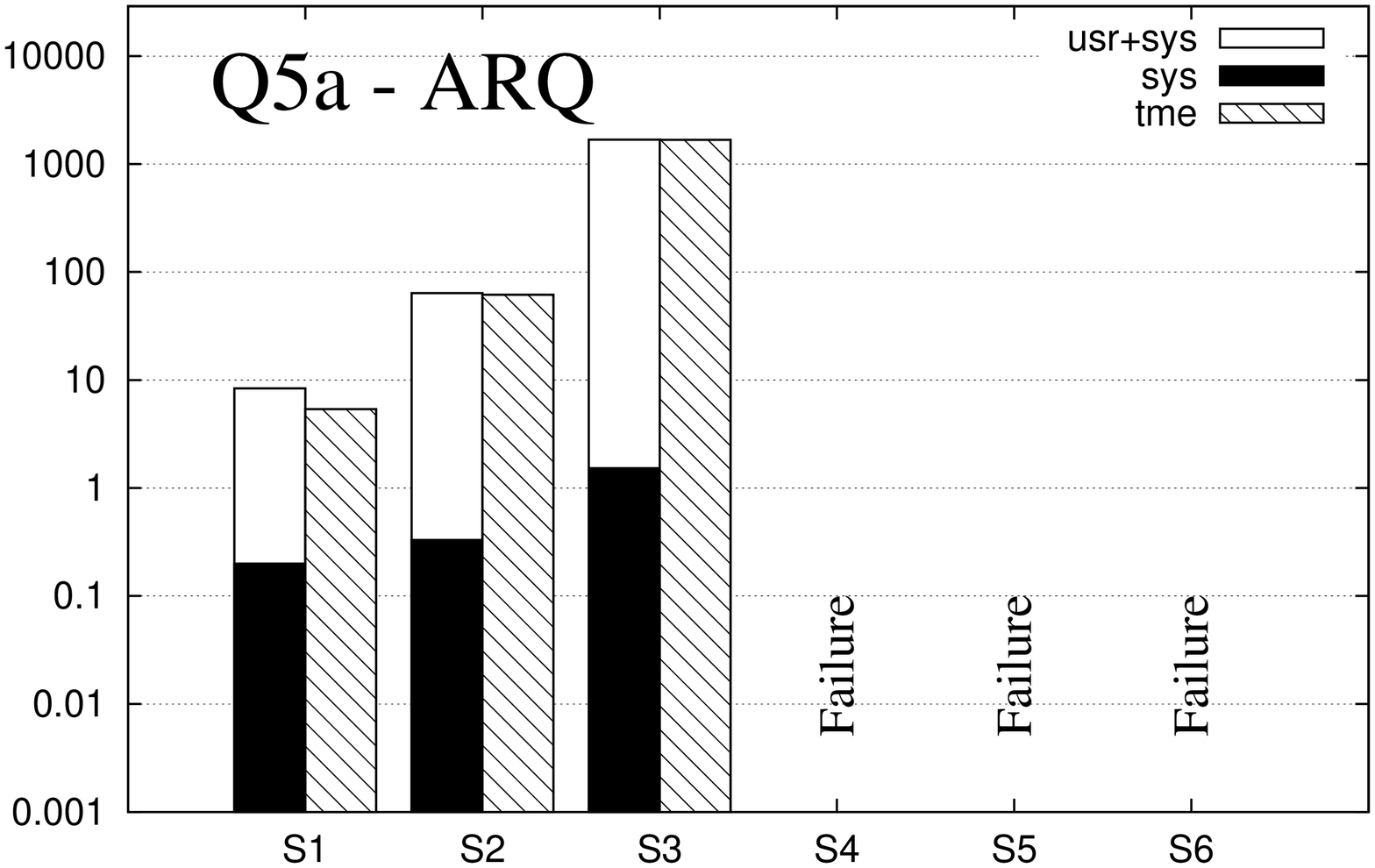}}
&
\hspace{-0.5cm}
\rotatebox{270}{\includegraphics[scale=0.182]{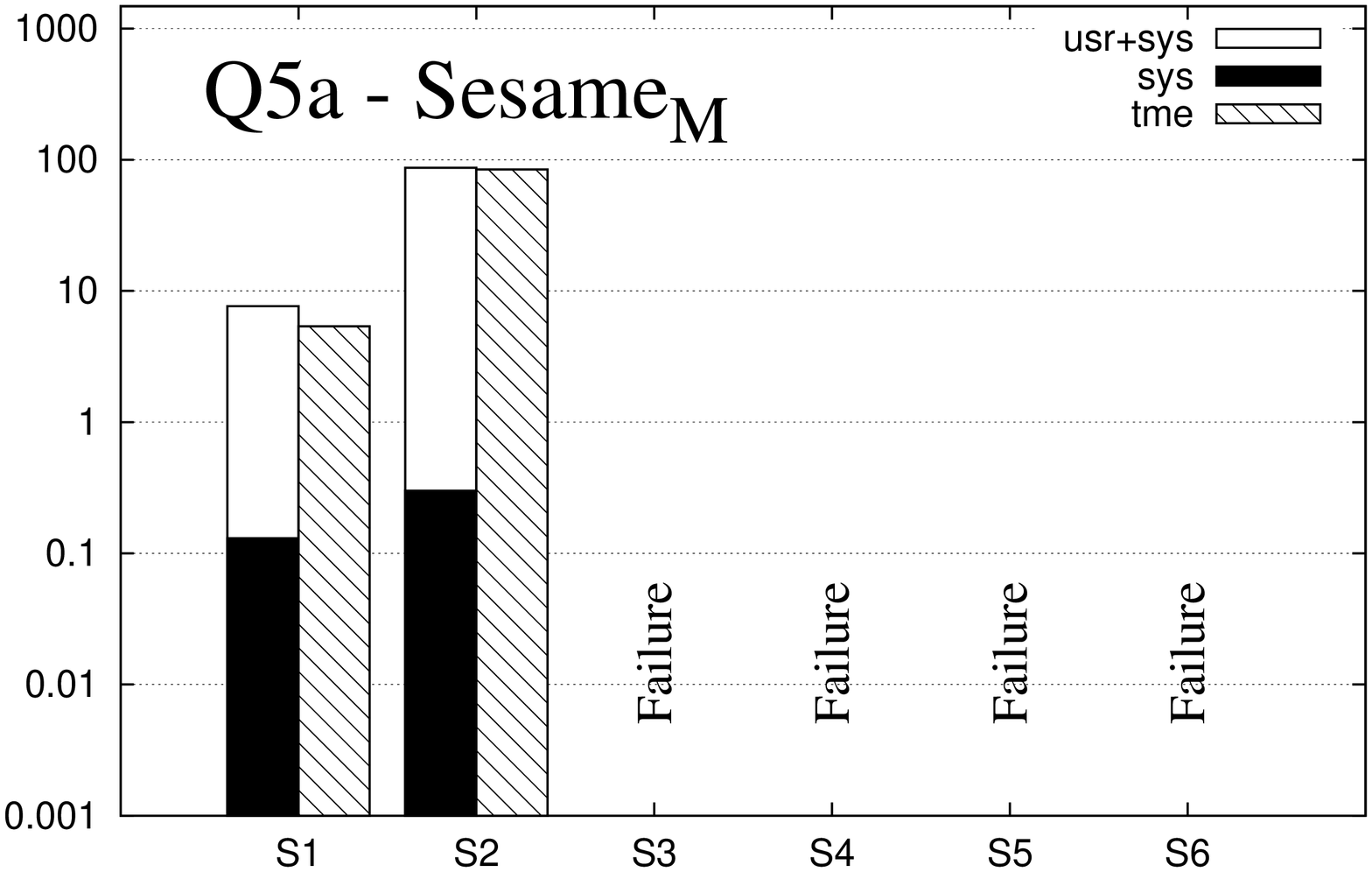}}
&
\hspace{-0.5cm}
\rotatebox{270}{\includegraphics[scale=0.182]{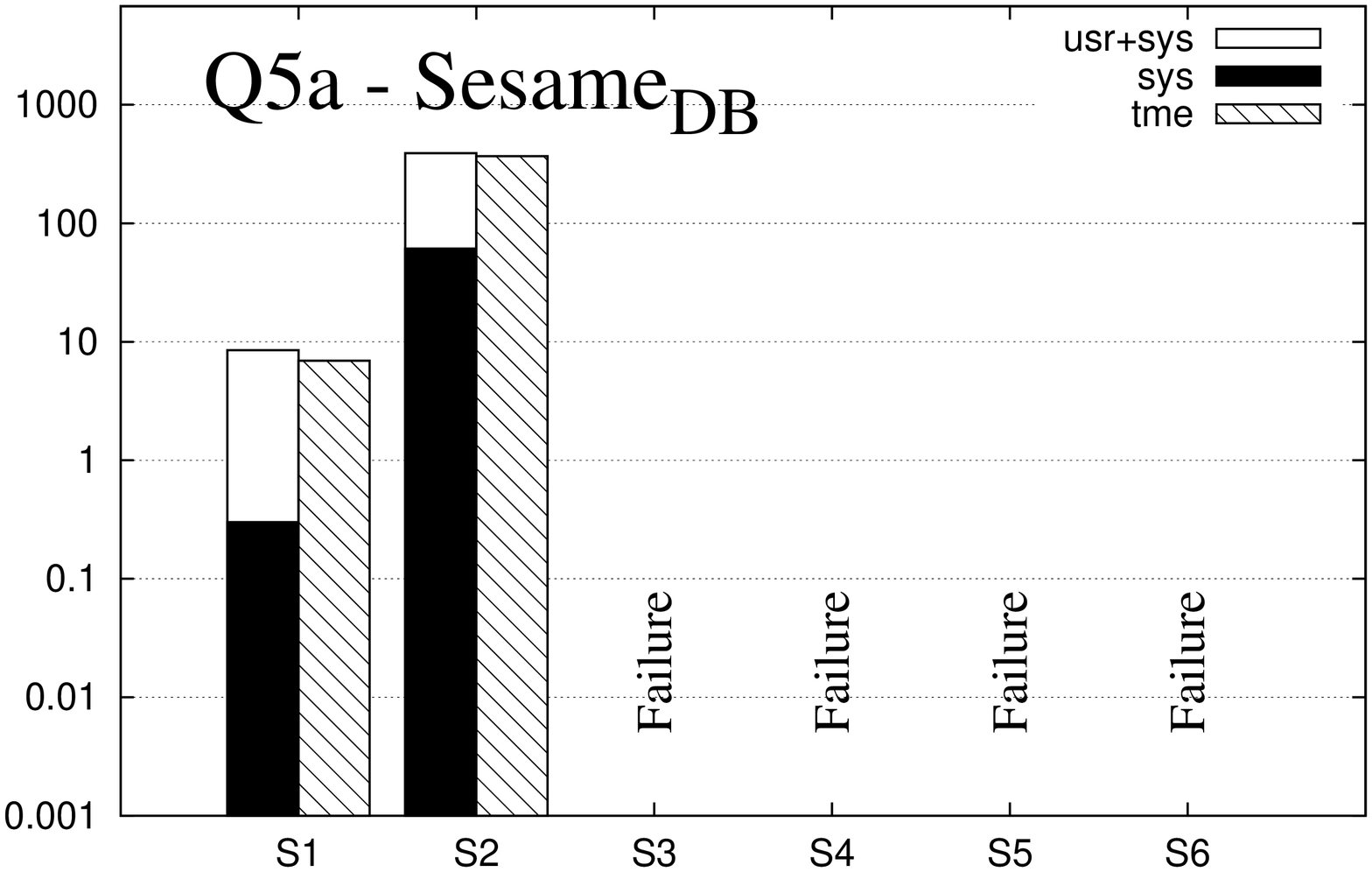}}
&
\hspace{-0.5cm}
\rotatebox{270}{\includegraphics[scale=0.182]{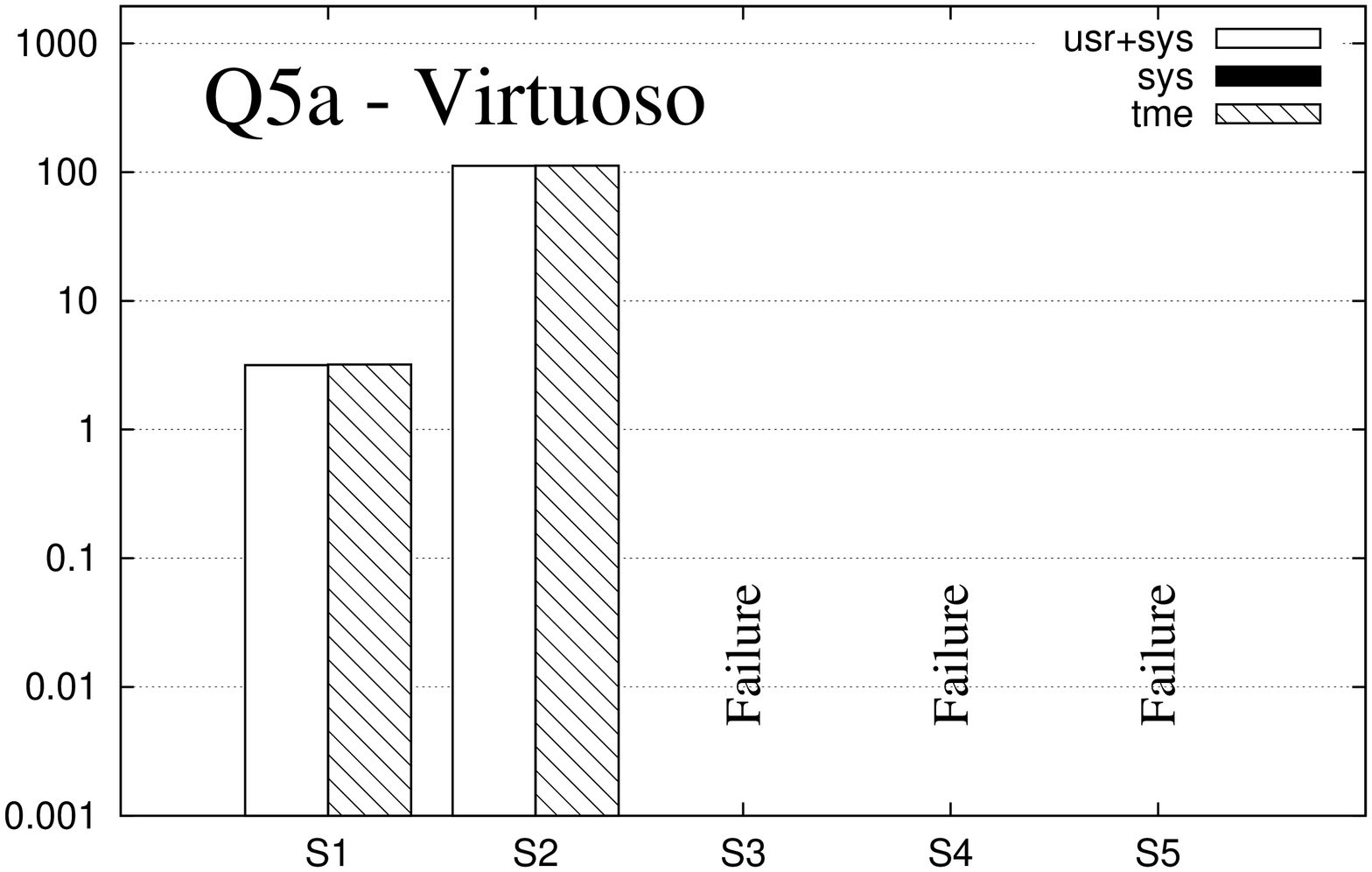}}
\\[-0.65cm]
\rotatebox{270}{\includegraphics[scale=0.18]{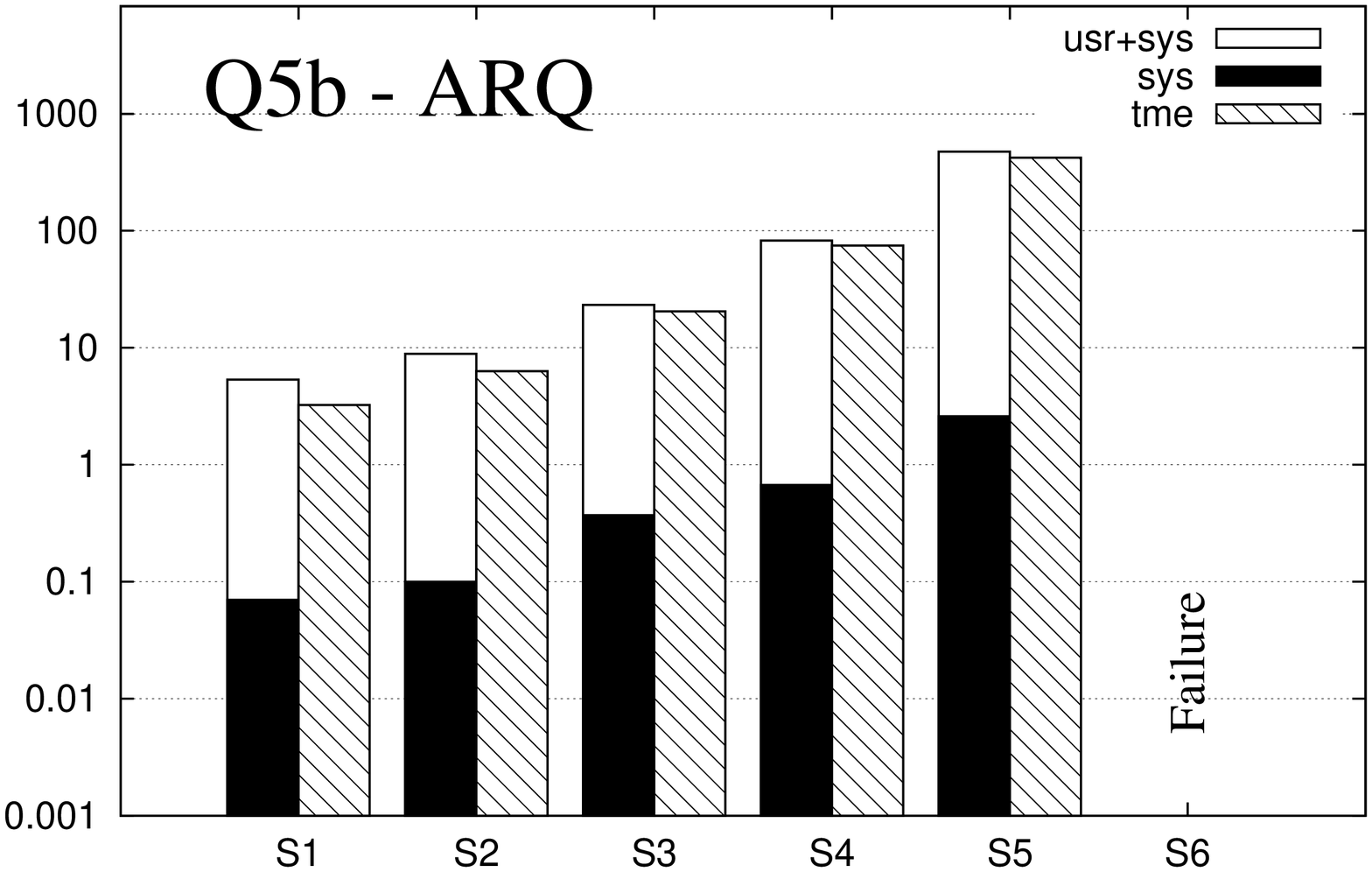}}
&
\hspace{-0.5cm}
\rotatebox{270}{\includegraphics[scale=0.18]{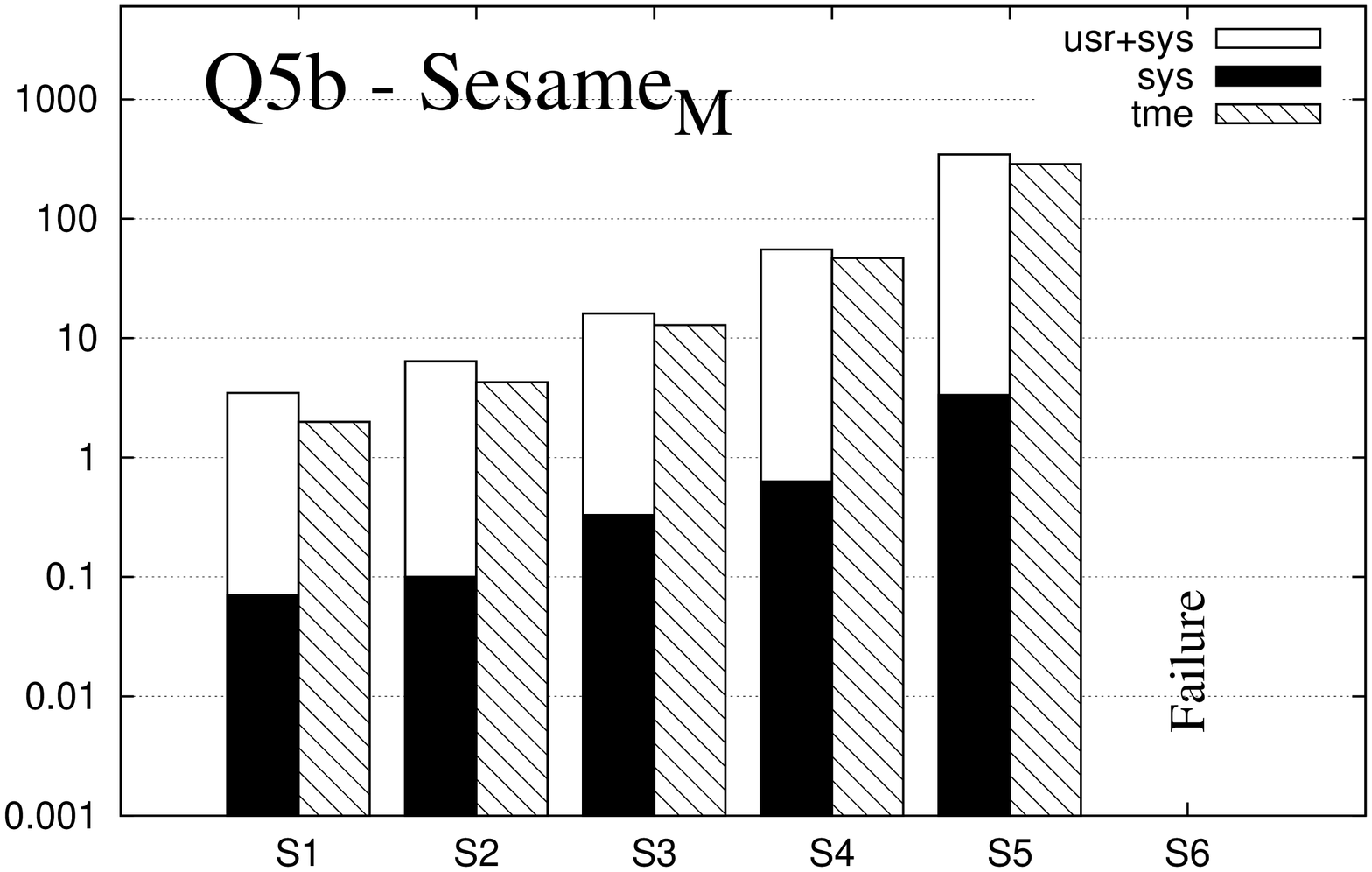}}
&
\hspace{-0.5cm}
\rotatebox{270}{\includegraphics[scale=0.18]{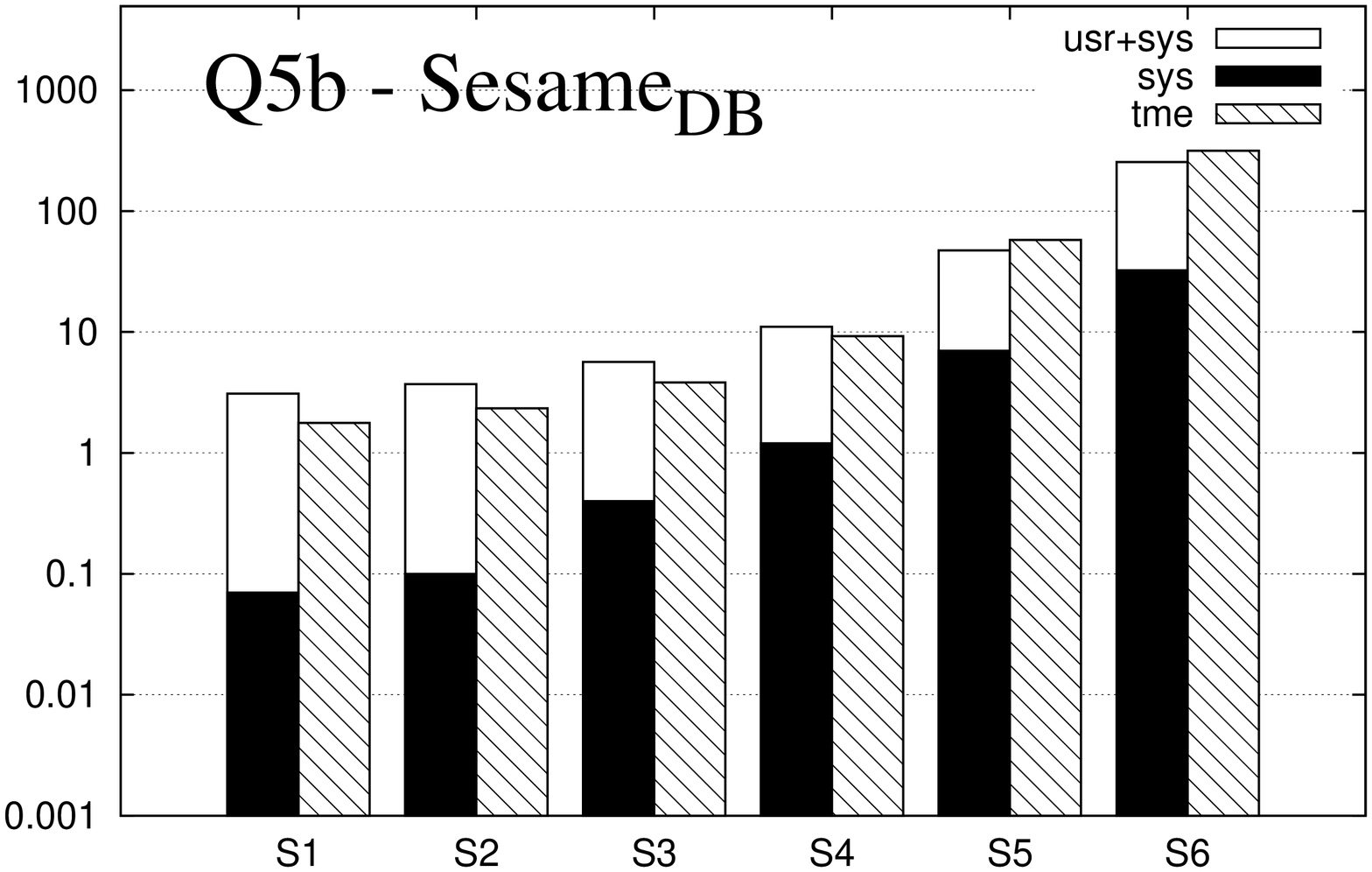}}
&
\hspace{-0.5cm}
\rotatebox{270}{\includegraphics[scale=0.18]{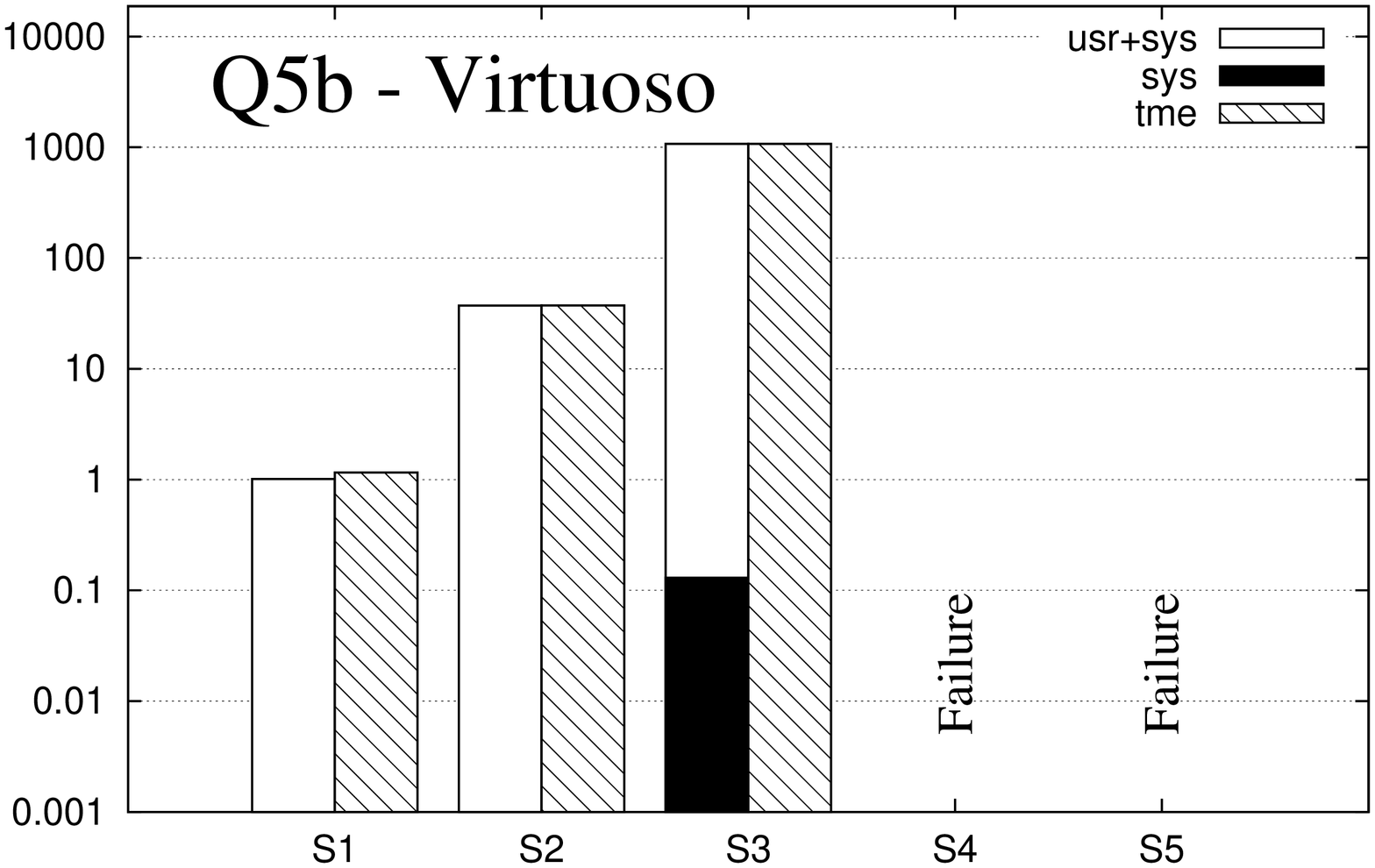}}
\vspace{-1.45cm}
\\[-0.65cm]
\rotatebox{270}{\includegraphics[scale=0.18]{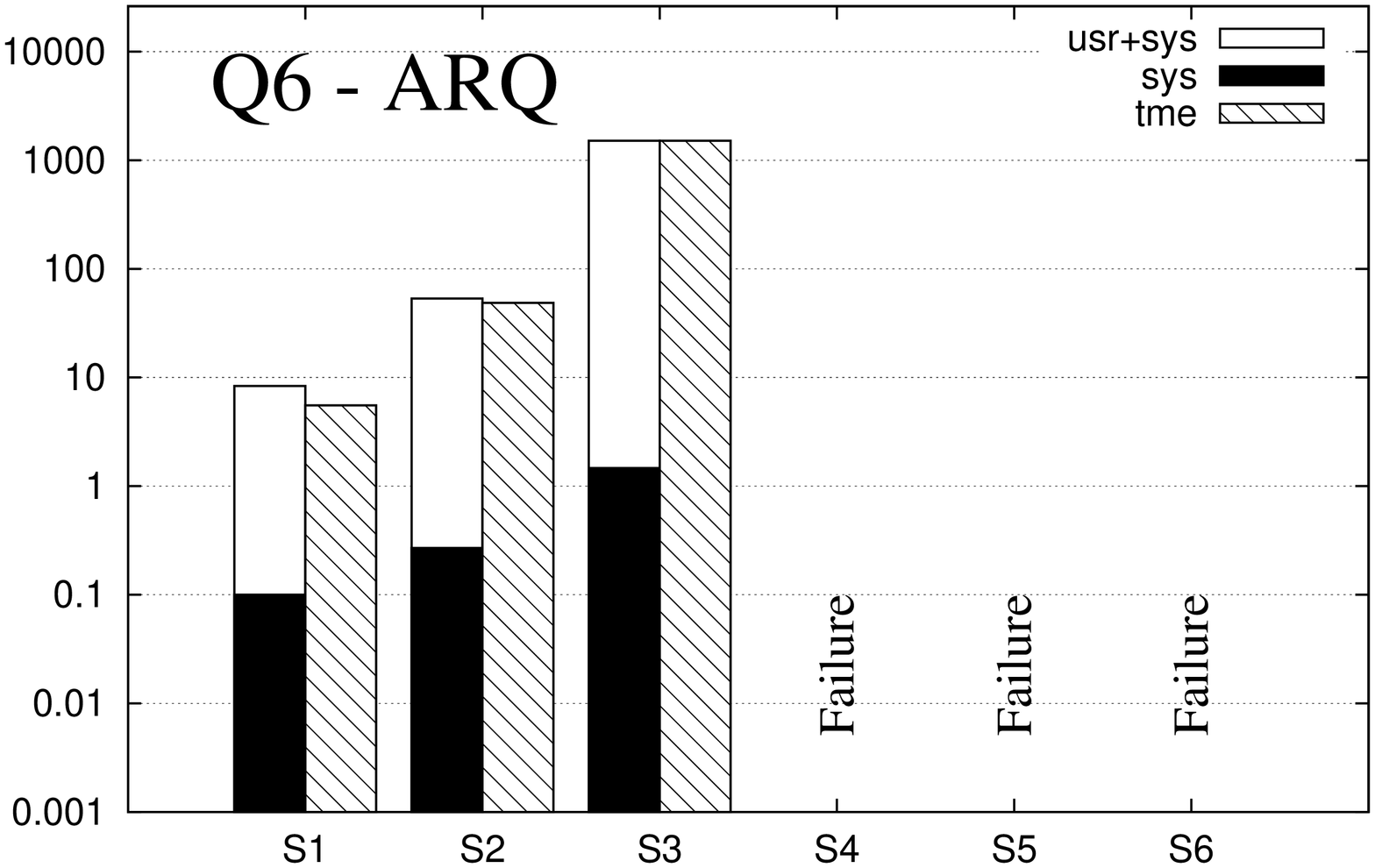}}
&
\hspace{-0.5cm}
\rotatebox{270}{\includegraphics[scale=0.18]{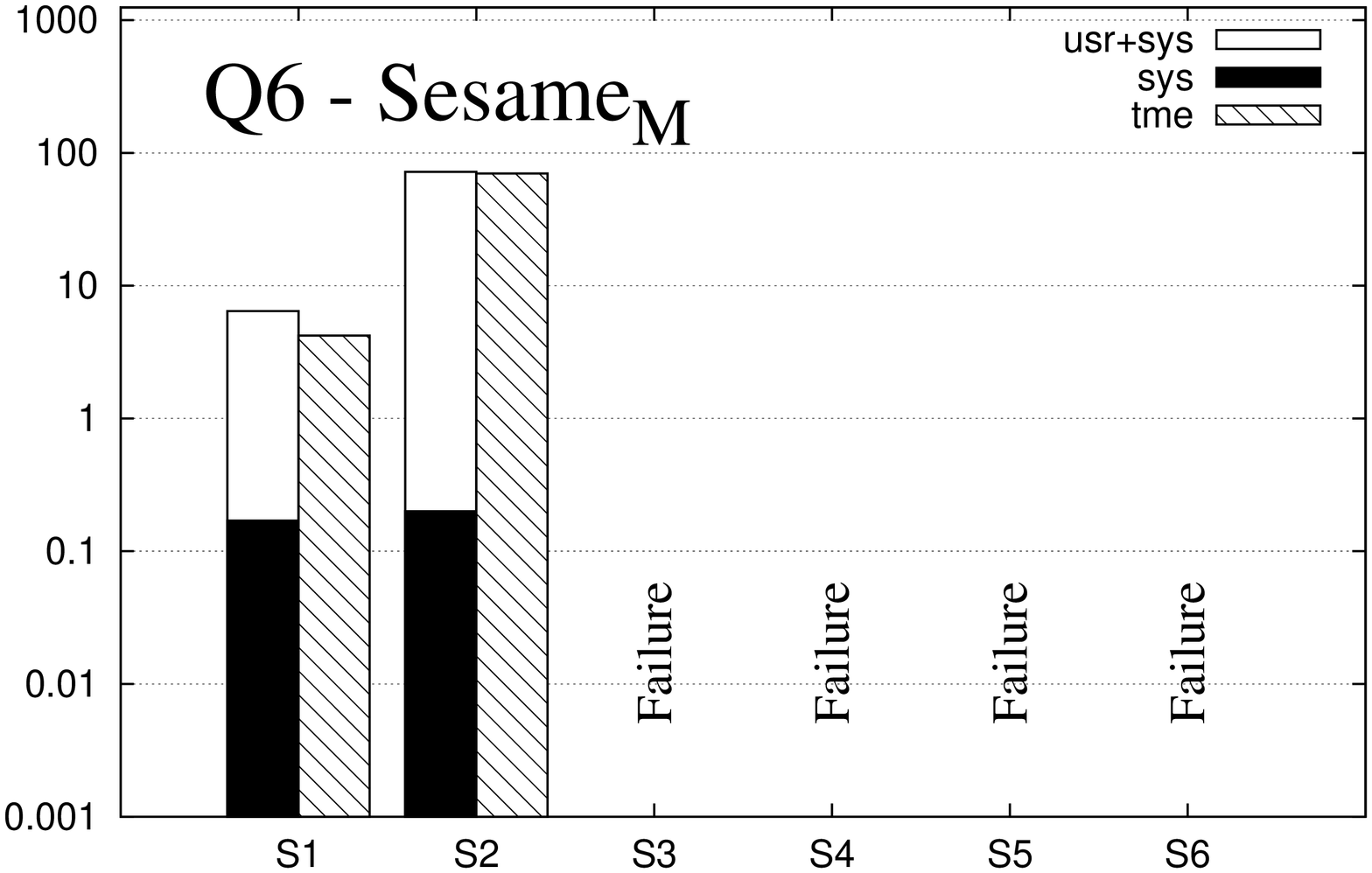}}
&
\hspace{-0.5cm}
\rotatebox{270}{\includegraphics[scale=0.18]{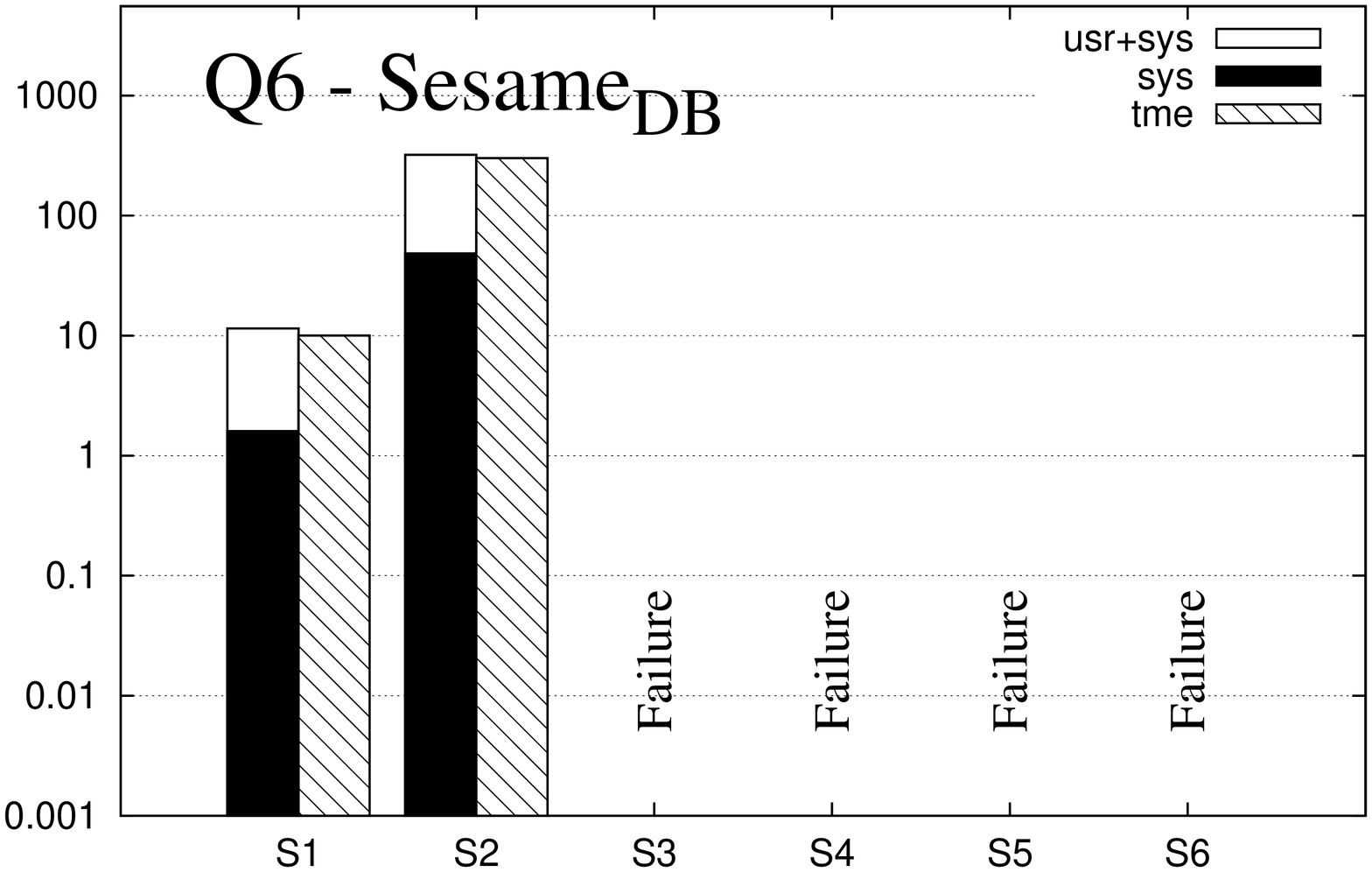}}
&
\begin{minipage}{3cm}
\begin{center}
\begin{tabbing}
\\
\\
\\
\\
\\
\\
\\
(query failed)
\end{tabbing}
\end{center}
\end{minipage}
\\[-0.65cm]
\rotatebox{270}{\includegraphics[scale=0.18]{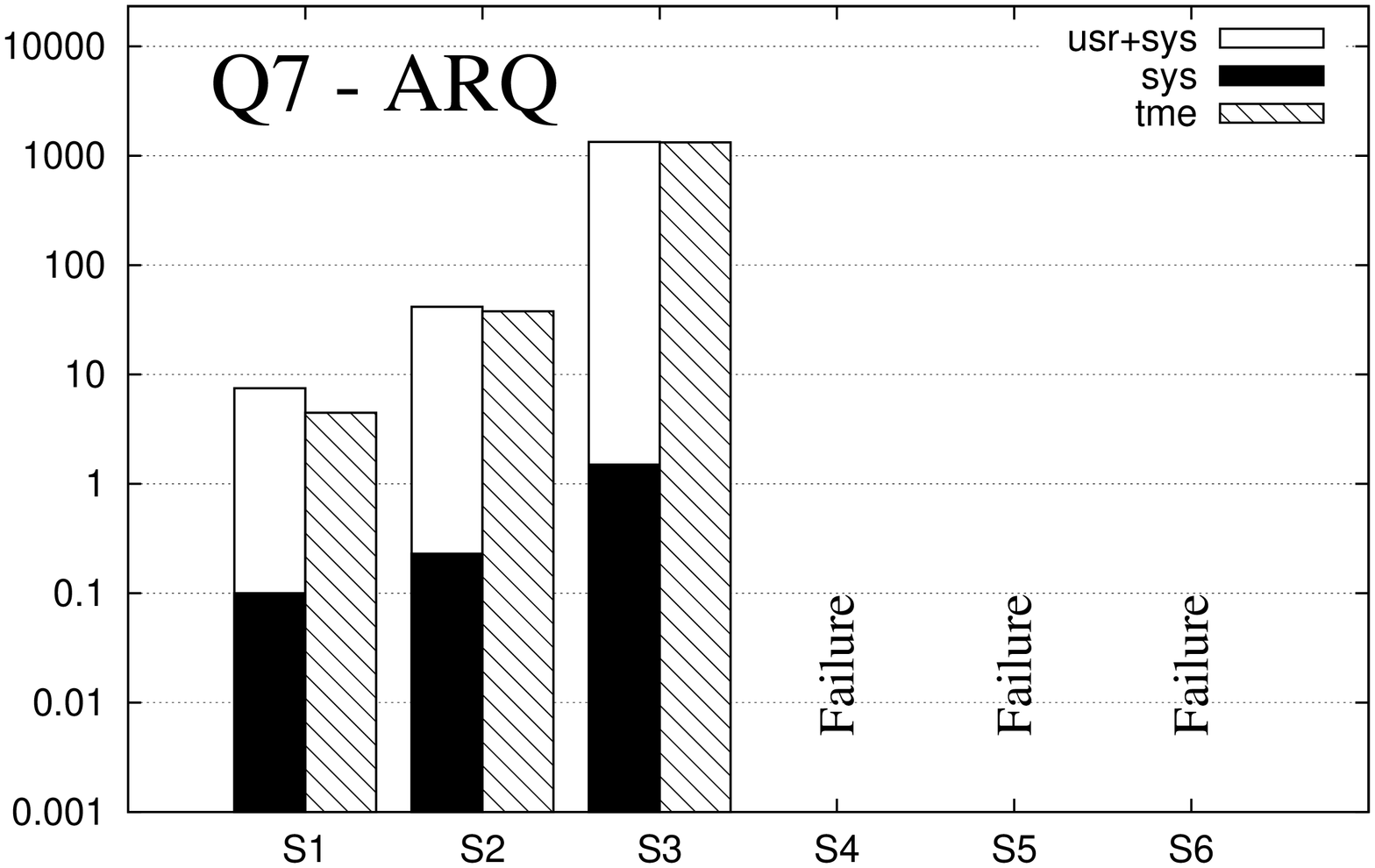}}
&
\hspace{-0.5cm}
\rotatebox{270}{\includegraphics[scale=0.18]{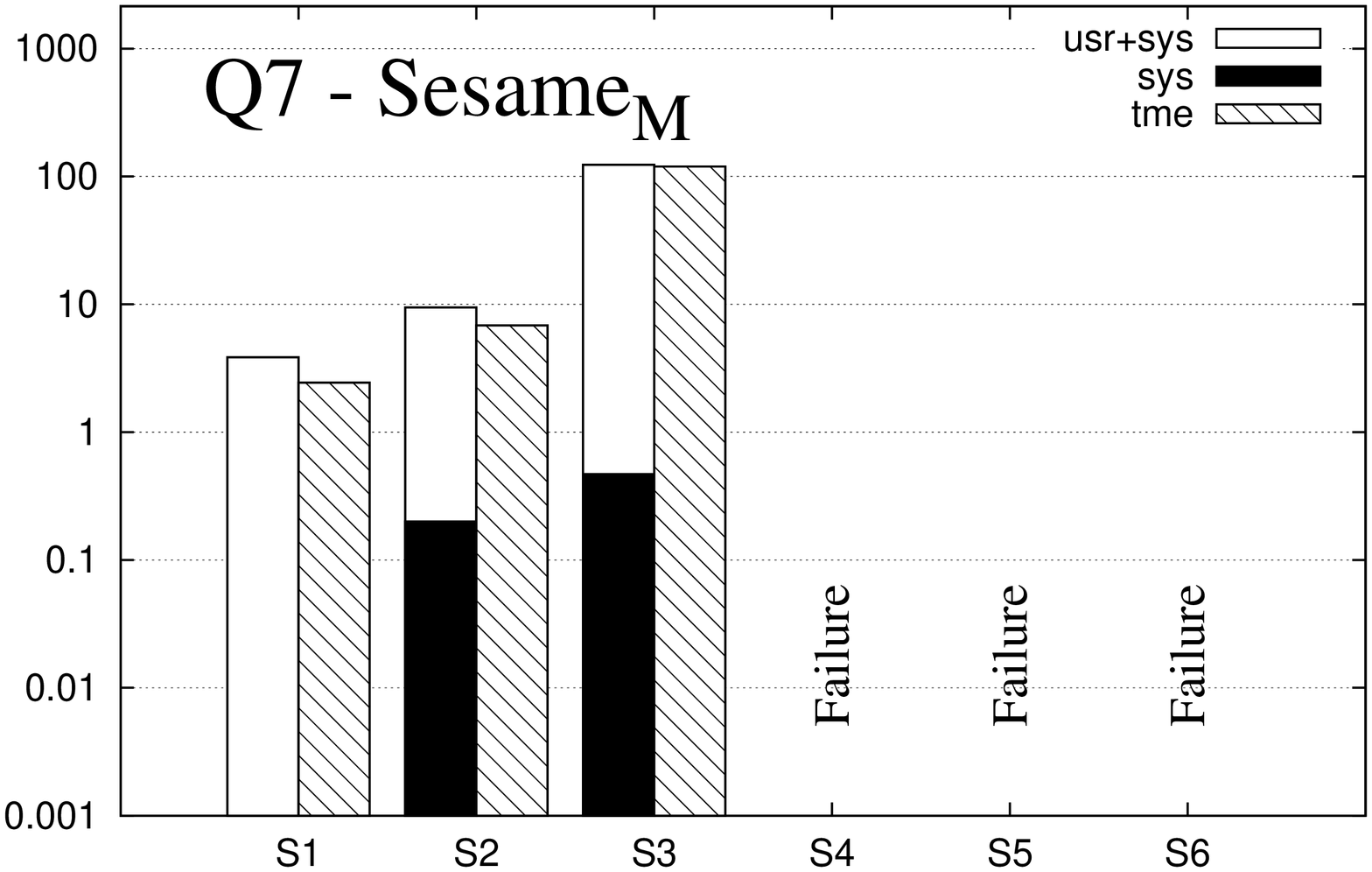}}
&
\hspace{-0.5cm}
\rotatebox{270}{\includegraphics[scale=0.18]{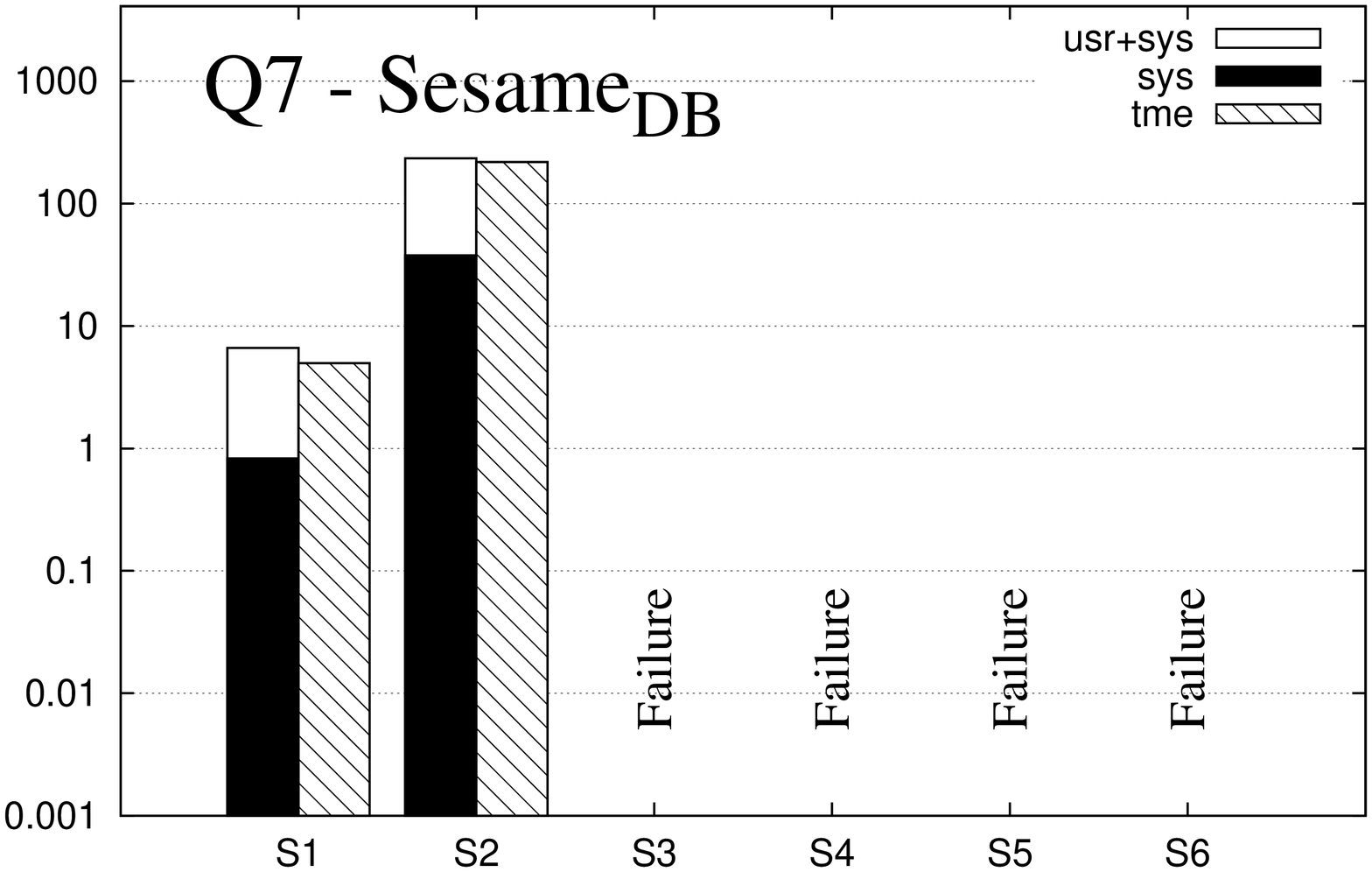}}
&
\hspace{-0.5cm}
\rotatebox{270}{\includegraphics[scale=0.18]{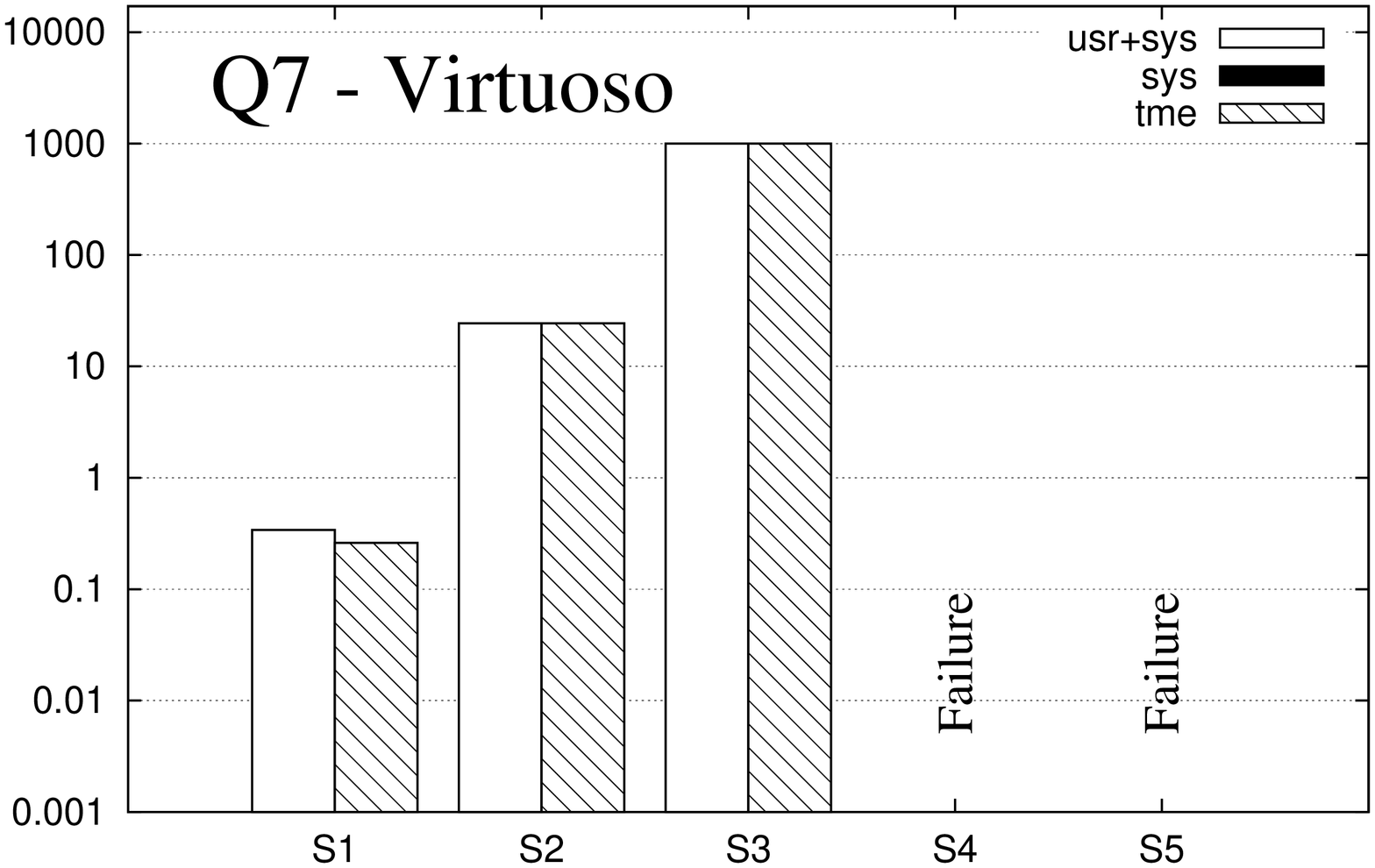}}
\\[-0.65cm]
\rotatebox{270}{\includegraphics[scale=0.18]{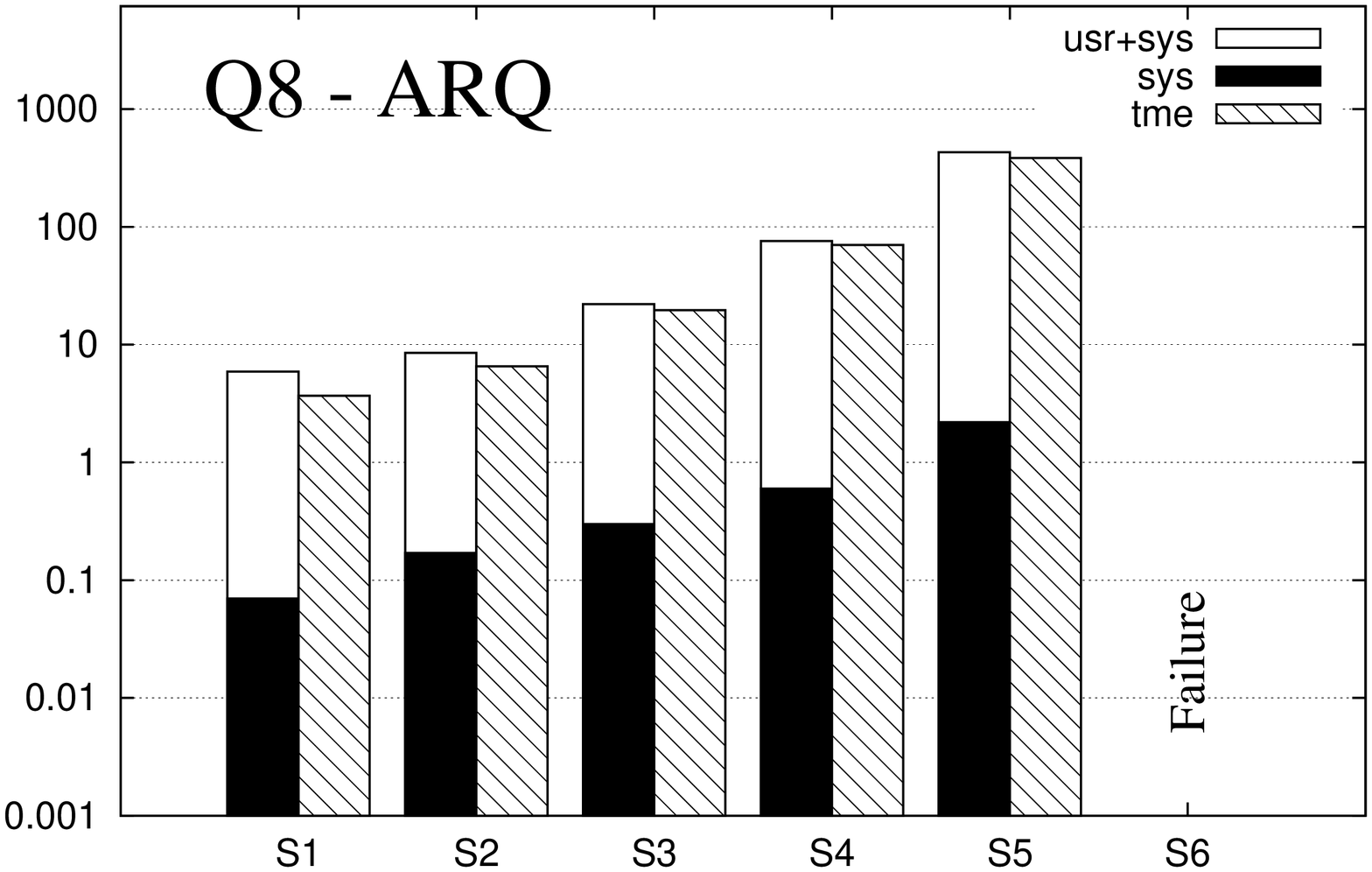}}
&
\hspace{-0.5cm}
\rotatebox{270}{\includegraphics[scale=0.18]{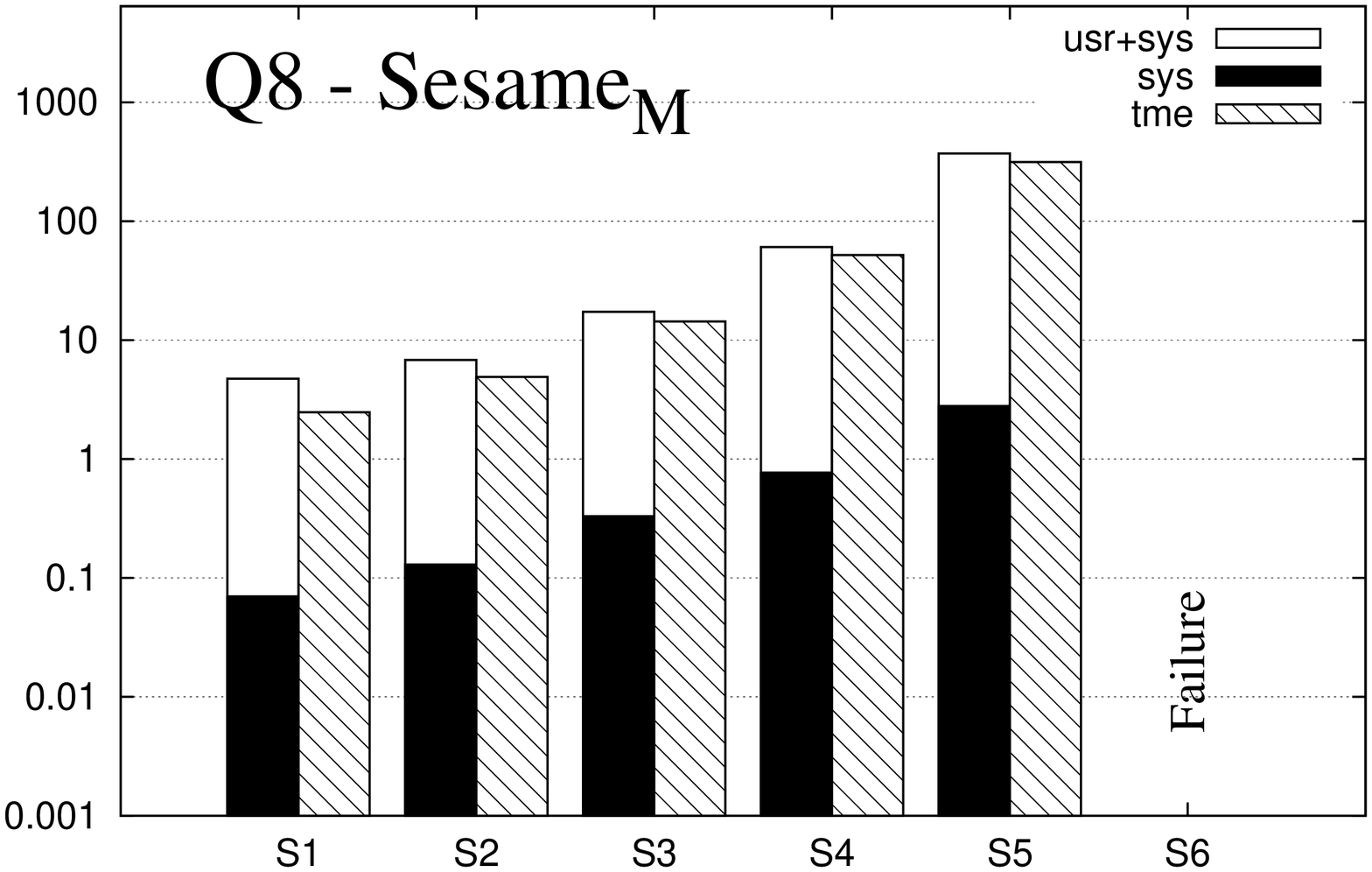}}
&
\hspace{-0.5cm}
\rotatebox{270}{\includegraphics[scale=0.18]{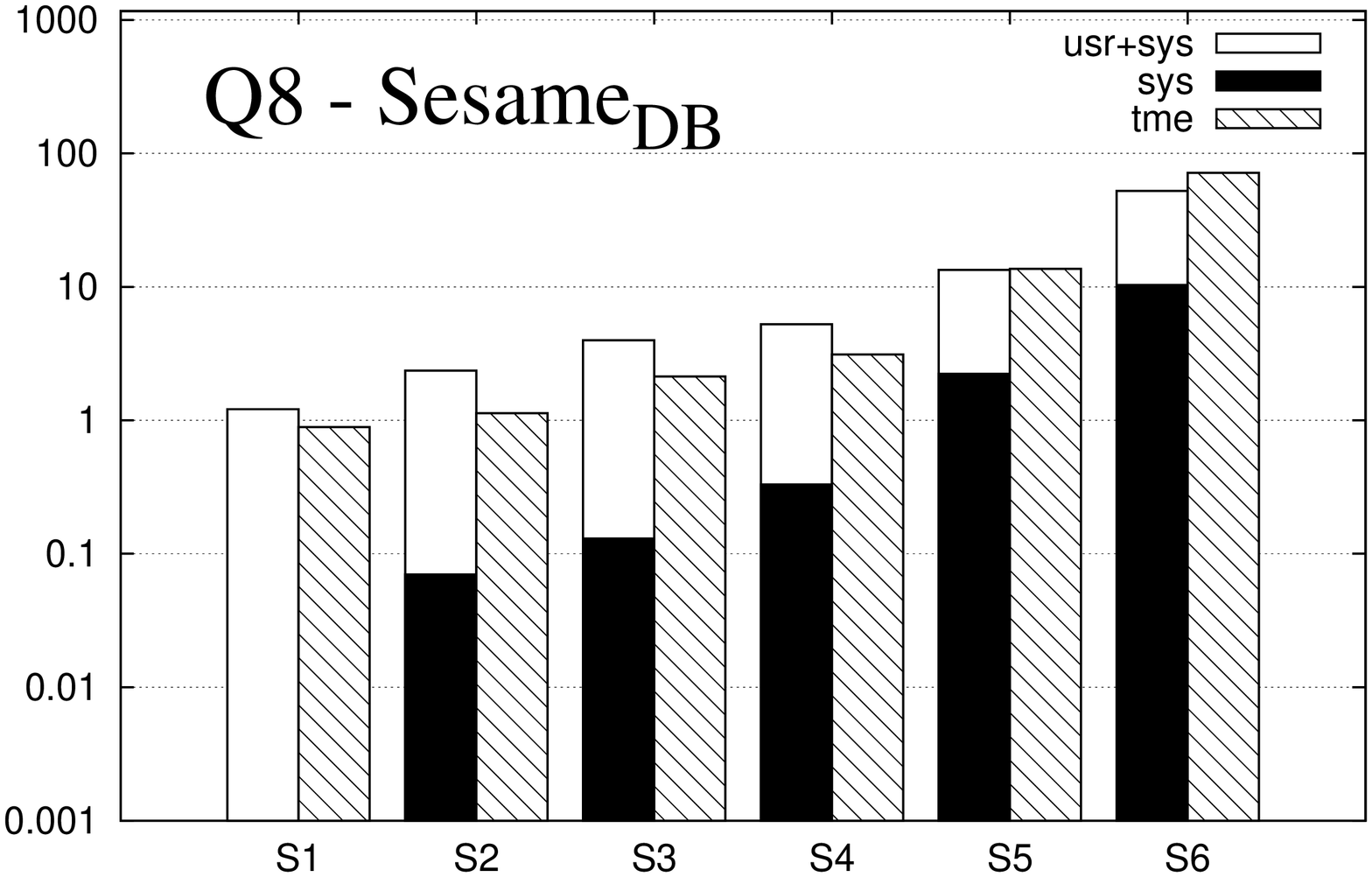}}
&
\hspace{-0.5cm}
\rotatebox{270}{\includegraphics[scale=0.18]{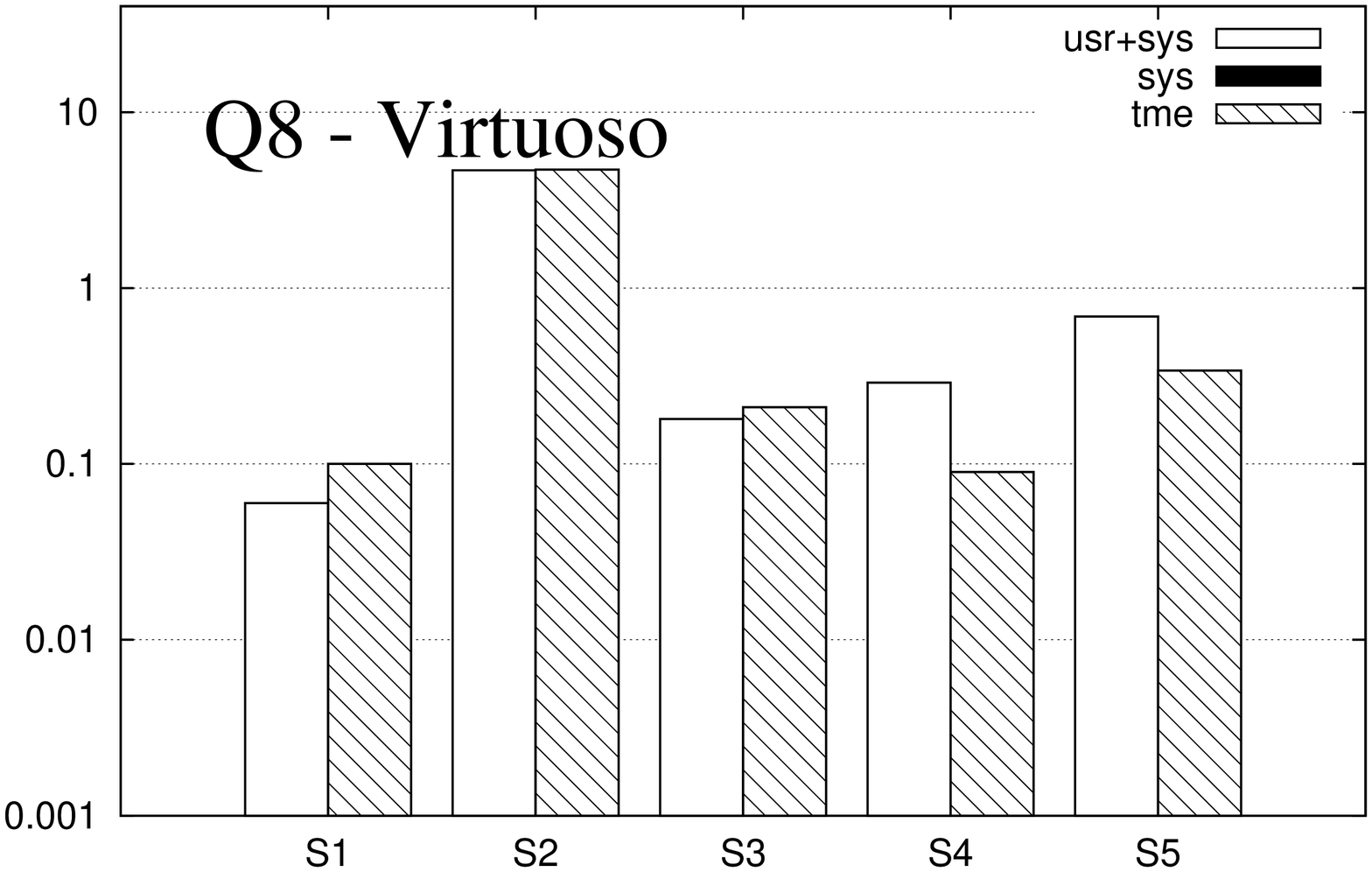}}
\\[-0.65cm]
\rotatebox{270}{\includegraphics[scale=0.18]{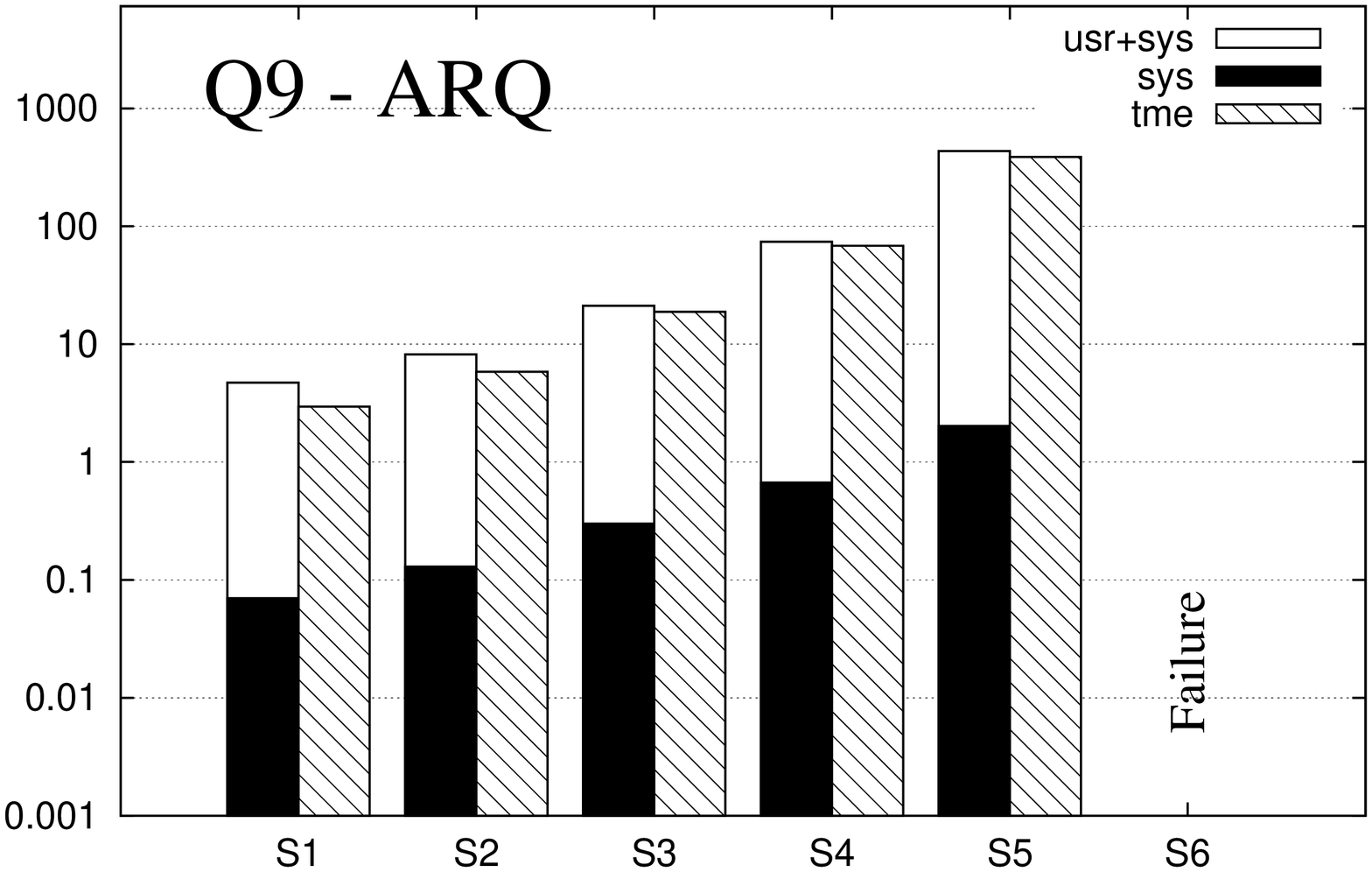}}
&
\hspace{-0.5cm}
\rotatebox{270}{\includegraphics[scale=0.18]{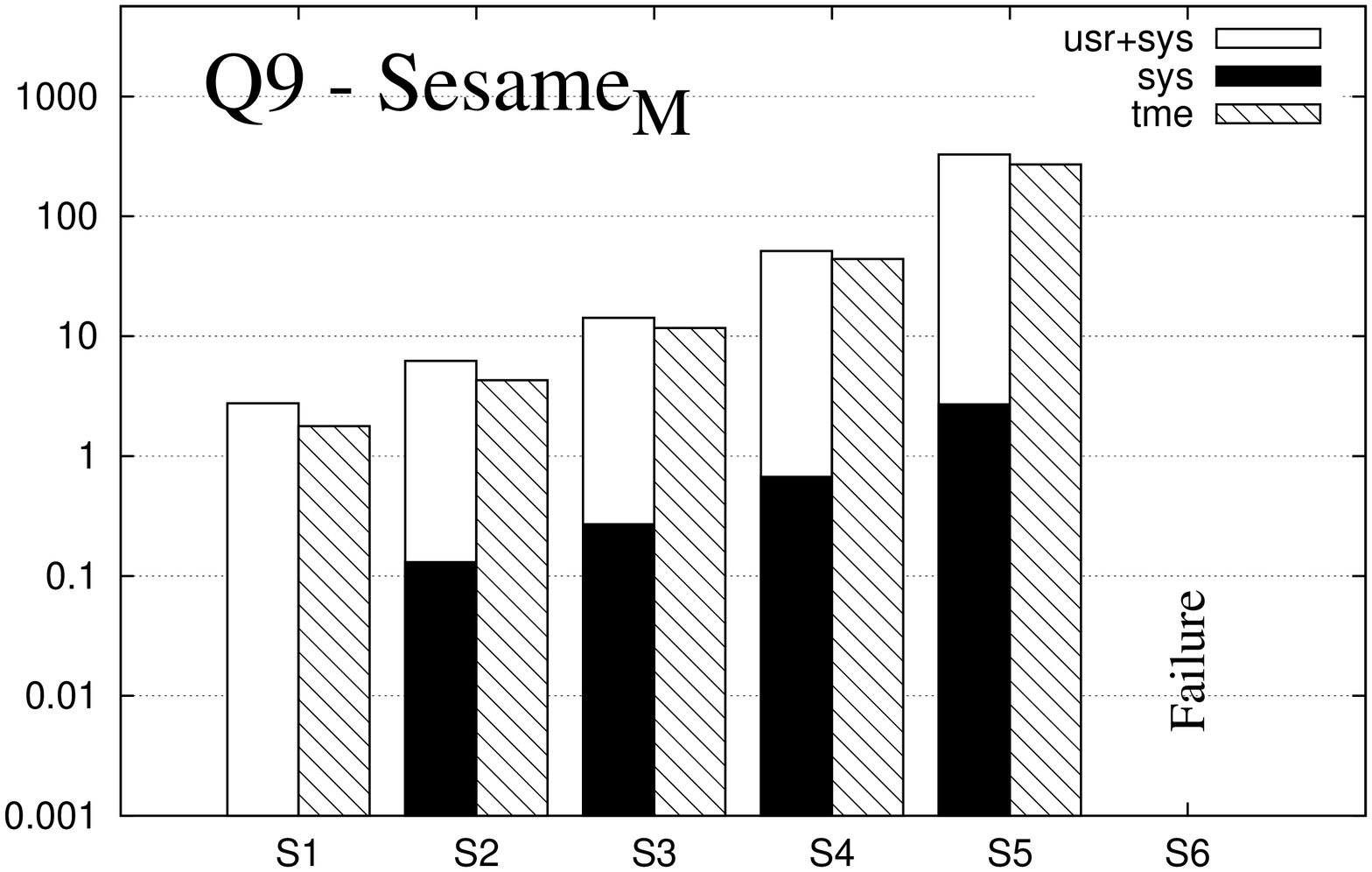}}
&
\hspace{-0.5cm}
\rotatebox{270}{\includegraphics[scale=0.18]{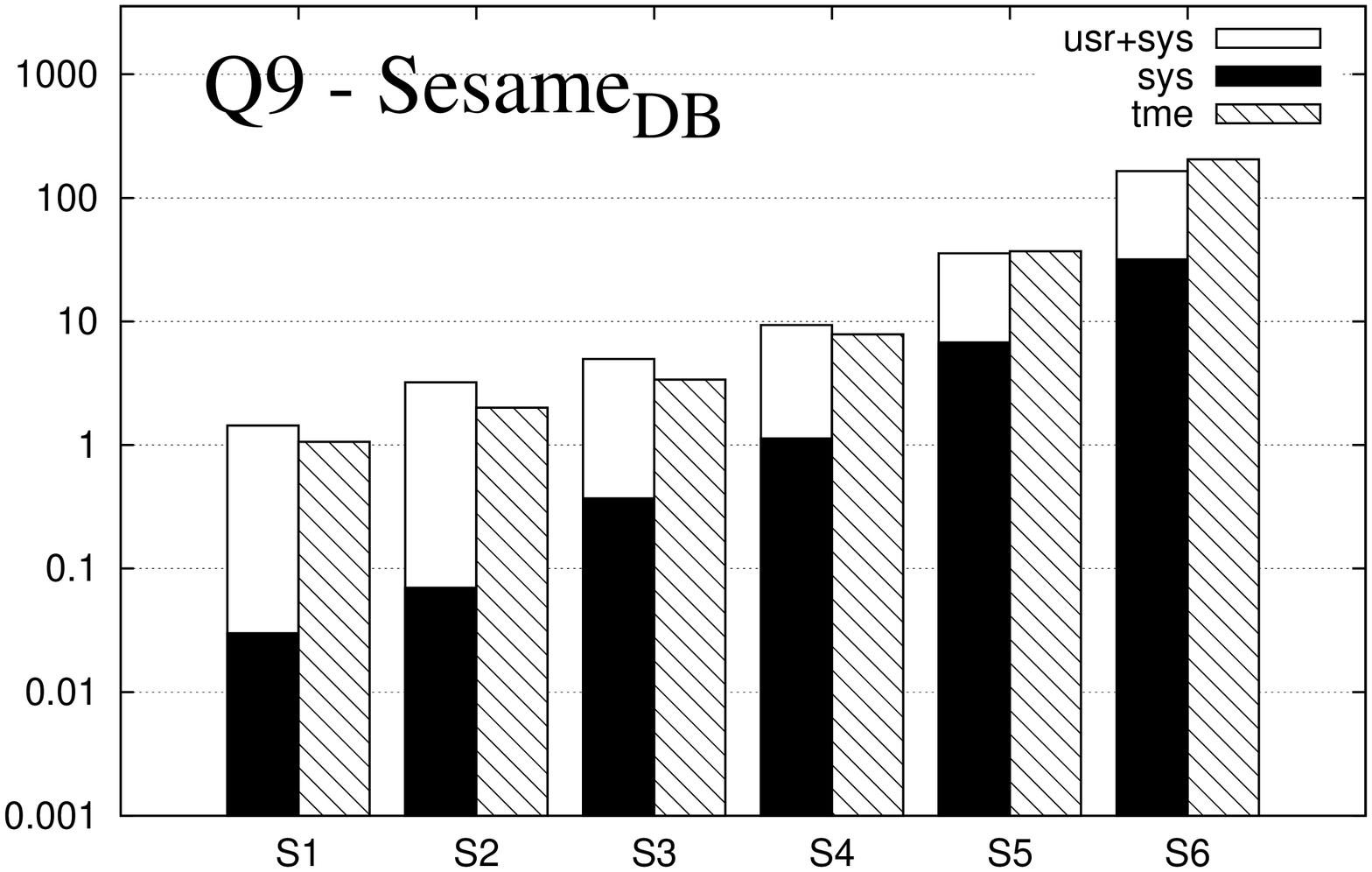}}
&
\hspace{-0.5cm}
\rotatebox{270}{\includegraphics[scale=0.18]{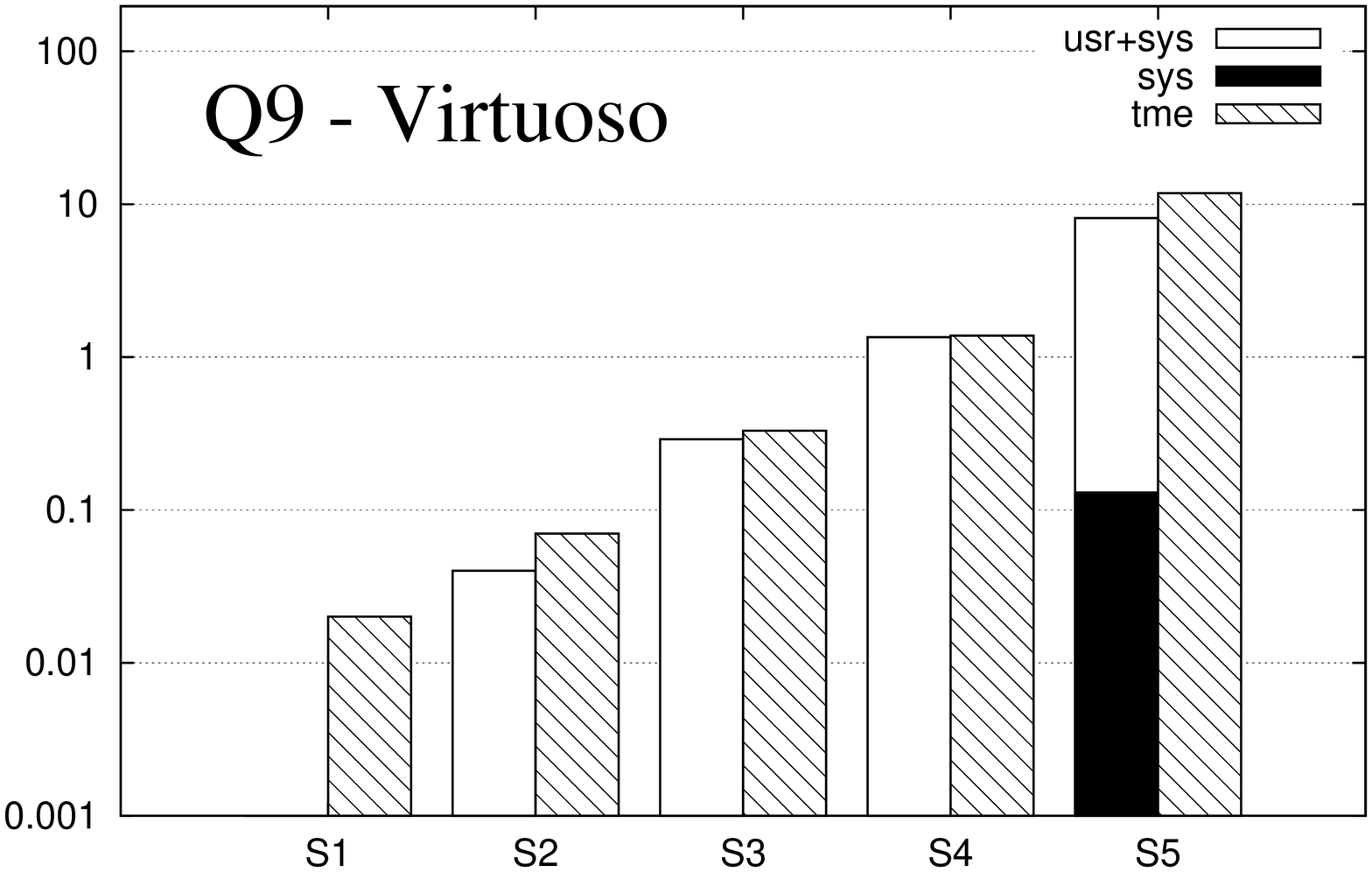}}
\end{tabular}
\caption{Query evaluation results on S1=10k, S2=50k, S3=250k, S4=1M, S5=5M, and S6=25M triples}
\label{fig:experiments2}
\end{figure*}

\begin{figure*}[t]
\hspace{-1.1cm}
\begin{tabular}{cccc}
\rotatebox{270}{\includegraphics[scale=0.182]{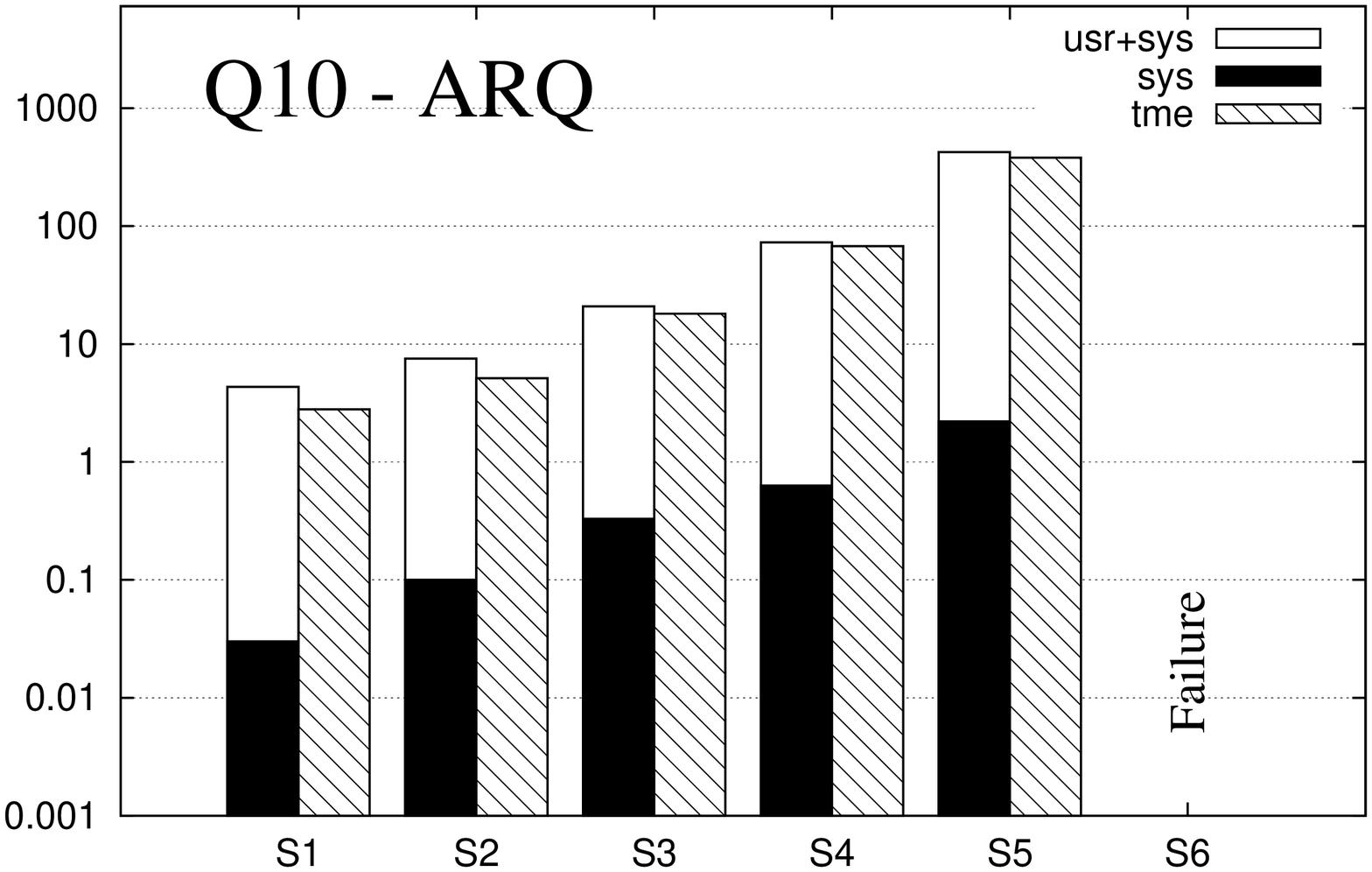}}
&
\hspace{-0.5cm}
\rotatebox{270}{\includegraphics[scale=0.182]{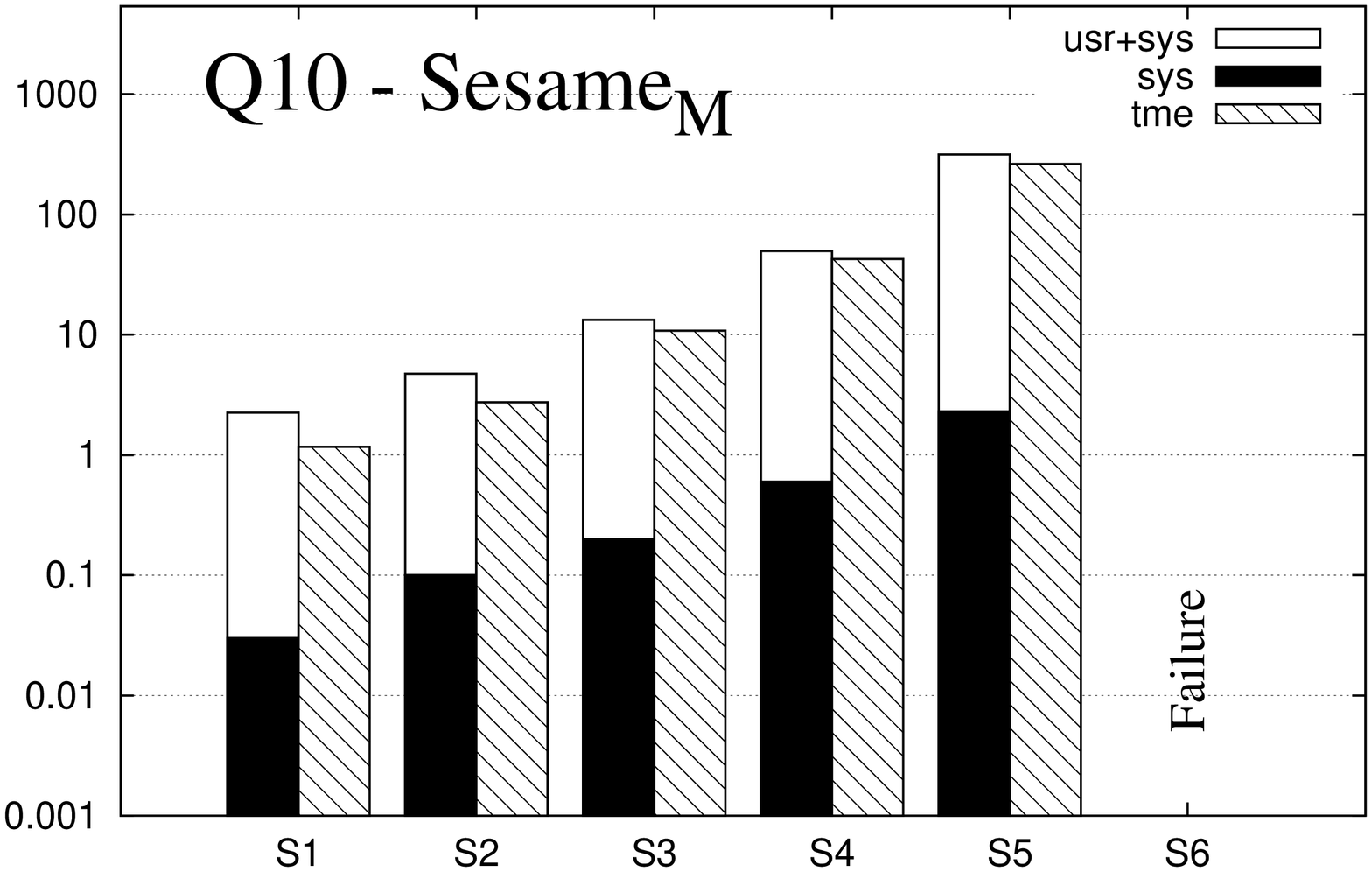}}
&
\hspace{-0.5cm}
\rotatebox{270}{\includegraphics[scale=0.182]{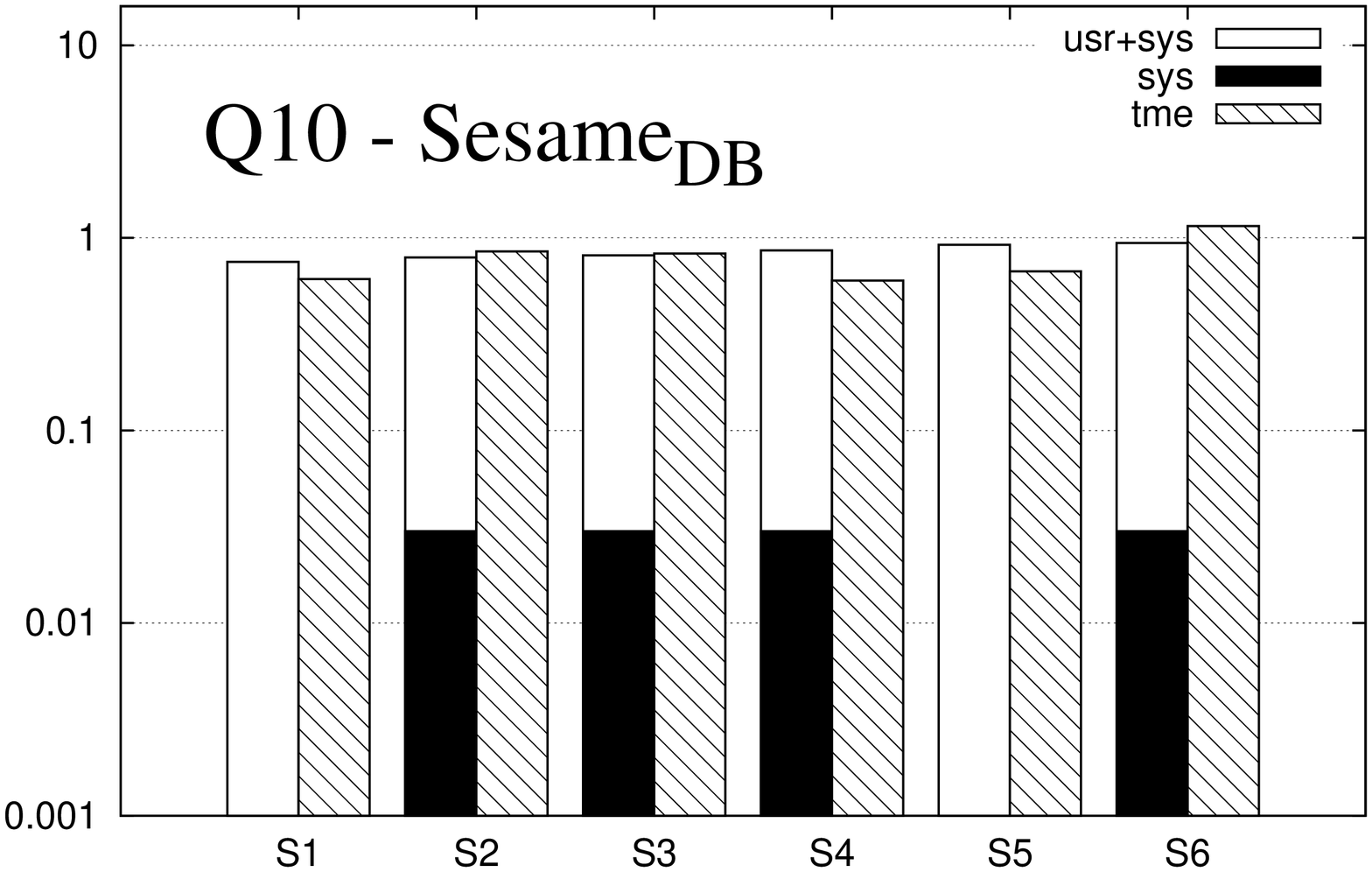}}
&
\hspace{-0.5cm}
\rotatebox{270}{\includegraphics[scale=0.182]{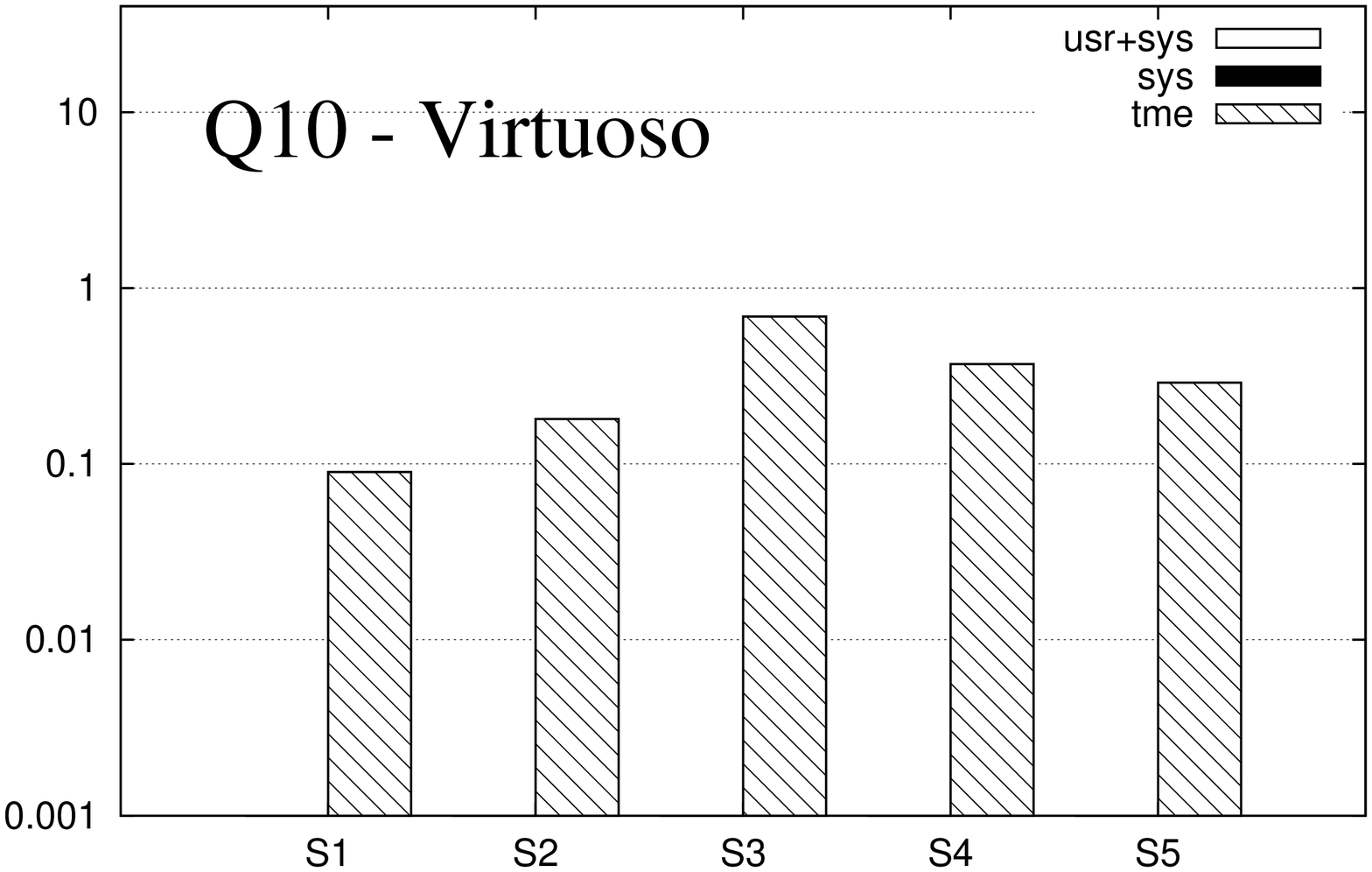}}
\\[-0.65cm]
\rotatebox{270}{\includegraphics[scale=0.18]{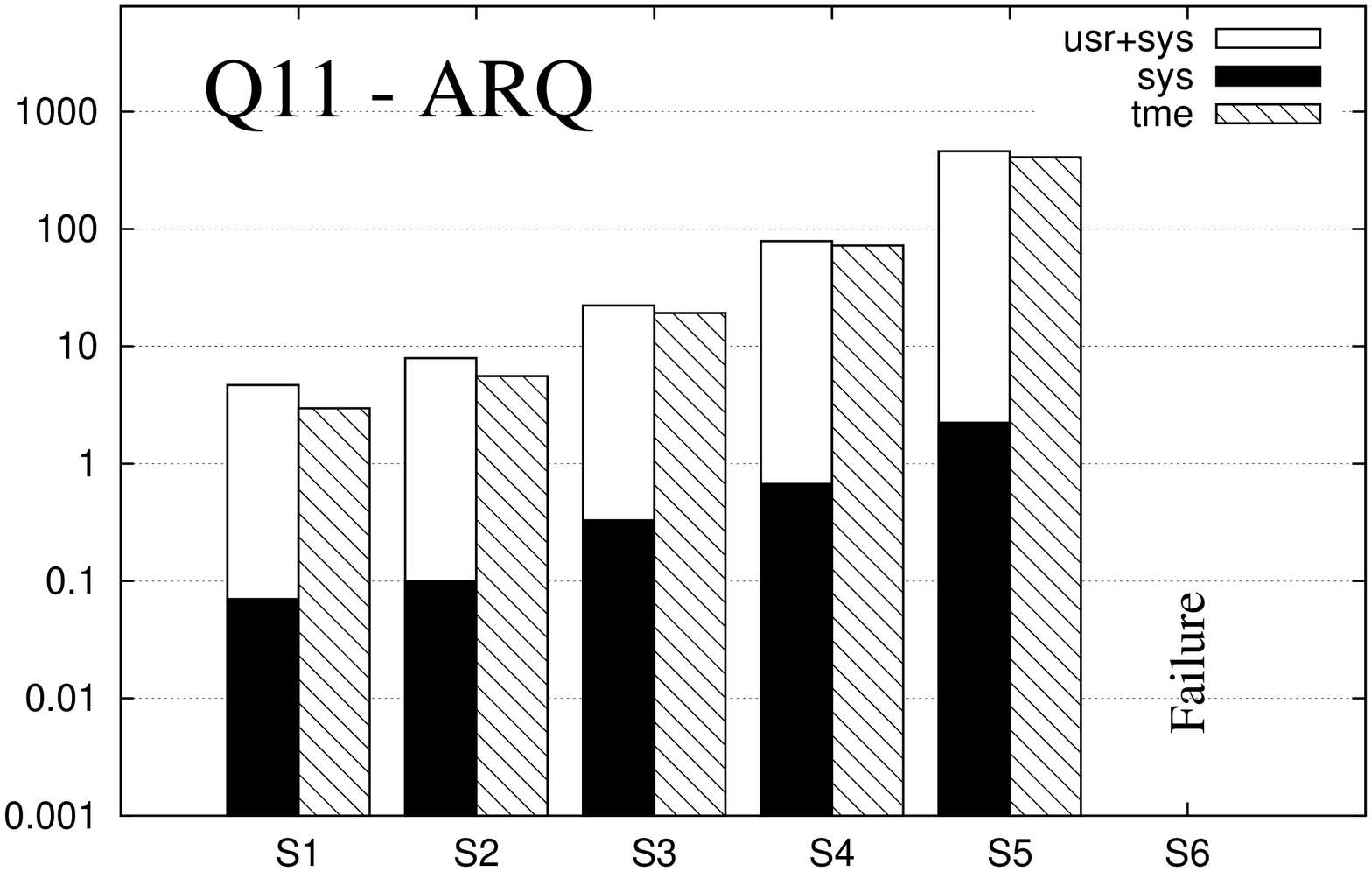}}
&
\hspace{-0.5cm}
\rotatebox{270}{\includegraphics[scale=0.18]{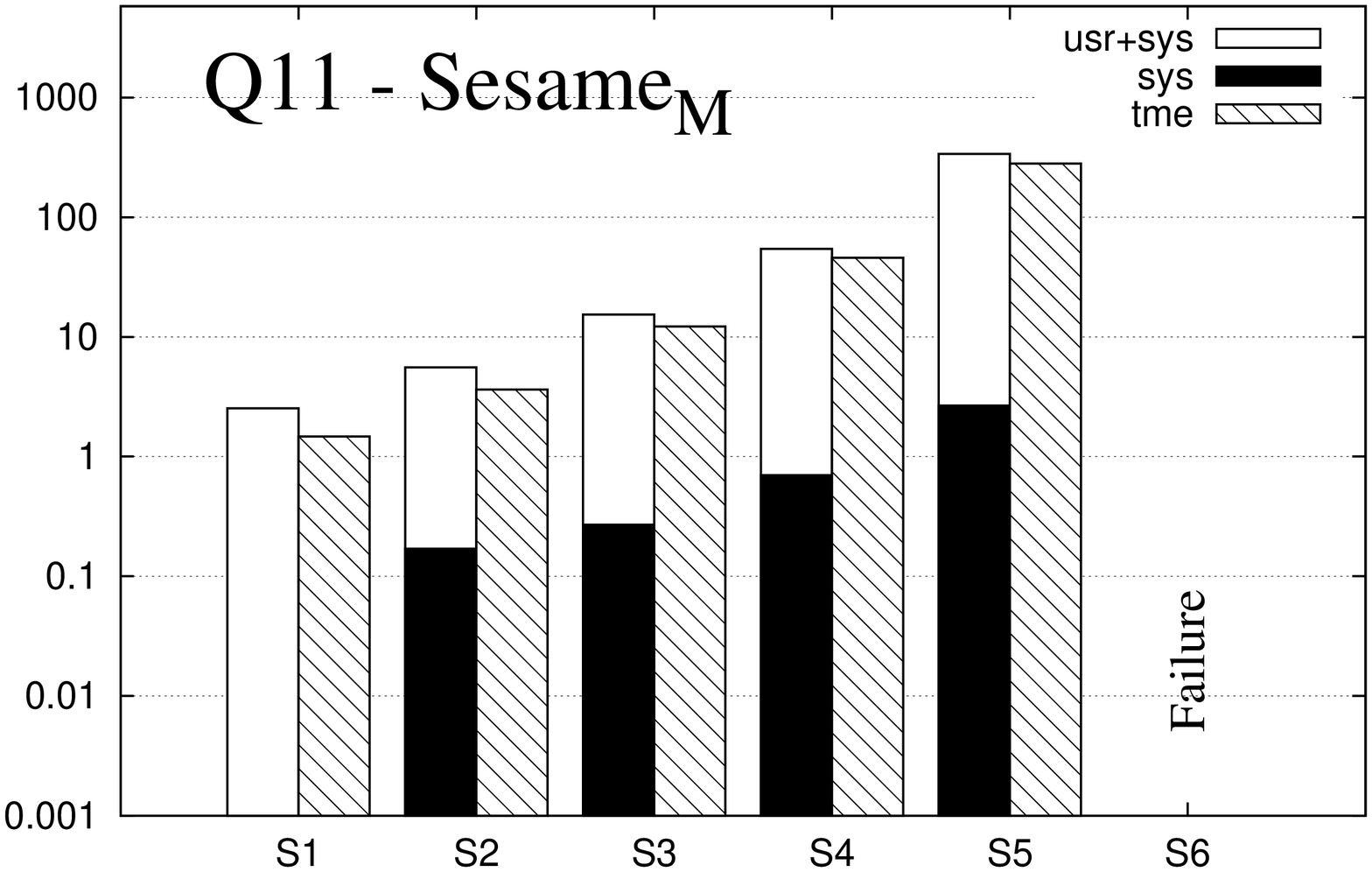}}
&
\hspace{-0.5cm}
\rotatebox{270}{\includegraphics[scale=0.18]{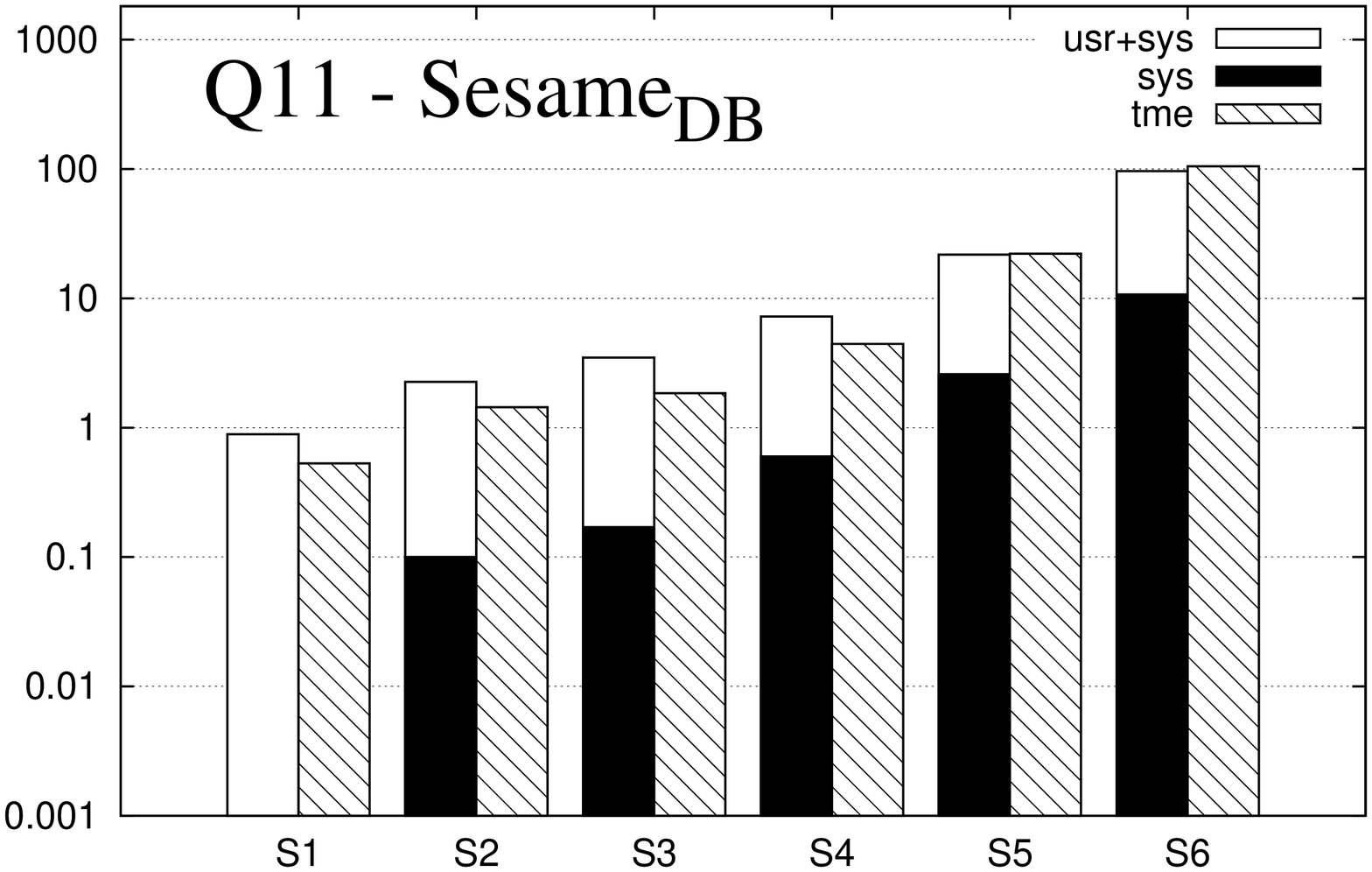}}
&
\hspace{-0.5cm}
\rotatebox{270}{\includegraphics[scale=0.18]{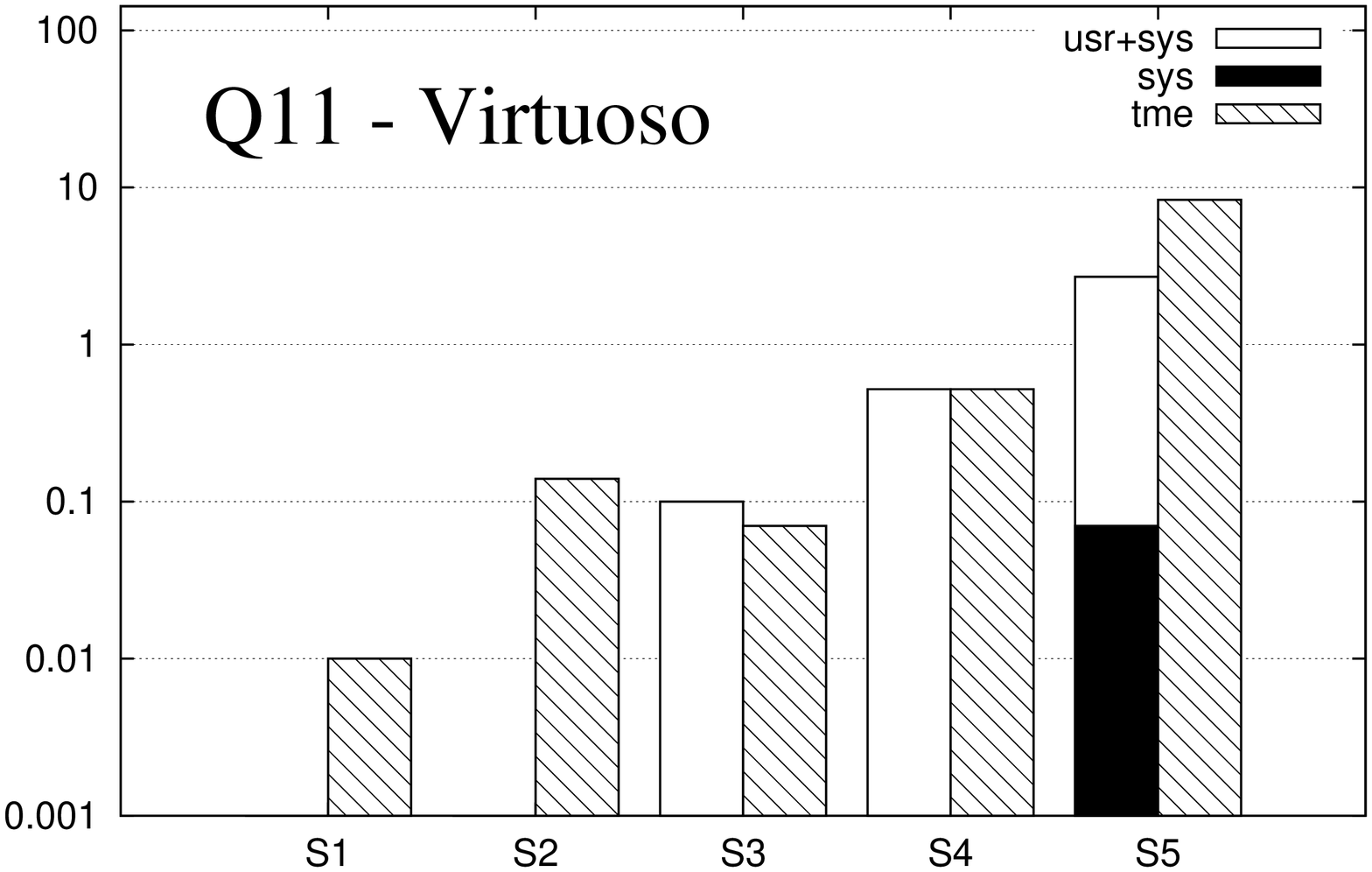}}
\\[-0.65cm]
\rotatebox{270}{\includegraphics[scale=0.18]{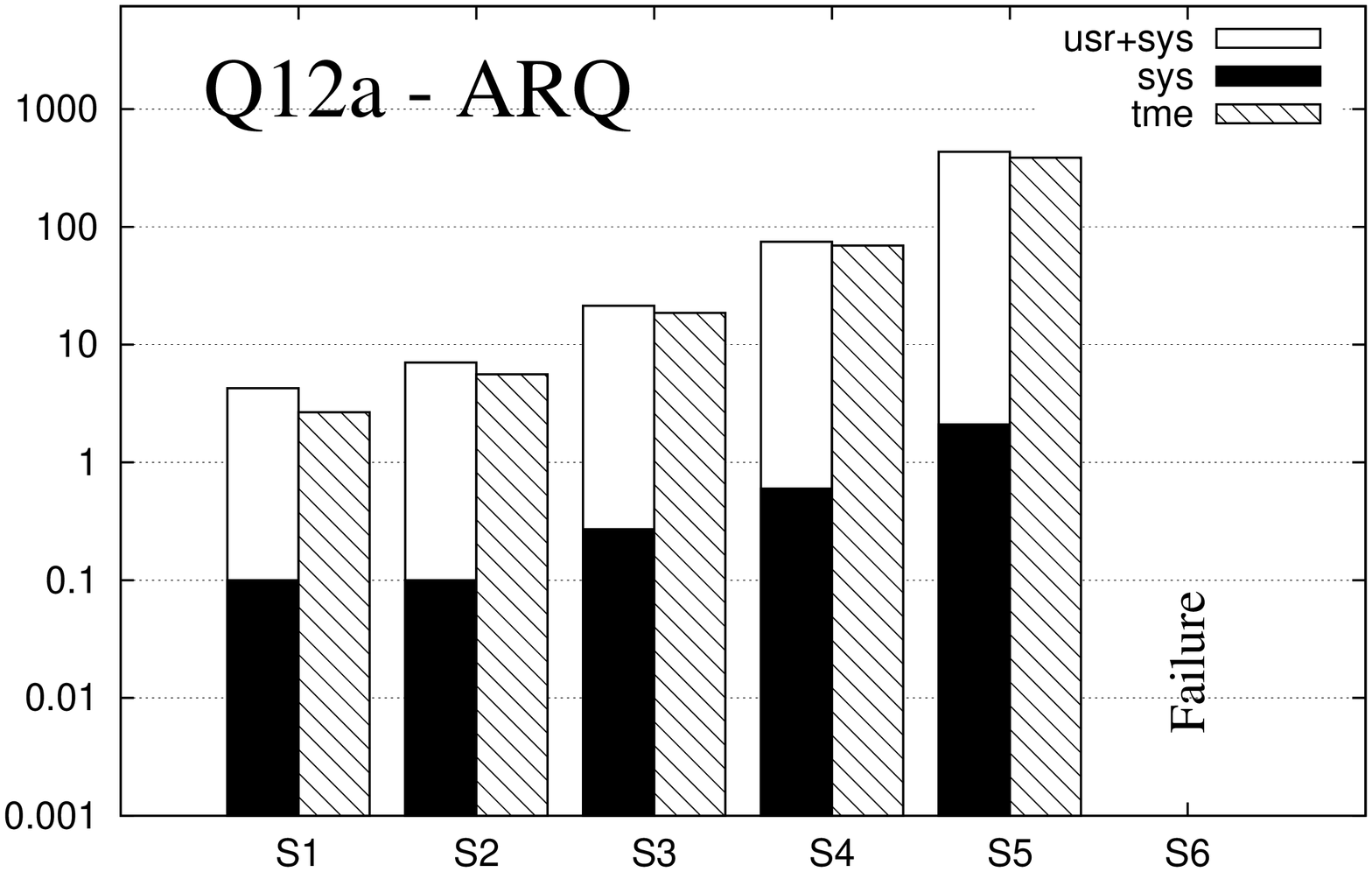}}
&
\hspace{-0.5cm}
\rotatebox{270}{\includegraphics[scale=0.18]{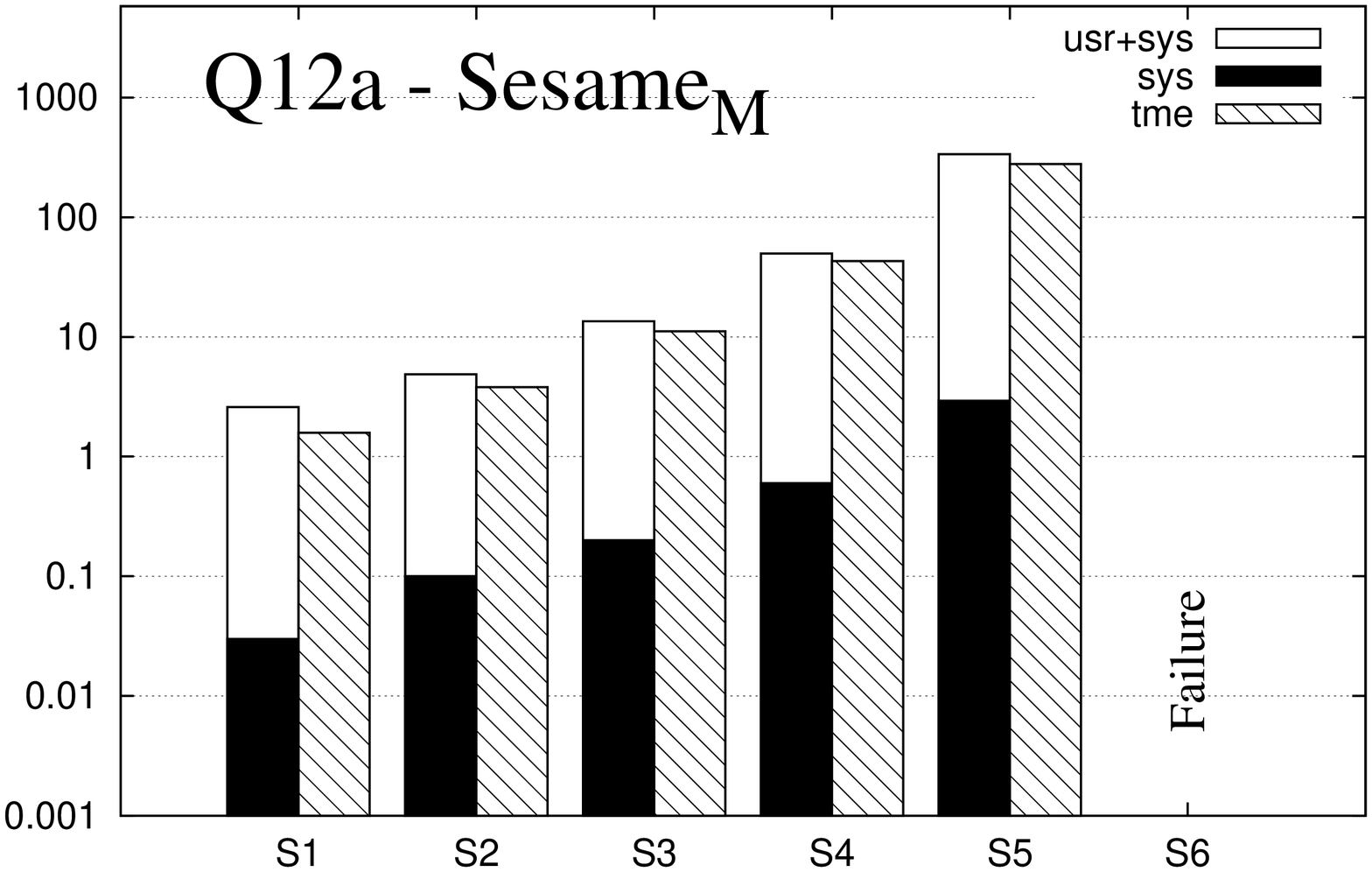}}
&
\hspace{-0.5cm}
\rotatebox{270}{\includegraphics[scale=0.18]{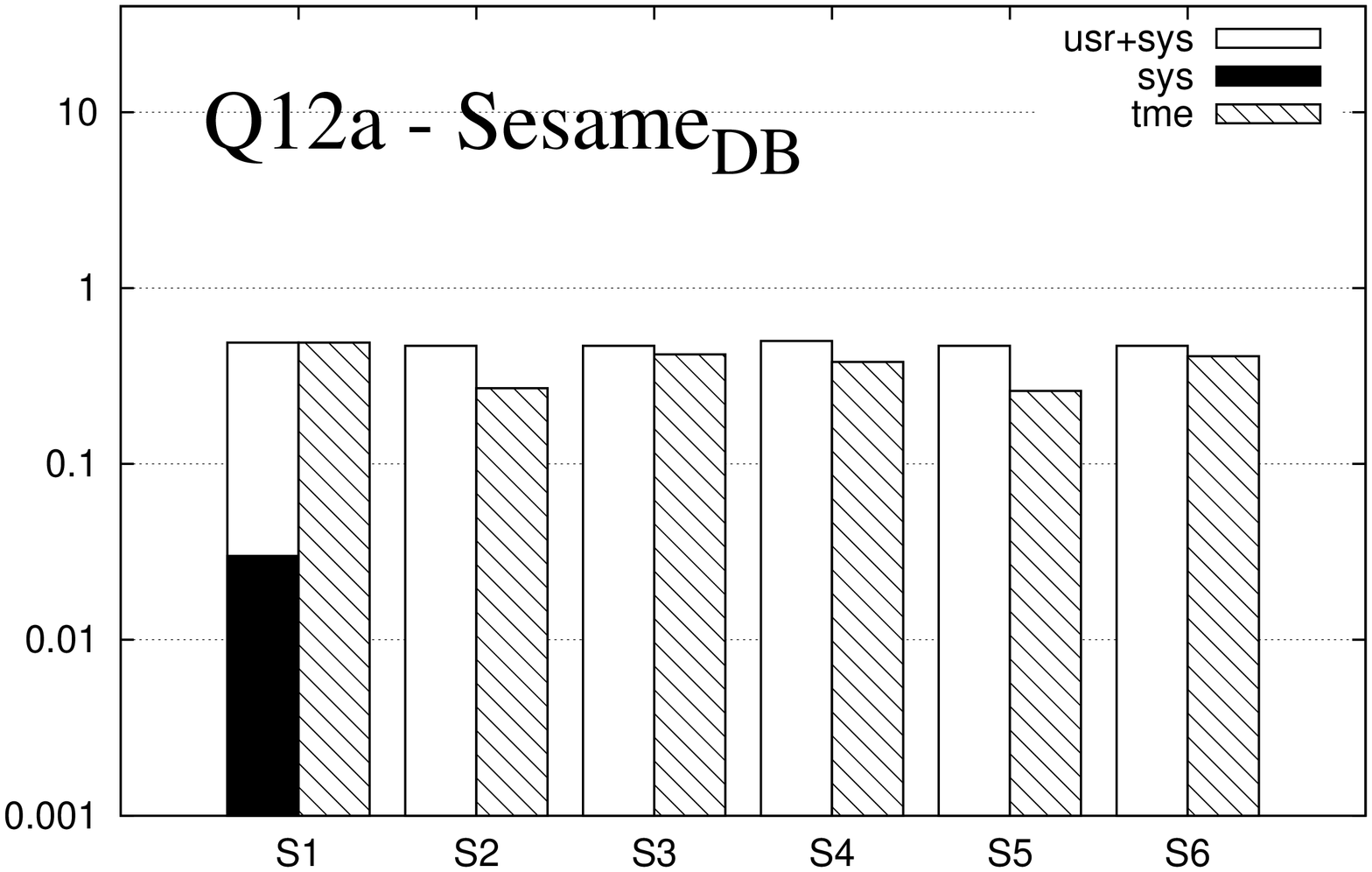}}
&
\hspace{-0.5cm}
\rotatebox{270}{\includegraphics[scale=0.18]{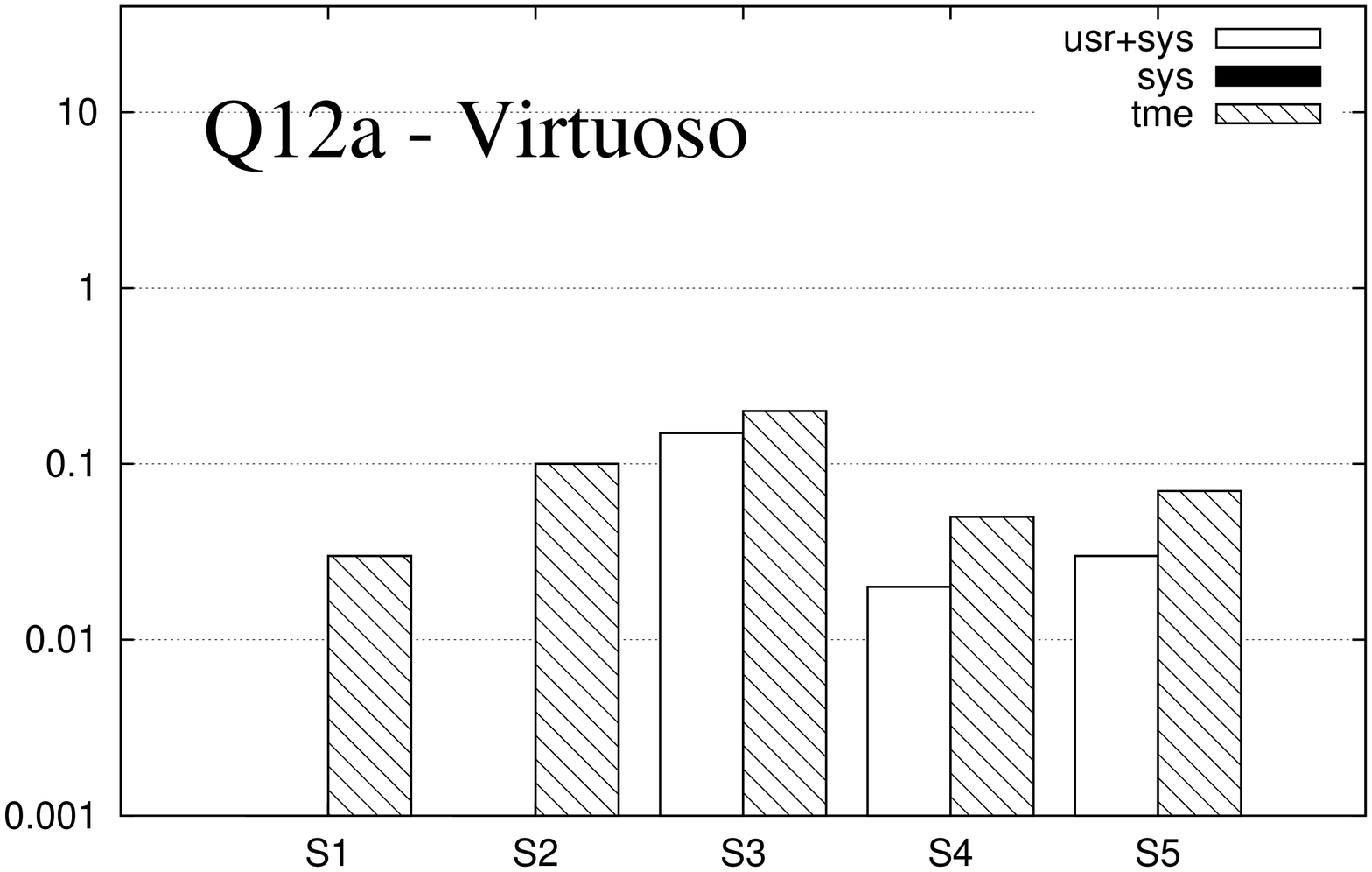}}
\\[-0.65cm]
\rotatebox{270}{\includegraphics[scale=0.18]{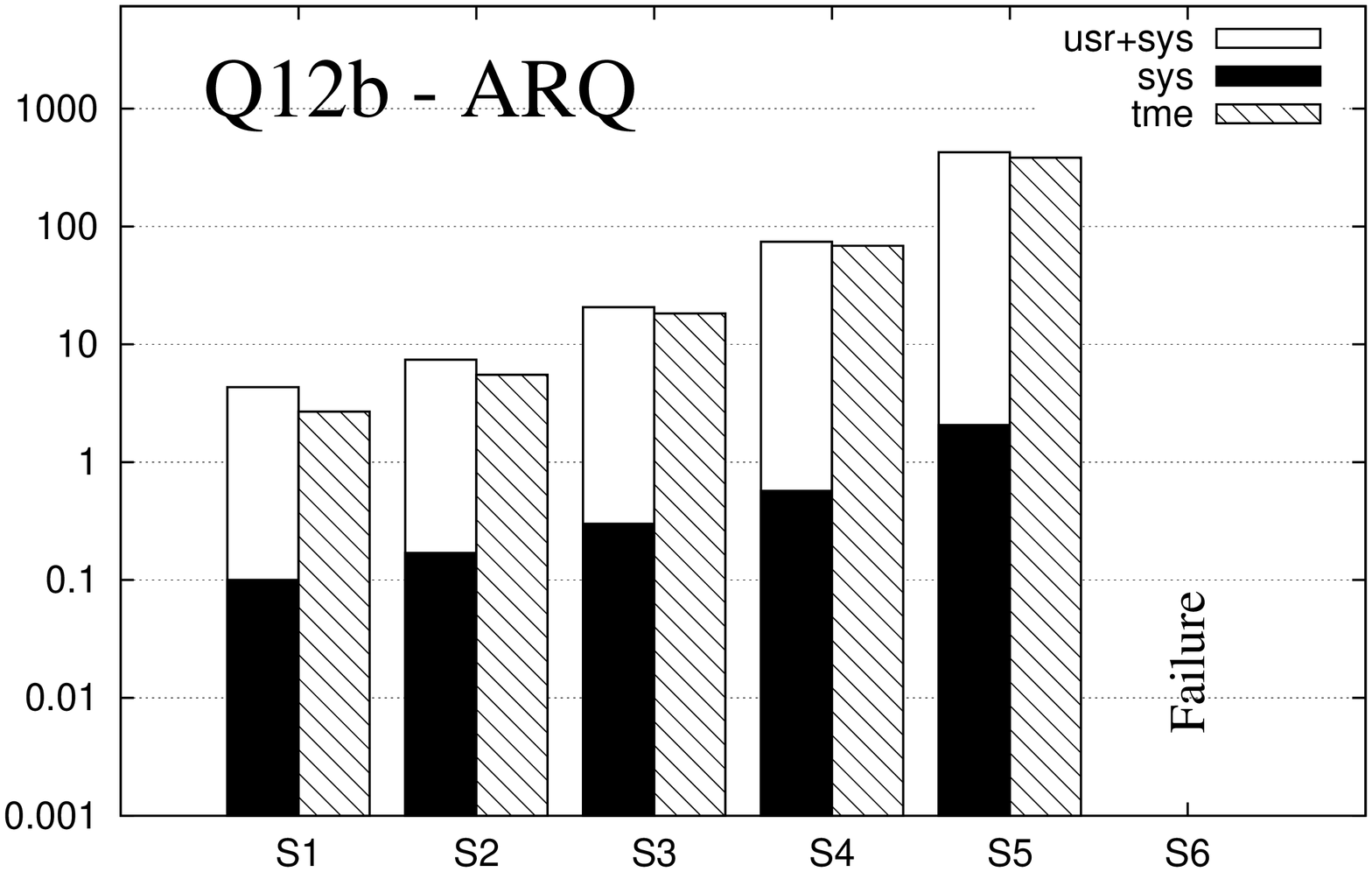}}
&
\hspace{-0.5cm}
\rotatebox{270}{\includegraphics[scale=0.18]{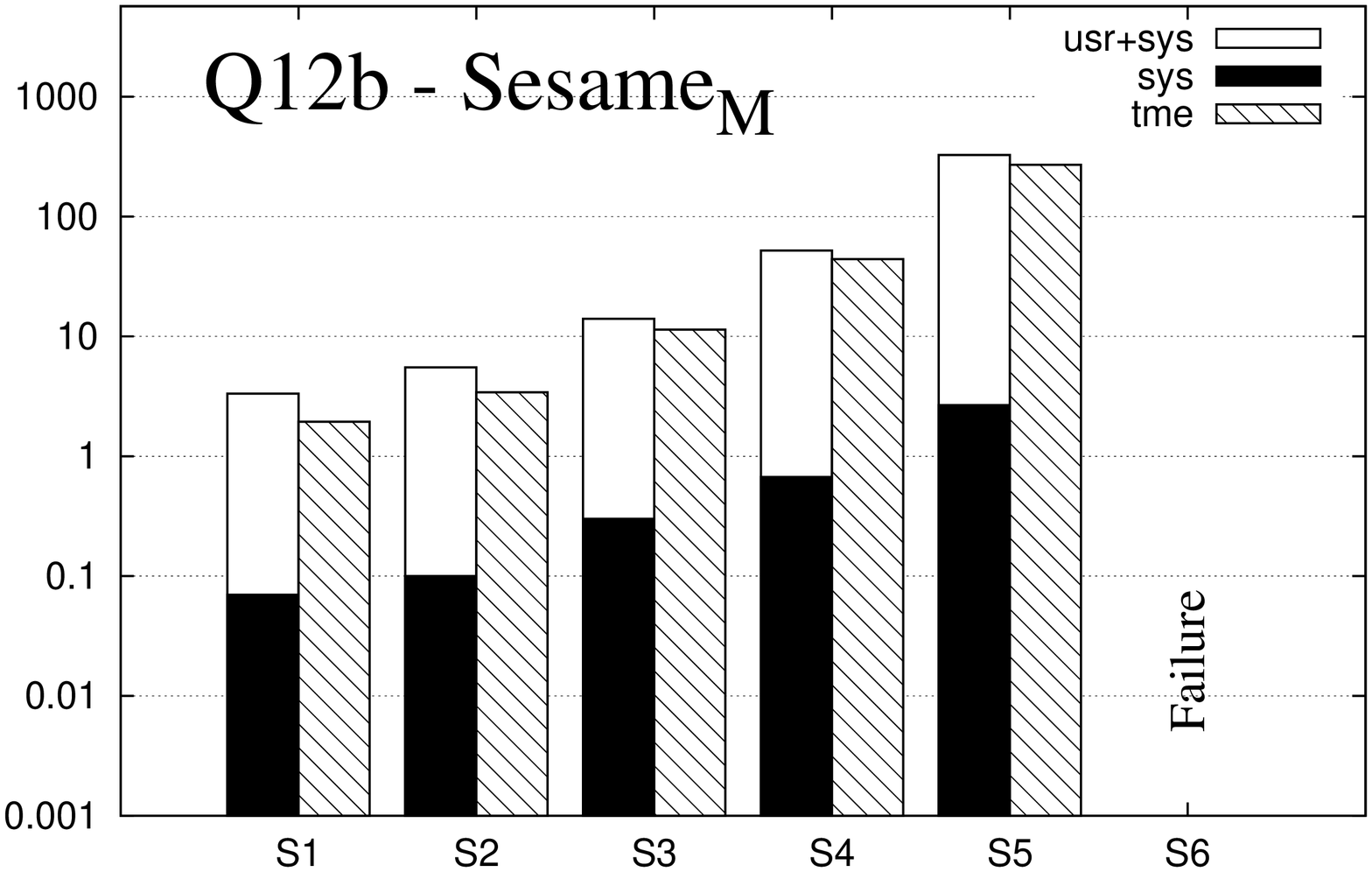}}
&
\hspace{-0.5cm}
\rotatebox{270}{\includegraphics[scale=0.18]{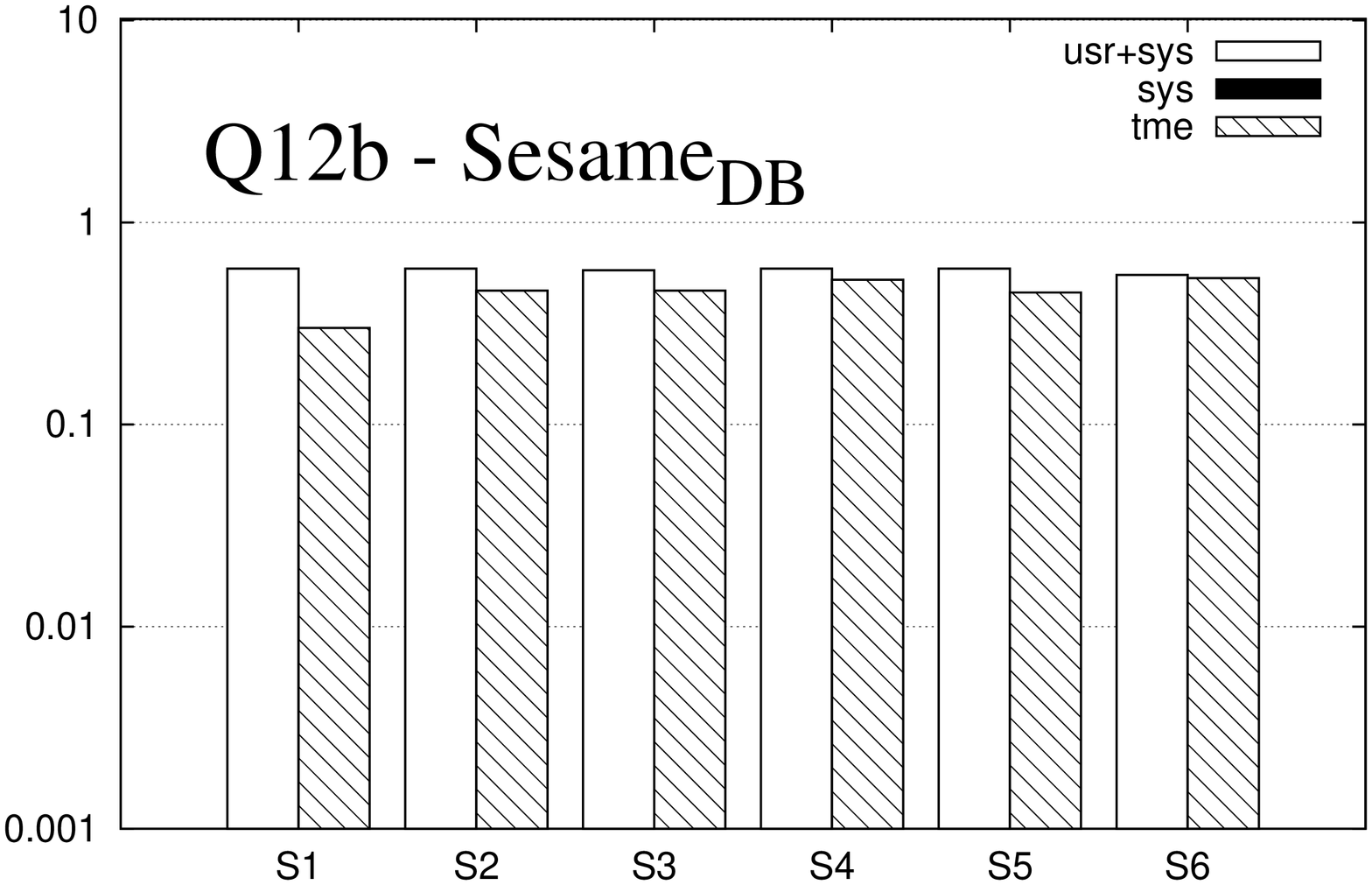}}
&
\hspace{-0.5cm}
\rotatebox{270}{\includegraphics[scale=0.18]{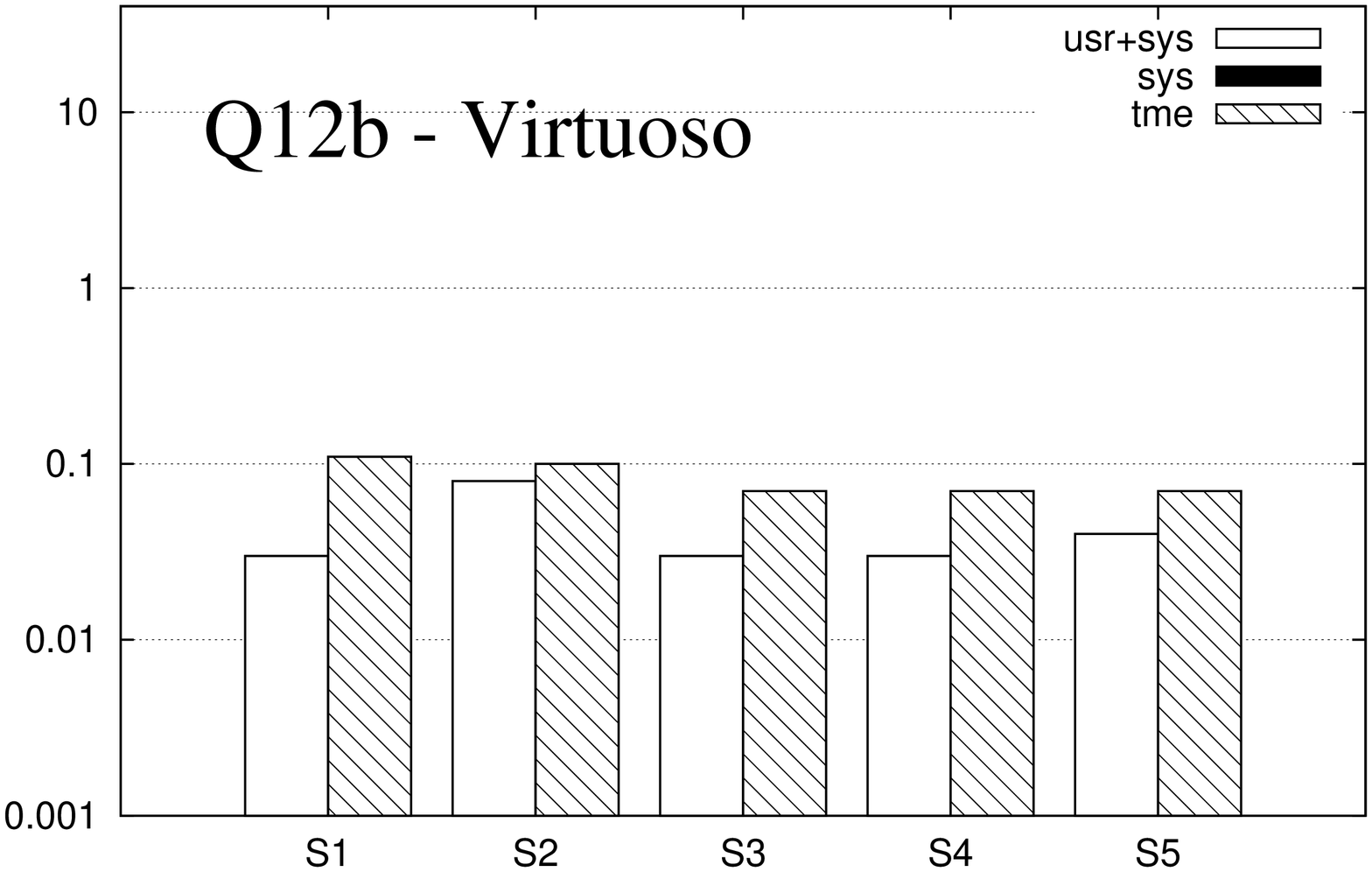}}
\\[-0.65cm]
\rotatebox{270}{\includegraphics[scale=0.18]{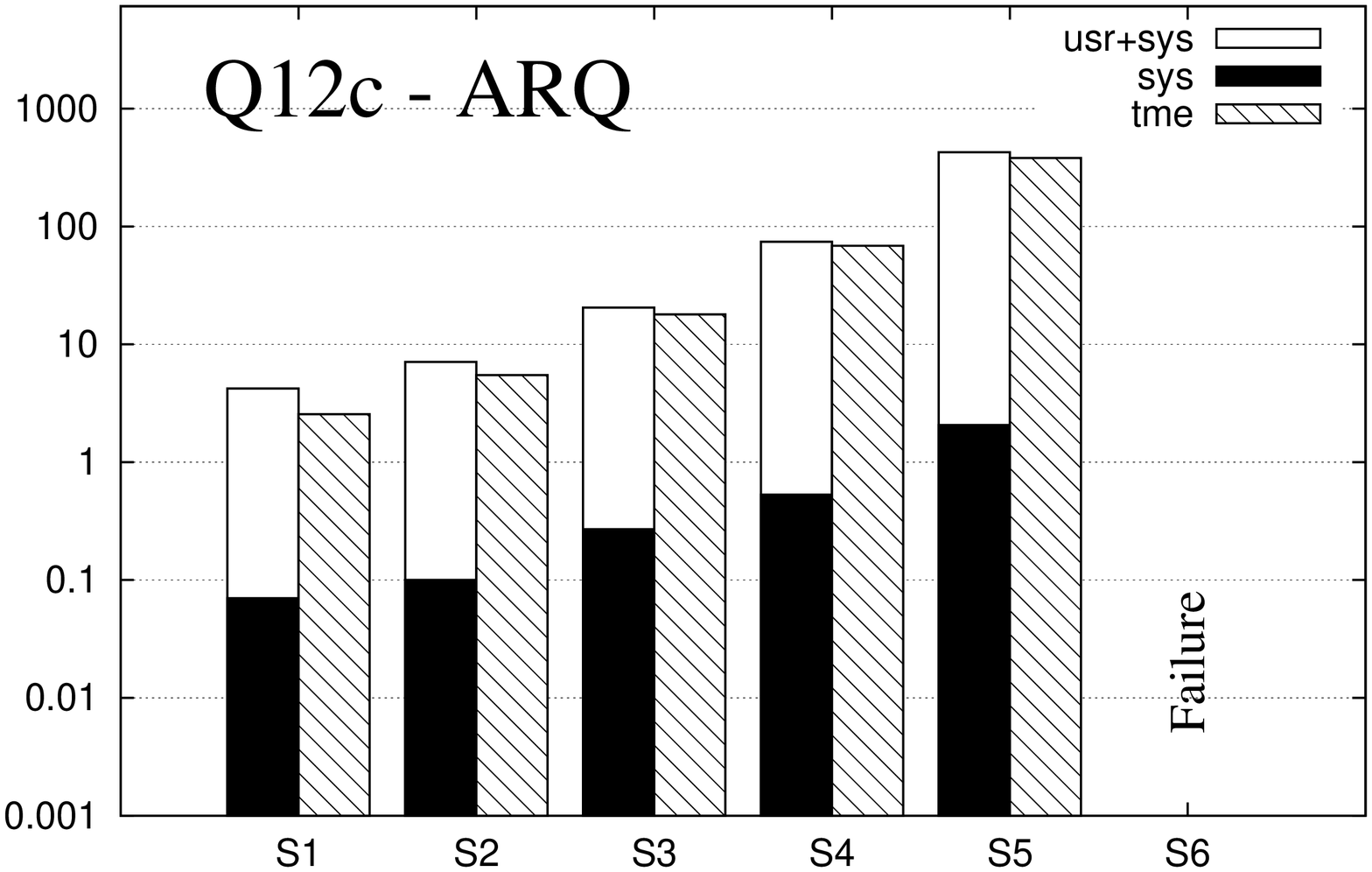}}
&
\hspace{-0.5cm}
\rotatebox{270}{\includegraphics[scale=0.18]{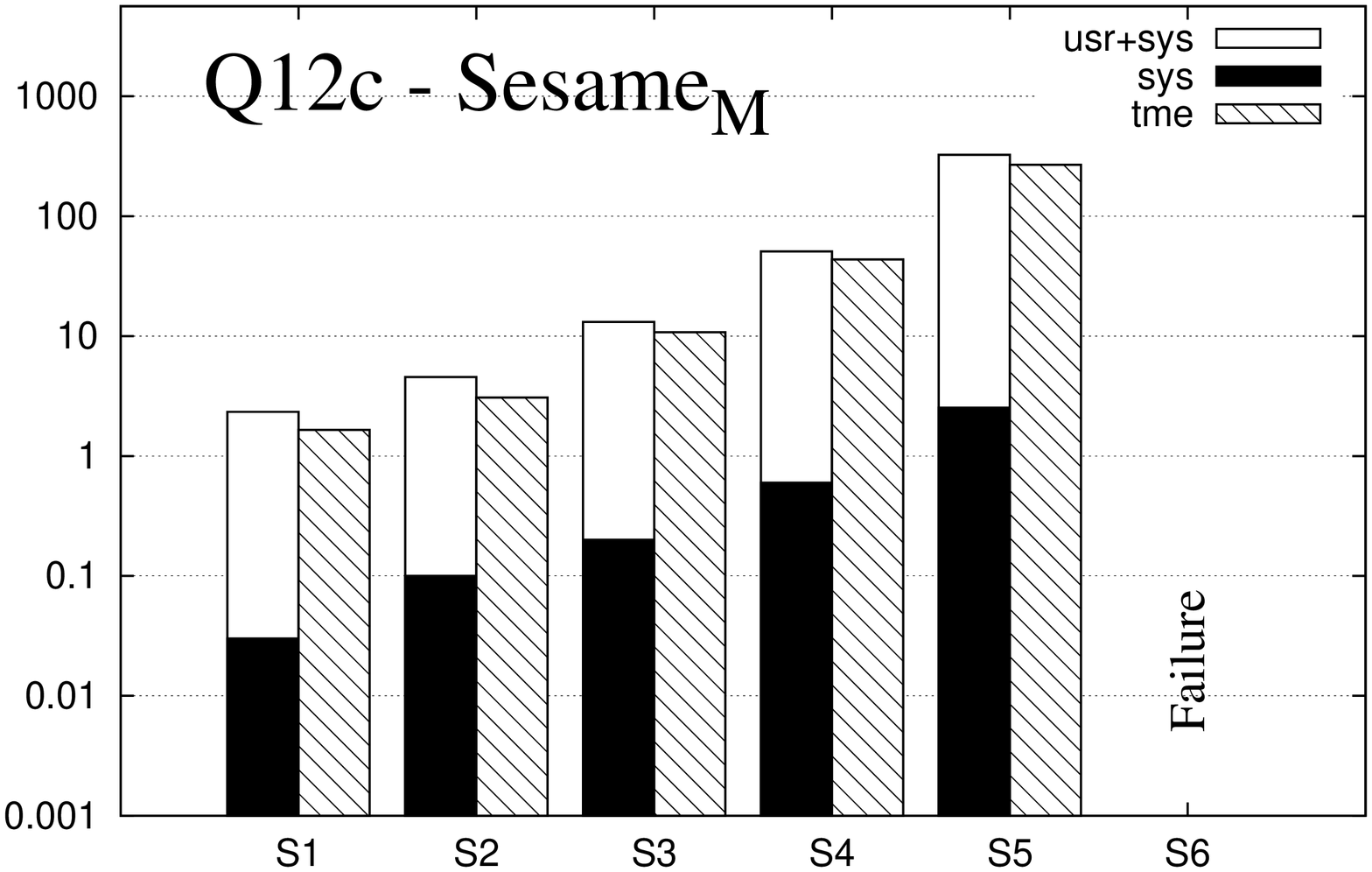}}
&
\hspace{-0.5cm}
\rotatebox{270}{\includegraphics[scale=0.18]{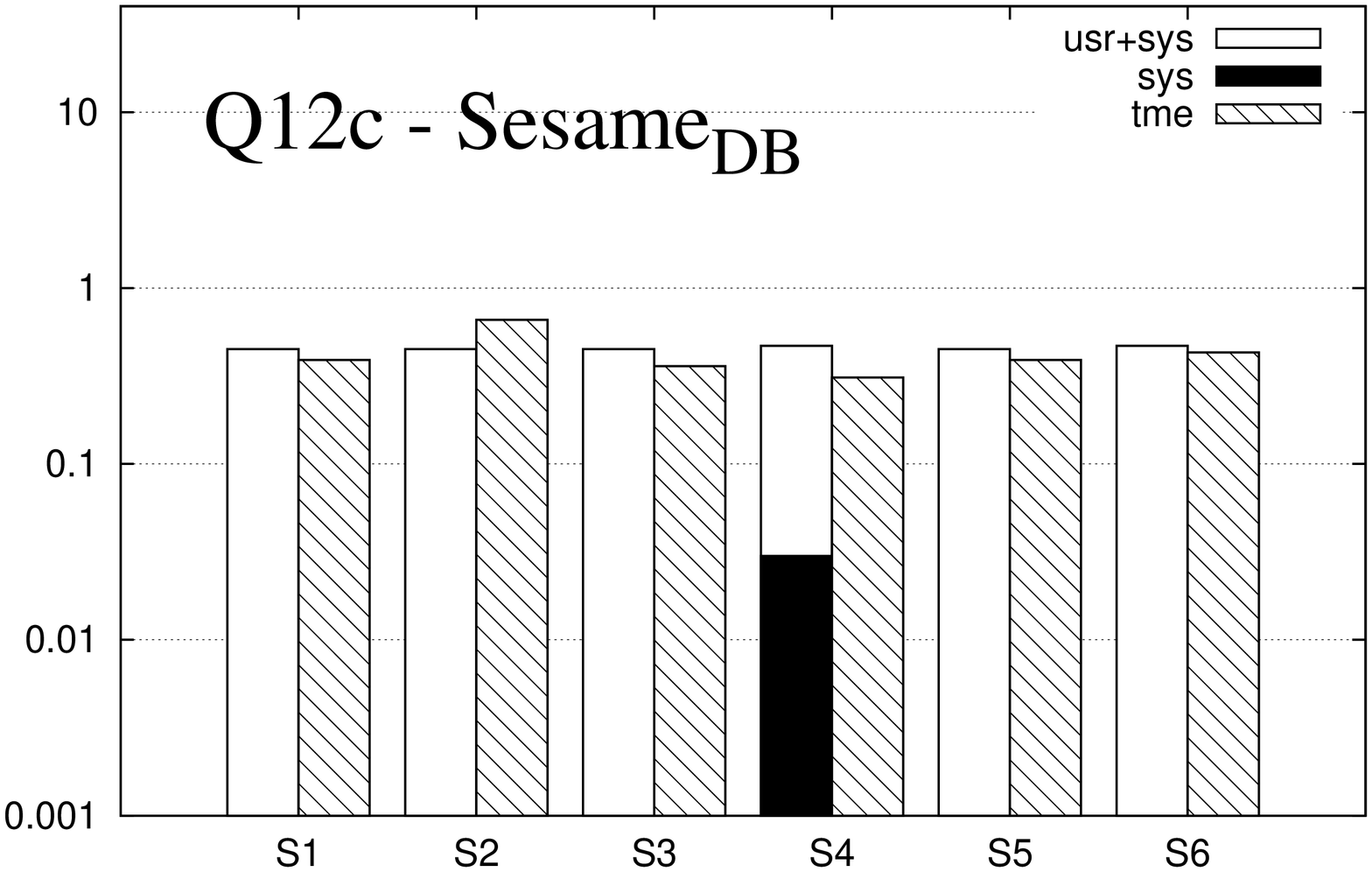}}
&
\hspace{-0.5cm}
\rotatebox{270}{\includegraphics[scale=0.18]{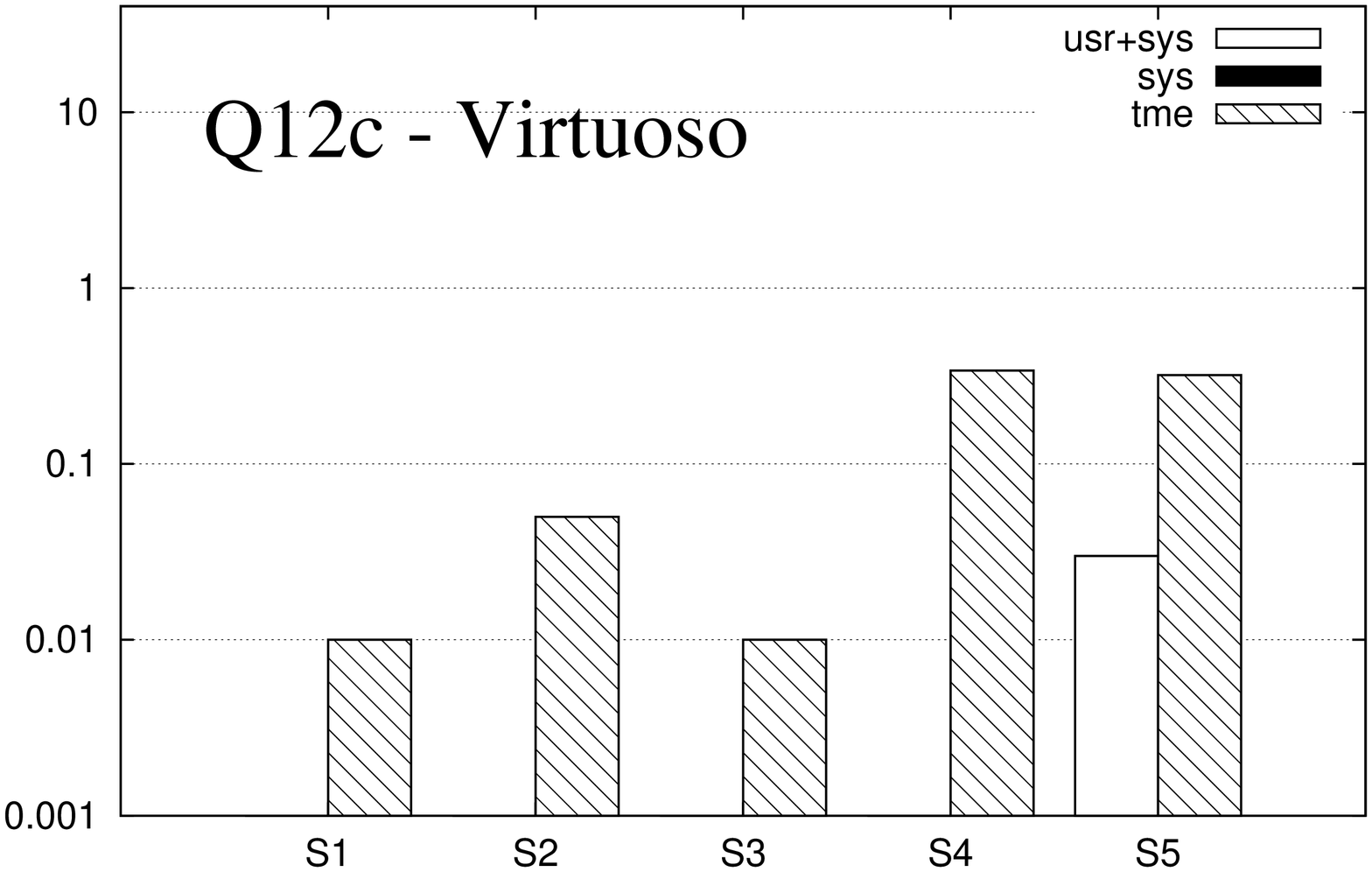}}
\end{tabular}
\caption{Query evaluation results on S1=10k, S2=50k, S3=250k, S4=1M, S5=5M, and S6=25M triples}
\label{fig:experiments3}
\end{figure*}

\section{Attribute Distribution (full table)}
\begin{table*}[ht]
\caption{Probability distribibution for attributes and document classes}
\begin{center}
\begin{tabular}{lcccccccc}
\toprule[.11em]
& {\bf Article} & {\bf Inproc.} & {\bf Proc.} & {\bf Book} & {\bf Incoll.} & {\bf PhDTh.} & {\bf MastTh.} & {\bf WWW}\\
\cmidrule{2-9}
{\bf address} & 0.0000 & 0.0000 & 0.0004 & 0.0000 & 0.0000 & 0.0000 & 0.0000 & 0.0000\\
\midrule
{\bf author} & 0.9895 & 0.9970 & 0.0001 & 0.8937 & 0.8459 & 1.0000 & 1.0000 & 0.9973\\
\midrule
{\bf booktitle} & 0.0006 & 1.0000 & 0.9579 & 0.0183 & 1.0000 & 0.0000 & 0.0000 & 0.0001\\
\midrule
{\bf cdrom} & 0.0112 & 0.0162 & 0.0000 & 0.0032 & 0.0138 & 0.0000 & 0.0000 & 0.0000\\
\midrule
{\bf chapter} & 0.0000 & 0.0000 & 0.0000 & 0.0000 & 0.0005 & 0.0000 & 0.0000 & 0.0000\\
\midrule
{\bf cite} & 0.0048 & 0.0104 & 0.0001 & 0.0079 & 0.0047 & 0.0000 & 0.0000 & 0.0000\\
\midrule
{\bf crossref} & 0.0006 & 0.8003 & 0.0016 & 0.0000 & 0.6951 & 0.0000 & 0.0000 & 0.0000\\
\midrule
{\bf editor} & 0.0000 & 0.0000 & 0.7992 & 0.1040 & 0.0000 & 0.0000 & 0.0000 & 0.0004\\
\midrule
{\bf ee} & 0.6781 & 0.6519 & 0.0019 & 0.0079 & 0.3610 & 0.1444 & 0.0000 & 0.0000\\
\midrule
{\bf isbn} & 0.0000 & 0.0000 & 0.8592 & 0.9294 & 0.0073 & 0.0222 & 0.0000 & 0.0000\\
\midrule
{\bf journal} & 0.9994 & 0.0000 & 0.0004 & 0.0000 & 0.0000 & 0.0000 & 0.0000 & 0.0000\\
\midrule
{\bf month} & 0.0065 & 0.0000 & 0.0001 & 0.0008 & 0.0000 & 0.0333 & 0.0000 & 0.0000\\
\midrule
{\bf note} & 0.0297 & 0.0000 & 0.0002 & 0.0000 & 0.0000 & 0.0000 & 0.0000 & 0.0273\\
\midrule
{\bf number} & 0.9224 & 0.0001 & 0.0009 & 0.0000 & 0.0000 & 0.0333 & 0.0000 & 0.0000\\
\midrule
{\bf pages} & 0.9261 & 0.9489 & 0.0000 & 0.0000 & 0.6849 & 0.0000 & 0.0000 & 0.0000\\
\midrule
{\bf publisher} & 0.0006 & 0.0000 & 0.9737 & 0.9992 & 0.0237 & 0.0444 & 0.0000 & 0.0000\\
\midrule
{\bf school} & 0.0000 & 0.0000 & 0.0000 & 0.0000 & 0.0000 & 1.0000 & 1.0000 & 0.0000\\
\midrule
{\bf series} & 0.0000 & 0.0000 & 0.5791 & 0.5365 & 0.0000 & 0.0222 & 0.0000 & 0.0000\\
\midrule
{\bf title} & 1.0000 & 1.0000 & 1.0000 & 1.0000 & 1.0000 & 1.0000 & 1.0000 & 1.0000\\
\midrule
{\bf url} & 0.9986 & 1.0000 & 0.986 & 0.2373 & 0.9992 & 0.0222 & 0.3750 & 0.9624\\
\midrule
{\bf volume} & 0.9982 & 0.0000 & 0.567 & 0.5024 & 0.0000 & 0.0111 & 0.0000 & 0.0000\\
\midrule
{\bf year} & 1.0000 & 1.0000 & 1.0000 & 1.0000 & 1.0000 & 1.0000 & 1.0000 & 0.0011\\
\bottomrule[.11em]
\end{tabular}
\label{tbl:attributes-tr}
\end{center}
\end{table*}


\end{appendix}

\end{document}